\documentclass[11pt,a4paper,twoside]{report}

\def\theequation{\thesection.\arabic{equation}}

\usepackage{axodraw,epsfig}
\usepackage[latin1]{inputenc}
\usepackage{graphicx}
\def\be{\begin{equation}}
\def\ee{\end{equation}}
\def\ba{\begin{array} }
\def\bac{\begin{array} {c}}
\def\bacc{\begin{array} {cc}}
\def\baccc{\begin{array} {ccc}}
\def\bacccc{\begin{array} {cccc}}
\def\ea{\end{array}}
\def\bea{\begin{eqnarray} }
\def\eea{\end{eqnarray}}
\def\bd{\begin{displaymath}}
\def\ed{\end{displaymath}}
\def\ha{\hat{\alpha}}
\def\hb{\hat{\beta}}
\def\D{\mathcal{D}}
\def\a{\alpha}
\def\b{\beta}
\def\c{\gamma}
\def\d{\delta}
\def\ob{\overline{\omega}}
\def\mv{\mathcal{V}}
\def\e{\mbox{det}\left(e^{\a}_m\right)}
\def\U{\mathcal{U}}
\def\D{\mathcal{D}}
\def\Do{\D^{\bot}}

\renewcommand{\theequation}{\thesection.\arabic{equation}}

\catcode`@=11
\def\marginnote#1{}
\newcount\hour
\newcount\minute
\newtoks\amorpm
\hour=\time\divide\hour by60 \minute=\time{\multiply\hour by60
\global\advance\minute by-\hour}
\edef\standardtime{{\ifnum\hour<12 \global\amorpm={am}%
        \else\global\amorpm={pm}\advance\hour by-12 \fi
        \ifnum\hour=0 \hour=12 \fi
        \number\hour:\ifnum\minute<10 0\fi\number\minute\the\amorpm}}
\edef\militarytime{\number\hour:\ifnum\minute<10 0\fi\number\minute}
\def\draftlabel#1{{\@bsphack\if@filesw {\let\thepage\relax
   \xdef\@gtempa{\write\@auxout{\string
      \newlabel{#1}{{\@currentlabel}{\thepage}}}}}\@gtempa
   \if@nobreak \ifvmode\nobreak\fi\fi\fi\@esphack}
        \gdef\@eqnlabel{#1}}
\def\@eqnlabel{}
\def\@vacuum{}
\def\draftmarginnote#1{\marginpar{\raggedright\scriptsize\tt#1}}
\def\draft{\oddsidemargin 0.0truein
        \def\@oddfoot{\sl preliminary draft \hfil
        \rm\thepage\hfil\sl\today\quad\militarytime}
        \let\@evenfoot\@oddfoot \overfullrule 3pt
        \let\label=\draftlabel
        \let\marginnote=\draftmarginnote
   \def\@eqnnum{(\theequation)\rlap{\kern\marginparsep\tt\@eqnlabel}%
\global\let\@eqnlabel\@vacuum}  } \catcode`@=12
\begin{document}

\thispagestyle{empty}

\setlength{\textwidth}{6.0 in}
\setlength{\evensidemargin}{0.0cm}
 \setlength{\oddsidemargin}{0.0cm}
\def\theequation{\thesection.\arabic{equation}}

\begin{center}
{\LARGE {\bf Aspects of Physics with Two Extra Dimensions}}

\vspace{6cm} {\Large PhD Thesis}

\vspace{2.5cm} {\Large Alberto Salvio
}

\vspace{2cm} {\Large Supervisor: Prof. Seif
Randjbar-Daemi
}

\vspace{3cm}{\Large Academic Year 2005/2006}

\vspace{1cm}{\Large{\it International School for Advanced Studies
(SISSA/ISAS)\\Via Beirut 2-4 34014, Trieste, Italy}}

\end{center}

\newpage

\setlength{\oddsidemargin}{0.0cm}
\newpage

\vspace*{\fill}
\newpage

\vskip 10cm $   \qquad \qquad \qquad \qquad \qquad \qquad \qquad
\qquad \qquad \qquad \qquad \qquad ${\it {\Large To my parents}}

\newpage

\vspace*{\fill}
\newpage

\chapter*{Abstract}

In this thesis we discuss some aspects concerning the construction
of a 4D effective theory derived from a higher dimensional model.
The first part is devoted to the study of how the heavy Kaluza-Klein
modes contribute to the low energy dynamics of the light modes. The
second part concerns the analysis of the spectrum arising from non
standard compactifications of 6D minimal gauged supergravities,
involving a warp factor and conical defects in the internal manifold.

To prepare the background for such topics, first we review standard
Kaluza-Klein theories and brane world models.

 Afterwards, in Part I, which contains original results, we
introduce the study of the heavy mode contribution. We do so by
discussing scalar models in arbitrary dimension and then by treating
in some detail a 6D Einstein-Maxwell theory coupled to a charged
scalar and fermions. The latter model has some interesting features
as it can lead to a chiral low energy 4D effective theory, which is
similar to the electroweak part of the Standard Model of Particle
Physics. In this first part of the thesis our main interest is in
the interaction terms. We point out that the contribution of the
heavy KK modes is generally needed in order to reproduce the correct
predictions for the observable quantities involving the light modes.
In the 6D Einstein-Maxwell-Scalar model the contribution of the
heavy KK modes are geometrically interpreted as the deformation of
the internal space.

In Part II we introduce 6D minimal gauged supergravities, which are
supersymmetric and non-Abelian extensions of the 6D model of Part I.
We begin by summarizing the main features and possible applications
of these models. Moreover we review warped brane solutions with 4D
Poincar\'e invariance, and a compact and axisymmetric internal
manifold, which, in a simple case, turns out to have conical
defects. Afterwards we present our original contribution to such a
scenario: we study fluctuations about these axisymmetric warped
brane solutions. Much of our analysis is general and could be
applied to other scenarios. We focus on bulk sectors that could give
rise to Standard Model like gauge fields and charged matter. We
reduce the dynamics to Schroedinger type equations plus physical
boundary conditions, and obtain exact solutions for the Kaluza-Klein
wave functions and discrete mass spectra.  The power-law warping, as
opposed to exponential in 5D, means that zero mode wave functions
can be peaked on negative tension branes, but only at the price of
localizing the whole Kaluza-Klein tower there.  However, the
codimension two defects allow the Kaluza-Klein mass gap to remain
finite even in the infinite volume limit.  In principle, in this
scenario, not only gravity, but Standard Model fields could `feel'
the extent of large extra dimensions, and still be described by an
effective 4D theory.

 \tableofcontents
\newpage

\phantom{}
\newpage

\addcontentsline{toc}{chapter}{Introduction}
\chapter*{Introduction}

The Standard Model (SM) of the electroweak and strong interactions,
including neutrino masses, and the Einstein's theory of gravity
successfully predict most of
the physical quantities that we are able to
measure. The former is the theoretical framework in which we usually
study elementary particles physics and it consistently
includes the principles of quantum mechanics and special relativity.
The latter describes large scale structures such as the expanding
universe and its history.

One of the main attractive features of the SM is the formulation of
the non gravitational interactions as a consequence of a local
invariance principle, generalizing the gauge invariance of
electrodynamics. This leads to two additional fine structure
constants associated to the weak and strong nuclear forces. In a
similar way the Einstein's theory, or General Relativity (GR),
includes gravitational interactions by requiring invariance under
general space-time coordinate transformations.

Despite this similarity so far we do not have a complete and well
understood quantum theory of gravity, which is valid at every energy
scale. The most promising attempt to quantize GR is the Superstring
Theory, which includes a graviton state in its physical spectrum.
Moreover, Superstrings, having just one independent parameter, give
hope to solve the theoretical problem of explaining the ad hoc
structure of the gauge group and the quantum numbers of the SM. A
consistent formulation of such theory requires a space-time
dimensionality equal to 10 (or 11 in the case of M-Theory),
necessarily leading to extra dimensions. The idea that we can live
in a world with more than 4 dimensions was proposed by Kaluza and
Klein before Superstring theory in order to unify gravitational and
electromagnetic interactions. Indeed they  analyzed a 5D Einstein's
theory of gravity and proved that in the low energy limit this
theory gives the 4D Einstein's theory and the Maxwell's theory of
electrodynamics, if the extra dimension is {\it compactified}.

 Now it is known that, remarkably, the presence of extra dimensions can help us to
solve two longstanding problems of theoretical physics, which are
related to the structure of the underlying quantum theory of
gravity: the hierarchy problem and the cosmological constant
problem. They both are fine-tuning problems because the former is
related to the huge difference between the Planck scale and the
electroweak scale, and the latter to the very small value of the
observed vacuum's energy compared to the other energy scales (apart
from neutrino masses).

Since we observe only 4 dimensions in our world, higher dimensional
theories require a mechanism to hide extra dimensions. In the standard Kaluza-Klein (KK) theories the extra dimensions are invisible
because they describe a smooth compact manifold with very small size,
naturally of the order of the Planck length. More recently it has
been proved that this is not actually necessary because SM
fields can be localized on a (1+3)-dimensional sub-manifold, called 3-brane.
However, every higher dimensional theory leads to an infinite number
of particles from the 4D point of view. It is of
course interesting to know whether or not it is possible to
construct a 4D effective theory for the lightest
particles, and, when this is possible, to study the role of the heavy particles in the low energy
dynamics.

In the standard KK theories a higher dimensional field can be
decomposed as an infinite but discrete sequence of 4D modes (KK
modes), due to the compactness of the internal space. This sequence
is called KK tower and the difference between two consecutive masses
is of the order of the inverse proper radius of the internal space.
However, in some cases a finite {\it mass gap} between the zero (or
lightest) modes and the heavy KK modes can emerges even if the
internal manifold is not compact. If this happens it seems clear
that a low energy theory for the zero modes can be constructed by
using the effective theory approach, namely {\it integrating out}
the heavy degrees of freedom.

In an original part of the present thesis we will study if the heavy
KK modes can actually give a measurable contribution to the SM
particles dynamics by assuming a finite mass gap. In particular we
shall analyze the broken phase of the 4D effective theory, when the
light modes acquire a vacuum expectation value (VEV), and the
corresponding low energy mass spectrum. We will call this method the
effective theory approach to spontaneous symmetry breaking (SSB). In
order to study the heavy mode contribution to these observable
quantities we shall also perform an alternative approach to SSB, in
which we go directly to the broken phase by means of a solution of
the higher dimensional EOM. We shall call the latter approach the
{\it geometrical} approach to SSB. The final result will prove that
in general the heavy mode contribution to the 4D effective theory is
not negligible, because this is actually needed to reproduce the
geometrical approach. Remarkably we will be able to prove that this
is true even if the heavy modes are as massive as the Planck mass,
contrary to the standard lore for which the physics at the Planck
scale is not relevant for the dynamics at a lower scale, for
instance at the electroweak scale.

In a second original part of the present thesis we will consider a
higher dimensional (in this case 6D) supergravity compactified on a
manifold with singularities, which are not usually introduced in
original KK theories. The specific model that we consider has
interesting features in relation with both the hierarchy and the
cosmological constant problem. Indeed recently Supersymmetric Large
Extra Dimensions have been proposed as a possible scenario in which
the cosmological constant can be {\it self-tuned} to the observed
value if the dimension of the internal space is equal to 2. This is
essentially due to a numerical coincidence between the inverse
invariant radius of the internal space and the observed vacuum
energy if the fundamental 6D Planck scale is at the electroweak
scale. However, in order to implement this mechanism supersymmetry
must be broken to allow a small but non vanishing cosmological
constant. The class of solutions that we will consider are
particularly interesting in this contest because they break
supersymmetry completely. Such configurations present a {\it warp
factor} and an internal manifold with 2 {\it conical singularities}
and with the sphere topology. We shall study exactly and in great
detail the KK towers for fermion and gauge field sectors that can
contain SM fields. Moreover we will study the effect of conical
singularities on such towers, finding that the mass gap is not
necessarily equal to the inverse proper radius as in ordinary KK
theories, with interesting application to Large Extra Dimensions.

The present thesis is organized as follows.

In Chapter \ref{KKbraneModels} we will review standard KK theories
and brane models by focusing on the details that are useful to
understand the original work. In particular we will construct the 4D
effective theory starting with the original 5D KK theory, we shall
introduce the brane world idea by discussing the simple {\it kink}
domain wall and finally we will summarize the main features of Large
Extra Dimensions and Randall-Sundrum scenarios, which are both
motivated by the hierarchy problem.

Afterwards we will present Part I of the thesis, which contains the
original results on the heavy mode contribution to the low energy
dynamics. This part is divided in Chapters \ref{GHBscalar} and
\ref{EMS}. In the former we will study the heavy mode contribution
to the 4D effective theory in a simple scalar set up. This will
allow us to explain in a rigorous way what we have proved and our
steps. In Chapter \ref{EMS} we will address the same problem but in
a physically interesting contest. We will start with a 6D
Einstein-Maxwell-Scalar model and consider the compactification over
the sphere with a monopole background. This will lead to a chiral
$SU(2)\times U(1)$ effective theory with an {\it Higgs field}
triggering SSB from $SU(2)\times U(1)$ to $U(1)$. In this framework,
which is similar to the electroweak part of the SM, we will show
that the heavy KK modes, with masses of the order of the Planck
scale, actually have effects on some measurable quantities of the
low energy physics.

Part II is divided in Chapters \ref{6DSUGRA} and
\ref{SUGRAspectrum}. In the former we will review 6D supergravity
focusing on the case in which we have the minimal number of
supersymetries (minimal supergravity) and part of the R-symmetry
group is gauged (gauged supergravity). Moreover we will review the
Supersymmetric Large Extra Dimensions scenario and some singular
solutions, which break completely supersymmetry. In Chapter
\ref{SUGRAspectrum} we will present our original contribution in
this scenario studying gauge field and fermion fluctuations around
non-smooth solutions. We will emphasize that these sectors can
contain SM fields and we shall study the effect of (conical)
singularities on the mass gap. Moreover we will analyze the
particular case in which both the internal proper radius and the
mass gap are large and the effect of this set up on the fundamental
(6D) constants.

There are also three appendices. In Appendix \ref{GR} we give our
conventions and notations, including recurrent abbreviations. In Appendix \ref{AppGHB} we give the
explicit computation of the 4D spectrum coming from the 6D
Einstein-Maxwell-Scalar model. In Appendix \ref{AppSugra} we provide
some extra calculations concerning the 6D minimal gauged
supergravity and we perform a stability analysis, considering all
the present known anomaly free models of this type.

\chapter{Kaluza-Klein and Brane Models}\label{KKbraneModels}

In this chapter we review {\it KK theories} and {\it brane models}
by focusing on the aspects that are relevant for the original
contribution of this thesis presented in Chapters \ref{GHBscalar},
\ref{EMS}, and \ref{SUGRAspectrum}. The composition is as follows.
In Section \ref{KK} we introduce KK theories by studying the simple
example of 5D Einstein gravity compactified over $S^1$. In Section
\ref{DW} we discuss the {\it kink} domain wall configuration that
localizes matter fields on a {\it 3-brane}. We will devote Section
\ref{LED} to models with large extra dimensions, discussing in
particular their relation with the hierarchy problem and their
phenomenological implications. In Section \ref{RS} we will discuss
5D {\it warped} brane world models and the possibility of creating
mass hierarchy in this contest. Furthermore, in Section
\ref{CCproblem} we review attempts that aim at a solution of the
cosmological constant problem in theories with extra dimensions.
Finally we will specifically consider 6D models in Section
\ref{C2B}, where we will study {\it codimension 2} branes and their
conical nature.

\section{Kaluza-Klein Theories}\label{KK}

The original motivation for studying field theories in space-time
with more than 4 dimensions is to obtain a geometrical
interpretation of internal quantum numbers such as the electric
charge, that is to place them in the same context as energy and
momentum \cite{Kaluza:1921tu,Klein:1926tv}. The latter observable
quantities are associated with translational symmetry in
$(Minkowski)_4$, the 4D Minkowski space-time, whereas the internal
observable quantities would be associated with symmetry motions in
the extra dimensions.

In theories of the standard KK type\footnote{For a review on this
topic see \cite{Appelquist:1987nr}.} one assumes a $D$-dimensional
($D>4$) generally covariant field theory and, by some dynamical
mechanism, obtains a partially compactified and {\it factorizable}
background geometry,
\be M_4\times K_d,\label{factorizable}\ee
where $M_4$ is a 4D pseudo euclidean manifold, and $K_d$ is a
$d$-dimensional euclidean smooth compact manifold. The proper volume
$V_d$ of $K_d$ must be sufficiently small to render the extra
dimensions invisible. For instance for $V_d^{1/d}<10^{-17}cm$, we
expect the effects of $K_d$ to be invisible up to energies of the
order\footnote{We use the following conversion relation:
$(TeV)^{-1}=10^{-17}cm$.} of $TeV$. However, in order to support
such compactification, extra matter fields in general are needed and
therefore a completely geometrical explanation of the fundamental
forces can be lost in this process. An interesting exception could
be the anomaly free higher dimensional supergravities, in which
supersymmetry and anomaly freedom can motivate the presence of
additional matter fields.

The original works by Kaluza and Klein analysed a particular example
of such a framework: the standard 5D Einstein-Hilbert theory
compactified on $M_4\times S^1$. The action of this model is
\be S=\frac{1}{\kappa^2}\int d^5 X \sqrt{-G}R,\label{EH}\ee
where $\kappa$ is a 5D Planck scale and other conventions are given in Appendix \ref{GR}.
Compactification on $S^1$ means the physical equivalence
\be y\sim y+L,\label{periodicity}\ee
where $y$ is the fifth coordinate and $L$ is the circumference of
$S^1$. The five-dimensional metric separates into $G_{\mu \nu}$,
$G_{\mu 5}$, and $G_{55}$. From the 4D point of view these are a
metric, a vector, and a scalar. We can parametrize the metric as
\be ds^2=G_{MN}dX^M dX^N= g_{\mu \nu}dx^{\mu} dx^{\nu}+G_{55}\left(dy + A_{\mu}dx^{\mu}\right)^2. \label{paramKK}\ee
If $g_{\mu \nu}$, $G_{55}$, and $A_{\mu}$ depend on both $x$ and $y$, (\ref{paramKK}) is the most general 5D metric,
but henceforth we will assume they depend only on the noncompact coordinates $x$ and, in this case,
(\ref{paramKK}) is the most general
metric invariant under $x$-dependent translations of $y$. That is, this form still allows the following reparametrizations
\bea x^{\mu}&\rightarrow & x'^{\mu}(x) \nonumber \\
y&\rightarrow & y + \Lambda(x), \eea
and under the latter
\be A_{\mu}\rightarrow A_{\mu}-\partial_{\mu}\Lambda.\label{KKgauge}\ee
So gauge transformations arise as part of the higher-dimensional coordinate group. This is the KK mechanism.

To see the effect of y-dependence, consider a massless complex scalar $\Phi$
in 5D.
Relation (\ref{periodicity}) can be implemented by requiring $\Phi$ to be periodic with respect to $y$.
Expanding the y-dependence of $\Phi$
in a complete set we have
\be \Phi(X)=\sum_m \Phi_m(x)e^{i 2\pi m y/L}, \label{FourierKK}\ee
where $m$ is an integer.
The momentum in the periodic dimension is quantized $p_y=2\pi m/L$.
The action for such a scalar is
\be S_{\Phi}=-\int d^5X \sqrt{-G}\partial_M\Phi^{\dagger}\partial^{M}\Phi. \ee
By using (\ref{FourierKK}), and $\sqrt{-G}=\sqrt{-g}e^{\phi/2}$,
where $g$ is the determinant of $g_{\mu \nu}$ and $e^{\phi}=G_{55}$,
we have
\be S_{\Phi}=-Le^{<\phi>/2}\sum_m\int d^4x\sqrt{-g}\left[\partial_{\mu}\Phi_m^{\dagger}\partial^{\mu}\Phi_m+
\left(\frac{2\pi m }{Le^{<\phi>/2}}\right)^2\Phi^{\dagger}_m\Phi_m\right]+...,\ee
where the dots represent interaction terms.
Therefore $\Phi$ contains an infinite tower (KK tower) of 4D fields with squared mass
\be M^2_m=\left(\frac{2\pi m}{V_1}\right)^2, \ee
where $V_1\equiv Le^{<\phi>/2}$ represents the invariant volume of
the internal space $S^1$. We observe that the gap between two
consecutive KK masses is fixed by the volume of the internal space.
This is a general property of standard KK theories. Therefore, it
seems that at energies small compared to $V_1^{-1}$ only the zero
modes ($M^2=0$) can be physically relevant. However, integrating out
the heavy modes in general gives a non-trivial contribution to the
low energy dynamics. We will clarify this point in Chapters
\ref{GHBscalar} and \ref{EMS}, which contains part of the original
work \cite{Randjbar-Daemi:2006gf}.

The charge corresponding to the KK gauge invariance (\ref{KKgauge})
is the $p_y$-momentum. In this simple example, all fields carrying
the KK charge are massive. More generally there can be massless
charged fields. We will provide an example in Chapter \ref{EMS},
where we will discuss a 6D Einstein-Maxwell-Scalar model
compactified on $(Minkowski)_4\times S^2$ with a non vanishing gauge
field background. In this model the KK gauge group is $SU(2)$ and
there are massless 4D fields in non-trivial $SU(2)$-representations.

We compute now the 4D effective action for the zero modes by putting the ansatz (\ref{paramKK}) in the Einstein-Hilbert
term (\ref{EH}). The 5D Ricci scalar can be expressed in terms of the scalar field $\phi$,
the field strength $F_{\mu \nu}$ of $A_{\mu}$, and the 4D metric $g_{\mu \nu}$:
\be R=R(g_{\mu \nu})-2e^{-\phi/2}\nabla^2 e^{\phi/2} -\frac{1}{4}e^{\phi}F_{\mu \nu}F^{\mu \nu}. \ee
Therefore the effective action for the zero modes is\footnote{The
kinetic term for $\phi$ appears if one performs the Weyl
transformation $g_{\mu \nu}\rightarrow e^{-\phi/2}g_{\mu \nu}$,
which converts the gravitational term in the standard
Einstein-Hilbert form.}
\be S_{eff}=\frac{L}{\kappa^2}
\int d^4 x \sqrt{-g}e^{\phi/2}\left(R(g_{\mu \nu})-\frac{1}{4}e^{\phi}
F_{\mu \nu}F^{\mu \nu}\right), \ee
where we have used the fact that $g_{\mu \nu}$, $\phi$, and $A_{\mu}$ do not depend on $y$.
We observe that the 4D Planck lenght\footnote{We define $\kappa_4$ in a way that the coefficient of $R(g_{\mu \nu})$
in the 4D lagrangian is $1/\kappa_4^2$.} $\kappa_4$ is given by
\be \frac{1}{\kappa_4^2}=\frac{V_1}{\kappa^2}. \ee
On the other hand the effective gauge constant $g_{eff}$ associated to $A'_{\mu}\equiv A_{\mu}/L$ is given
by\footnote{We define $g_{eff}$ in a way that the coefficient of $F'_{\mu \nu}F'^{\mu \nu}$
in the 4D lagrangian is $-1/(4 g_{eff}^2)$.}
\be \frac{1}{g_{eff}^2}=\frac{V_1^2}{\kappa_4^2}.\ee
Therefore the gauge constant is determined in terms of the 4D
gravitational coupling and the volume of the internal space. If we
require $g_{eff}\sim 1$, which is a natural choice as the SM gauge
constants are of the order of 1, we get that the size of the
internal space is naturally of the order of the 4D Planck length.

It is worth mentioning that this original KK theory can be
generalized to include a smooth internal space of the form
$\mathcal{G}/\mathcal{H}$ and develop an harmonic expansion,
analogous to (\ref{FourierKK}), on such coset space. In this way it
is possible to show that a conventional Einstein-Yang-Mills model
(realizing local $\mathcal{G}$-symmetry) emerges at the leading
approximation \cite{SS}. Moreover it can be proved that this
framework emerge as a 4D effective theory of a higher dimensional
Einstein-Yang-Mills system \cite{Randjbar-Daemi:1982rm}.

Besides these attractive properties the original KK theory suffers
from some phenomenological problems. It is not clear how to
interpret the {\it radion} $\phi$, because it represents a massless
scalar particle that is not observed in nature. Moreover, it is not
possible to get a 4D chiral fermionic spectrum in this framework,
which is of course needed if one requires to reproduce the SM in the
low energy limit.

The situation gets better if one includes bulk (D-dimensional) gauge
fields \cite{RSS} and considers generalizations of the original KK
theory. A non-trivial example will be given in Chapter \ref{EMS},
based on a 6D Einstein-Maxwell-Scalar model, where the internal
space will be taken to be $S^2=SU(2)/U(1)$, and chiral fermions are
obtained in the 4D effective theory.

Alternative solutions will be described in Sections \ref{DW}, \ref{LED}, and \ref{RS} where we will introduce the concept
of {\it 3-brane}. As we shall see the latter is a useful tool to address relevant
problems of high energy physics like the hierarchy problem and the cosmological constant problem in higher dimensional
models.

\section{Localized Wave Functions}\label{DW}

The original KK idea assumes a compact internal space with a very
small size to render the extra dimensions invisible. An interesting
alternative can arise when this hypothesis is relaxed, but ordinary
particles are confined inside a potential well, which is flat along
the ordinary 4 dimensions and sufficiently steep along the $d$ extra
dimensions. As we shall see in this section the origin of such a
potential can be purely dynamical, in the sense that it can emerge
as a solution of the equations of motion (EOM). Therefore in this
scenario we have spontaneous breaking of translation invariance. The
ordinary matter can propagate in the $D$-dimensional space-time if
it acquires high enough energy (basically if its energy exceeds the
depth of the well).

\subsection{The Domain Wall}\label{DWsub}

The original idea of confining particles on a (1+3)-dimensional
submanifold ({\it 3-brane}) was proposed in \cite{DomainWall}
  and independently in \cite{Akama:1982jy}. Now we illustrate the main idea by introducing the simplest higher dimensional
model, which can give rise to matter field localization on a 3-brane
({\it brane world}). This is a 5D field theory with one real scalar
living on $(Minkowski)_5$: the action is
\be S=\int d^5 X \left[
-\frac{1}{2}\partial_{M}\varphi\partial^{M}\varphi
-\lambda(\varphi^2-v^2)^2\right], \label{domainaction}\ee
where $\lambda$, and $v$ are real
parameters and we assume $\lambda \geq 0$ in order to have a bounded from below 5D potential. The internal
symmetry of this theory is a $Z_2$ group:
$\varphi\rightarrow \pm \varphi.$

To derive the EOM from (\ref{domainaction}) through an action principle we require that the
boundary terms in the integration by parts vanish. This leads to the conservation of current $J_M=\varphi \partial_M \varphi$,
as explained in
\cite{Nicolai:1984jg,Gibbons:1986wg,Kehagias:1999aa}:
\be \int d^5 X \partial_M\left(\varphi \partial^M \varphi \right)=0.\label{boundarydomain} \ee
Actually we impose that for every pair of fields $\varphi$ and
$\varphi '$ the condition $\int d^5X\partial_{M}\left(\varphi
\partial^{M}\varphi' \right)=0$ is satisfied but in
(\ref{boundarydomain}) the prime is understood. As usual we assume
that the dependence on the 4D coordinates is such that Condition
(\ref{boundarydomain}) reduces to
\be \left(\lim_{y\rightarrow +\infty}-\lim_{y\rightarrow
-\infty}\right)\varphi\partial_y \varphi =0, \label{DWHC}\ee
which involves only the dependence of $\varphi$ on the extra dimension.
A condition like (\ref{boundarydomain}) is usually used in brane world model to project out non physical
modes\footnote{We will use analogous conditions in Chapter \ref{SUGRAspectrum}.}.
Provided that (\ref{boundarydomain}) is satisfied, the EOM is
\be
\partial_M\partial^M\varphi-4\lambda(\varphi^2-v^2)\varphi=0,\ee
and we consider the following {\it kink} domain wall
solution\footnote{For extended discussions on this solution see for
example \cite{Dashen,Morse,Kink}.} \cite{DomainWall}:
\be <\varphi>=v\tanh(\sqrt{2\lambda} v y)\equiv
\varphi_c(y). \label{domain0}\ee
The VEV in (\ref{domain0}) spontaneously breaks the internal
symmetry $Z_2$, and the translation invariance along the extra
dimension.

To see how solution (\ref{domain0}) produces a potential well along the extra dimension we consider perturbations
around such a background solution. We define $\d\varphi=\varphi -\varphi_c$ and the EOM at the linear level with respect to
$\d \varphi$ reads
\be \eta^{\mu \nu}\partial_{\mu}\partial_{\nu}\d\varphi+\partial_y^2\d
\varphi-4\lambda\left(3\varphi_c^2-v^2\right)\d\varphi=0.\label{fluctKink0}\ee
The fluctuation $\d \varphi$ must satisfy Condition
(\ref{boundarydomain}) as well as $\varphi$ itself. This will give
us relevant information to construct the space of physical modes. We
consider now a 4D plane wave solution of (\ref{fluctKink0}), that is
we assume
\be \d\varphi(x,y)= \D(y) e^{i k_{\mu}x^{\mu}}\label{4Dmomentum}\ee
where $k$ represents the 4D momentum of an ordinary particle with
squared mass $M^2=-k^2$. In this case the probability density along
our 4D world is completely flat but we allow a non-trivial
probability density $|\D|^2$ of finding the particle in an interval
$[y,y+dy]$. Inserting (\ref{4Dmomentum}) in Equation
(\ref{fluctKink0}) we get
\be -\partial_y^2 \D
+4\lambda\left(3\varphi_c^2-v^2\right)\D=M^2 \D.\ee
The latter equation  is extensively studied in the literature
\cite{DomainWall,Dashen,Morse,Kink}. It is a 1D Schroedinger
equation with a potential
\be V(y)=4\lambda v^2[3\tanh^2(\sqrt{2\lambda} v y)-1]. \ee
This is exactly the potential well that we mentioned before. It is
easy to see that the Condition (\ref{DWHC}) ensures that the
hamiltonian $-\partial^2_{y}+V(y)$ is hermitian with respect to the
inner product $(\mathcal{D}',\mathcal{D})=\int
dy\mathcal{D'}(y)\mathcal{D}(y)$. To show that this potential
effectively localizes particles on the brane we present the spectrum
of the fluctuations. In the energy range $M\in [0,2\sqrt{2
\lambda}v]$ the spectrum is discrete and it consists of 2 states.
There is a normalizable wave function with $M^2=0$:
\be \D_1(y)=\frac{N_1}{\cosh^2(\sqrt{2\lambda}vy)}, \ee
where $N_1$ is a normalization constant. This wave function must
represent the ground state for such a quantum mechanical problem
because it does not have intersections with the $y$-axis. This
automatically ensures that there are not tachyonic fluctuations.
Moreover there is another normalizable wave function describing a
bound state with mass $M=\sqrt{6\lambda}v$, as explained in
\cite{Dashen}. This couple of normalizable states describes
particles being confined inside the wall. For $M\geq 2\sqrt{2
\lambda}v$ the spectrum is continous and represents perturbations
that are not confined.

We will turn to the kink domain wall in Section \ref{DomainWall}
where we will introduce an additional scalar field whose lightest 4D
mode will be interpreted as an Higgs field. We will study such model
to provide a brane world example for the relevance of the heavy
modes in the 4D low energy effective theory for the light modes,
which is one of the main topics of the present thesis.

\subsection{Fermion Zero Modes}

In the simple model of Subsection \ref{DWsub} one can introduce
fermions and study the corresponding localization problem, as for
the $\d \varphi$-fluctuations. If standard Yukawa couplings are
present, a chiral low energy 4D fermion spectrum emerges, in the
sense that only fermion zero modes with one chirality are localized
on the wall.

To illustrate this point we introduce one 5D fermion with the following action \cite{DomainWall}
\be S_F=\int d^5X\left( \overline{\Psi}\Gamma^M\partial_M\Psi+h\varphi \overline{\Psi}\Psi\right),\label{DWferm}\ee
where $h$ is a real constant and $\Gamma^{\mu}=\gamma^{\mu}$,
$\Gamma^5=\gamma^5$ and our conventions on $\gamma^{\mu}$ and
$\gamma^5$ are given in Appendix \ref{GR}. The second term in
(\ref{DWferm}) is a simple example of Yukawa coupling and it can
provide a chiral localized zero mode. In order to show this, now we
derive the EOM for $\Psi$. Similarly to the fluctuations coming from
$\varphi$, which we have analysed in the previous section, in order
to derive the EOM from (\ref{DWferm}) by means of an action
principle we have to impose\footnote{Actually we impose that for
every pair of fields $\Psi$ and $\Psi '$ the condition $\int
d^5X\partial_{M}\left(\Psi \Gamma^{M}\Psi' \right)=0$ is satisfied
but in (\ref{HCDMf}) the prime is understood.} the conservation of
current $\overline{\Psi}\Gamma^M \Psi$ \cite{Wetterich:1984dz}:
\be \int d^5 X\partial_M\left(\overline{\Psi}\Gamma^M\Psi\right)=0. \label{HCDMf}\ee
Also in the fermion sector we assume that the dependence on the 4D coordinates is such that (\ref{HCDMf}) reduces
to
\be \left(\lim_{y\rightarrow +\infty}-\lim_{y\rightarrow
-\infty}\right)\overline{\Psi}\gamma^5\Psi=0. \label{HCDMf2}\ee
Thus the EOM reads
\be \Gamma^M\partial_M\Psi + h \varphi \Psi=0. \ee
Now we decompose $\Psi$ as follows: $\Psi=\Psi_R+\Psi_L$, where $\gamma^5\Psi_R=\Psi_R$ and $\gamma^5\Psi_L=-\Psi_L$, and
therefore the EOM linearized around the kink background (\ref{domain0}) are
\bea  \gamma^{\mu}\partial_{\mu}\Psi_L+\partial_y\Psi_R+h\varphi_c \Psi_R&=&0, \nonumber\\
\gamma^{\mu}\partial_{\mu}\Psi_R-\partial_y\Psi_L+h\varphi_c \Psi_L&=&0.\eea
In general this is a set of coupled differential equations, but in
the case of zero modes ($\gamma^{\mu}\partial_{\mu}=0$) they
decouple. For instance $\Psi_R$ satisfies
\be \partial_y \Psi_R=-h\varphi_c \Psi_R, \label{RDM} \ee
which can be easily solved:
\be \Psi_R(x,y)=\Psi_R^{(4)}(x)e^{-h\int_0^y dy' \varphi_c(y')},\label{DWFprof} \ee
where $\Psi_R^{(4)}$ is a 4D chiral fermion, which does not depend
on $y$. By using the explicit expression of $\varphi_c$ given in
(\ref{domain0}), it is easy to see that (\ref{DWFprof}) represents a
bounded state, in the sense that its wave function profile along the
extra dimension is peaked on $y=0$ and rapidly goes to zero as
$y\rightarrow \pm \infty$. On the other hand, the fermion with
opposite chirality is unbounded, as can be seen by obtaining the
corresponding wave function from (\ref{DWFprof}) with the
substitution $h\rightarrow-h$. The Condition (\ref{HCDMf2}) is
automatically satisfied by these zero modes, but one chirality has
to be projected out because it is not normalizable and therefore it
has infinite kinetic energy from the 4D point of view. Therefore we
conclude that the zero mode spectrum is effectively chiral as
required by the SM.

These results, which we have obtained in a very simple model, can be
generalized to include an arbitrary number of space-time dimensions
and Yang-Mills and scalar backgrounds. In this more general
framework, the conditions under which localized chiral fermions
emerge are given in \cite{Randjbar-Daemi:2000cr}.

\subsection{Gauge Field Localization}\label{Gauge Field
Localization}

We conclude this section by examining gauge fields. Unlike the
spinless and the spin-1/2 case, localizing gauge field wave
functions is not simple, at least for massless non-Abelian fields.
The reason is that we can have phenomenological problems when we
construct the effective 4D gauge constants. Indeed, if we denote the
zero mode wave function of a gauge field by $A_0(y)$ and the
corresponding quantity for a fermion (or, in general, for  a charged
field) by $\Psi_0(y)$, usually the gauge constant in the 4D
effective theory turns out to be proportional to an overlap integral
of the form
\be \int dy \Psi_0^*(y)A_0(y)\Psi_0(y).\ee
 On the other hand, in
the previous subsection we have seen that the fermion wave functions
can be different for different types of particles as they depend on
various parameters, for instance the constant $h$ in (\ref{DWferm}).
This can create a problem, as in non-Abelian gauge theories the
gauge charges are quantized and fixed by group theory arguments
(this is referred to as {\it charge universality}).

We can imagine some ways out. A first possibility is finding a
theory that produces the same wave function profile for all the zero
modes $\Psi_0(y)$. However, this is not the general case because, as
we will see, in the explicit examples of Chapters \ref{EMS} and
\ref{SUGRAspectrum} this does not happen. Therefore, a physical
mechanism that ensures such an equality is needed, if one tries to
solve the problem in this way.

A second possibility is obtaining a constant profile for the zero
mode gauge fields, namely $A_0$ independent of $y$. Indeed, in this
case the overlap integral defining the gauge constants becomes
$A_0\int dy \Psi_0^*(y)\Psi_0(y)$, and it is proportional to the
normalization constant of the fermion kinetic term in the 4D
effective theory. Since we can normalize the fields in a way that
\be \int dy \Psi_0^*(y)\Psi_0(y)=1,\ee
we obtain charge universality. An explicit example is the framework
analyzed in Chapter \ref{SUGRAspectrum}, where the gauge field
profiles are dynamically predicted to be constant.

Finally, in the literature there exists a more sophisticated
mechanism to obtain localized gauge fields \cite{Dvali:1996xe}.
There it has been proposed to consider a gauge theory that is in
{\it confinement phase} outside the brane, whereas it is in the
Abelian Coulomb phase on the brane. This is achieved by means of a
scalar field that acquires a kink VEV, such as $\varphi$ in
(\ref{domain0}). Massless "quarks" and a U(1) gauge field are
present on the brane and they cannot escape far away from the brane,
since the lightest state of the confining theory of the bulk has a
non-vanishing mass of the order\footnote{The parameter $\Lambda$ is
the analogous of $\Lambda_{QCD}$ in Quantum Chromodynamics.}
$\Lambda$. It is interesting to note that this mechanism provides
both gauge field and fermion localization.

\section{Large Extra Dimensions}\label{LED}

So far we have discussed KK theories, compactified on internal space
with very small size, and brane world models with infinite extra
dimensions, where a physical localization mechanism is needed to
render the low energy physics effectively 4-dimensional. In this
section we study an intermediate set up, in which the internal space
is compact, like in original KK theories, but with {\it large size}.
As we will show, this scenario, originally proposed by Arkani-Hamed,
Dimopoulos and Dvali (ADD) \cite{ADD}, leads to an interesting
reformulation of the hierarchy problem.

Contrary to what happen in original KK theories of Section \ref{KK},
where the size of the internal space is naturally of the order of
the Planck length, in the Large Extra Dimensions (LED) scenario the
effects of extra dimensions are constrained by accessible
experimental tests. In particular the behaviour of gravitational
interactions should change at length scales below $r$, where
\be r\equiv (V_d)^{1/d}. \ee
Moreover several possible collider
experiments can detect extra dimensions if the KK mass gap (the mass
gap between the zero modes and the first KK excited states) is of
the order of
\be M_{GAP}\sim \frac{1}{r}, \label{MGAPV}\ee
as expected.

\subsection{General Idea of ADD}

To illustrate the main idea of ADD we consider an action containing
the standard D-dimensional Einstein-Hilbert term\footnote{See
Appendix \ref{GR} for conventions and notations.}
\be S_{EH}=\frac{1}{\kappa^2}\int d^D X \sqrt{-G}R, \ee
and we make the following ansatz:
\be ds^2=e^{A(y)}g_{\mu \nu}(x)dx^{\mu}dx^{\nu}+g_{mn}(y)dy^{m}dy^n. \label{LEDansatz}\ee
Namely we neglect 4D vectors and scalars coming from the
D-dimensional metric but we keep the 4D metric $g_{\mu \nu}(x)$. The
latter represents the complete 4D dynamical metric, including the
VEV and the fluctuations. If one requires 4D Poincar\'e invariance
one has also to impose $<g_{\mu \nu}>=\eta_{\mu \nu}$, but this is
not necessary for our argument. In Eq. (\ref{LEDansatz}) we allow
the presence of a non-trivial {\it warp factor} $e^{A(y)}$ to make
the argument more general. This will be useful in Chapter
\ref{SUGRAspectrum} when we will discuss 6D warped brane worlds.
This kind of space-time are called {\it non-factorizable geometries}
because they cannot be interpreted as products of manifolds like
(\ref{factorizable}), as in the standard KK compactifications.

An explicit calculation leads to
\be R=e^{-A}R(g_{\mu \nu})+..., \label{RR4}\ee
where $ R(g_{\mu \nu})$ is the Ricci scalar computed with $g_{\mu
\nu}$, and the dots represent extra terms containing the warp factor
and the metric components $g_{mn}$. So we obtain a 4D effective
action for the gravitational field $g_{\mu \nu}$ as follows
\be S_{EH}=\frac{1}{\kappa^2}\int d^D X\sqrt{-G} e^{-A}R(g_{\mu
\nu})+...=\frac{1}{\kappa_4^2}\int d^4x \sqrt{-g}R(g_{\mu \nu})
+...,\ee
where $g$ is the determinant of $g_{\mu \nu}$ and the 4D Planck length $\kappa_4$ is given by
\be \frac{1}{\kappa_4^2}=\frac{V_d}{\kappa^2}, \label{kk4}\ee
where
\be V_d=\int d^{d} y \sqrt{-\bar{G}}e^{-A},\label{Vd}\ee
and $$\bar{G}_{MN}dX^MdX^N=e^{A(y)}\eta_{\mu
\nu}dx^{\mu}dx^{\nu}+g_{mn}(y)dy^{m}dy^n.$$ We observe that $V_d$ is
a d-dimensional volume, which reduces to the volume of the internal
space in the unwarped case ($e^A=1$). Indeed the latter is defined
by
\be \tilde{V}_d= \int d^d y\sqrt{g_d}, \ee
where $g_d$ is the determinant of the metric $g_{mn}$ of the
internal space. In terms of mass scales $M= 1/\kappa^{2/(D-2)}$, and
$M_{Pl}=1/\kappa_4$, Eq. (\ref{kk4}) takes an illuminating form:
\be \left(\frac{M_{Pl}}{M}\right)^2=\left(M r\right)^d. \label{PplM}\ee
If $r$ is large compared to the fundamental length $M^{-1}$, the 4D
Planck mass is much larger than the fundamental gravity scale $M$.
One may push this line of reasoning to extreme and suppose that the
fundamental gravity scale is of the same order as the electroweak
scale, $M\sim TeV$. Then the hierarchy between $M_{Pl}$ and the
electroweak scale is entirely due to the large value of $r$. This is
an interesting reformulation of the hierarchy problem because now it
becomes the problem of explaining why $r$ is large. We observe that
$r$ can be large because the volume $\tilde{V}_d$ of the internal
space is large or because of a non-trivial contribution of the warp
factor. In this section we will assume that only the former
contribution is active. We will discuss the effect of the warp
factor in Section \ref{RS}.

\subsection{Phenomenological Implications}

In the ADD scenario new physics should emerge in the gravitational
sector when the length scale $r$ is reached. However, it is hard to
test gravity at very short distances because it is a much weaker
interaction than all the other forces. Over large distances gravity
is dominant, however, as one starts going to shorter distances,
electromagnetic forces are dominant and completely overwhelm the
gravitational forces. This is the reason why the Newton-law of
gravitational interactions has only been tested down to about a
fraction of a millimeter. Today the bound on the size of the extra
dimensions is $r\leq 0.1 mm$, if only gravity propagates in the
extra dimensions \cite{Adelberger:2003zx}.
   On the other hand, assuming $M\sim TeV$, we can calculate from (\ref{PplM}) the value of $r$ as a function of $d$,
\be r=M^{-1}\left(\frac{M_{Pl}}{M}\right)^{2/d}\sim 10^{32/d} \,10^{-17} cm, \ee
where we used $M_{Pl}\sim 10^{16}TeV$. For one extra dimensions one
obtains unacceptably large value of $r$. The case $d=2$ is
particularly interesting because it corresponds to $r\sim 1 mm$. By
increasing the fundamental mass scale of less than one order of
magnitude ($M\sim TeV,10 TeV$) one gets a value of $r$ close to the
present bound ($r\leq 0.1 mm$). Therefore for $d=2$, one can hope to
detect effects of the extra dimensions but still have a model, which
is not ruled out by experiments. If $d>2$ the size of the extra
dimensions is less than $10^{-6}cm$, which is unlikely to be tested
directly via gravitational measurements any time soon. For $d=6$
(full dimensionality of space-time, as suggested by superstring
theory), one has $r\sim 10^{-12}cm$, which is still much larger than
the electroweak scale, $TeV\sim 10^{-17}cm$. In order the extra
dimensions to be so large one has to find a physical mechanism to
make the SM matter and gauge fields effectively 4-dimensional. The
most popular way is to localize such fields on a 3-brane, but a
possible alternative is to relax relation (\ref{MGAPV}) and increase
$M_{GAP}$ considering compactifications on non-smooth space. We will
show that this is possible in Chapter \ref{SUGRAspectrum}, which
contains part of the original work \cite{SUSHA-RD-AS}, even if in
this case the tuning of ADD scenario ($r\gg M^{-1}$) becomes a
tuning on the fundamental bulk parameters.

Another interesting phenomenological consequence of this scenario is
that extra dimensions should start to show up in collider
experiments at energies approaching the $TeV$ scale. If we assume
(\ref{MGAPV}) and that the only interactions, which can propagate in
the bulk, is the gravitational interactions, the most distinctive
feature of this scenario is the possibility to emit gravitons into
the bulk. This process has strong dependence on the center of mass
energy of particles colliding on the brane and has large probability
at energies comparable to the fundamental gravity scale. Indeed even
though the coupling of every KK graviton is weak, the total emission
rate of KK gravitons is large at energies approaching $M$ due to
large number of KK graviton states. These particles will not be
detected, so the typical collider processes will involve missing
energy. For example the cross section of production of a KK graviton
in the process
\be e^+\,e^-\rightarrow \gamma + P_T,\label{ADDprocess}\ee
 which involves a transverse
missing particle $P_T$, is of the order of $\alpha/M^2_{Pl}$, so the
total cross section is of order $\sigma \sim \alpha N(E)/M^2_{Pl}$,
where $E$ is the center of mass energy, and $N(E)$ is the number of
species of KK gravitons with mass below $E$. By using (\ref{MGAPV})
we expect
\be N(E)\sim (Er)^d, \label{N(E)}\ee
 which is indeed correct for
the simple compactification on a $d$-dimensional torus. Therefore the total cross section becomes
$$\sigma \sim \frac{\alpha}{E^2}\left(\frac{E}{M}\right)^{d+2}$$, which rapidly increases with $E$, and becomes comparable
with the electromagnetic cross section at $E\sim M$. Processes like
(\ref{ADDprocess}) have been analyzed in detailed in Ref.
\cite{Giudice:1998ck}. It has been found that both a 1 TeV e+e-
collider and the CERN LHC will be able to reliably and
perturbatively probe the fundamental gravity scale up to several
TeV, with the precise value depending on the number $d$ of extra
dimensions.

It is interesting to note that relation (\ref{N(E)}) is relaxed if
the KK mass gap is not of the order $r^{-1}$, which is possible if
one considers non-smooth compactification. This can lead to a less
stringent bound on the KK modes production.

\section{Randall-Sundrum Models}\label{RS}

Until now we have considered general warped geometries without
giving an explicit example in a specific brane world scenario and
without discussing the role of the warp factor. In this section we
shall describe the most popular example of warped brane world,
originally proposed by Randall and Sundrum (RS)
\cite{Randall:1999ee, Randall:1999vf}. The original RS model is a 5D
gravitational model whose action is the sum of the standard
Einstein-Hilbert action, a 5D cosmological constant term, and a {\it
3-brane action}:
\be S=\int d^5 X \sqrt{-G}\left(\frac{1}{\kappa^2}R-\Lambda\right)-T\int d^4x\sqrt{-g}-T'\int d^4x\sqrt{-g'},\label{SRS} \ee
where $\Lambda$ is the 5D cosmological constant, $T$ and $T'$ are the brane tensions (energy densities) of two branes placed
at $y=0$ and $y=\pi r_c$ respectively, and $g$ and $g'$ are the determinants of the metrics
$g_{\mu \nu}$ and $g'_{\mu \nu}$ induced on the branes:
\be g_{\mu \nu }(x)=G_{\mu \nu}(x,0),\quad g'_{\mu \nu }(x)=G_{\mu \nu}(x,\pi r_c).\ee
We assume that the branes are located at the boundaries of the extra
dimensions, that is $0\leq y\leq \pi r_c$. One can interpret one
brane as our 4D world and the other one as an additional 4D world.

We consider now the most general background metric compatible with 4D Poincar\'e invariance:
\be ds^2=e^{A(y)}\eta_{\mu \nu}dx^{\mu}dx^{\nu}+dy^2. \ee
The explicit expression for the warp factor can be found by putting
this metric ansatz in the EOM, which follows from (\ref{SRS}). The
explicit calculation is given in the original work
\cite{Randall:1999ee}, so here we only give the final result. The
existence of 4D flat solution requires fine-tunings between
$\Lambda$, $T$, and $T'$.  Indeed these constants are related in
terms of a single scale $k$,
\be T=-T'=\frac{12 k}{\kappa^2},\quad \Lambda= -\frac{12 k^2}{\kappa^2},\label{RSconstraint}\ee
and the warp factor is given by
\be e^{A(y)}=e^{-2k|y|}.\ee
This fine-tuning is analogous to fine-tuning of the cosmological
constant to zero in conventional 4D gravity.

Given the background we can now compute the volume $V_1$ defined in
(\ref{Vd}). The result is
\be V_1=\frac{1}{2k}\left(1-e^{-2k\pi r_c}\right).\ee
In this case we can have an hierarchy between $M$ and $M_{Pl}$ but
only for large value of $r_c$, that is for a large internal space
volume. However, we observe that $M_{Pl}$ remains finite when $r_c$
goes to infinity. This is an important property of the RS model, in
which it is possible to have the ordinary 4D gravity even if we take
the non-compact limit ($r_c\rightarrow \infty$) for the internal
space. Moreover, although the exponential has very little effect in
determining the Planck scale, it plays a crucial role in the
determination of the observable  masses. Indeed since the induced
brane metrics are related by $g'_{\mu \nu}=e^{-2k\pi r_c}g_{\mu
\nu}$, any mass parameter $m'$ on the $y=\pi r_c$ brane will
correspond to a mass $m\equiv e^{-kr_c\pi}m'$ when measured on the
$y=0$ brane \cite{Randall:1999ee}. If $e^{kr_c \pi}$ is of order
$10^{16}$, this mechanism produces $TeV$ physical mass scales from
fundamental mass paramenters of the order of Planck mass. Because
this geometric factor is an exponential, we do not require very
large hierarchies among the fundamental parameters.

In the RS scenario one can localize ad hoc SM fields on a brane and
exploit the mechanism that we have discussed so far in order to
``solve'' the hierarchy problem. However, after the original RS work
models were developed in which SM fields originate from the
bulk\footnote{Although the Higgs field should be confined to the
brane in order not to lose the gauge hierarchy.}
\cite{Chang:1999nh,5DSM}. These authors tried to implement a
localization mechanism similar to the one explained in section
\ref{DW} but it is difficult to obtain a phenomenologically viable
effective theory, because the various bulk fields are not always
localized on a brane. We shall perform a similar study of bulk
fields but in the contest of 6D warped brane worlds in Chapter
\ref{SUGRAspectrum}.

\section{Addressing the Cosmological Constant Problem}\label{CCproblem}

In Sections \ref{LED} and \ref{RS} we have discussed higher
dimensional models, which give hope to solve the hierarchy problem
or can reformulate it in an interesting way. The purpose of this
section is to perform a similar study, but concerning the
cosmological constant problem. In theories with extra dimensions
such a problem can be reformulated as a problem of why the vacuum
energy density has (almost) no effect on the observable quantities
predicted by the 4D effective theory, which is valid at the energy
range of present experiments. In particular this reformulation
sounds suggestive in brane worlds scenarios, as it implies that the
vacuum energy density may affect the bulk geometry, and that this
may occur in such a way that the metric induced on our brane is
(almost) flat. Roughly speaking, it seems plausible that, in the
case of non-factorizable geometry, the vacuum energy density can
induce a non-trivial warp factor, while the 4D Poincar\'e invariance
remains unbroken. This possibility may exist irrespectively of the
brane world picture \cite{Rubakov:1983bz}.

In RS models with one compact or non-compact extra dimension, the
warped solution, that we discussed in Section \ref{RS}, requires a
fine-tuning between the bulk cosmological constant and the brane
tensions, which is explicitly given in (\ref{RSconstraint}). From
this point of view the cosmological constant problem is not solved
in these models, as the standard 4D fine-tuning is replaced by a
similar 5D one.

A different scenario with non-compact internal manifold and
vanishing cosmological constant in the bulk has been proposed in
\cite{Dvali:2000hr,Dvali:2000xg}. In this works, besides a standard
D-dimensional Einstein-Hilbert action, an additional 4D
Einstein-Hilbert term and a 4D cosmological constant $\Lambda_b$ are
introduced on a 3-brane, giving rise to the following gravitational
action
\be S=\frac{1}{\kappa^2}\int d^DX \sqrt{-G}R+ \int d^4 x
\sqrt{-g}\left(\frac{1}{\kappa_{b}^2}R(g_{\mu \nu})-\Lambda_b
\right).\ee
Here we have two different Planck scales, $\kappa$ in the bulk and
$\kappa_b$ on the brane. It has also been argued \cite{Dvali:2000hr}
that the 4D Einstein-Hilbert term is a natural ingredient, as
quantum corrections generate it anyway. In \cite{Dvali:2000hr} the
above model has been formulated in 5D, but generalizations involving
more than one extra dimension, which are relevant for the
cosmological constant problem, have been considered in
\cite{Dvali:2000xg}.

In this scenario the graviton is a
metastable state and the essential feature of the model is the large distance modification of gravity. More precicely, the graviton
propagator manifests a modified behaviour at length scales larger than $r_c$, where
\be r_c=\frac{M_{Pl}}{M^2},\ee
and $M_{Pl}=1/\kappa_b$ and $M= 1/\kappa^{2/(D-2)}$. As a
consequence, gravity does not necessarily react to sources that are
relevant at length scale of the order $r_c\sim H_0^{-1}$, where
$H_0$ is the Hubble constant, as is the case for the cosmological
constant on the brane. As explained in
\cite{Dvali:2000xg,Dvali:2002pe}, this property could lead to a
solution of the cosmological constant problem.

Other frameworks, which could be relevant for the cosmological
constant problem, are 6D models with 2 compact extra dimensions and,
more generally, self tuning models. These frameworks will be
briefly reviewed in the next section and in Chapter \ref{6DSUGRA}, where we
will deal with supersymmetric and large extra dimensions.

\section{Codimension 2 Branes}\label{C2B}

As we discussed in Section \ref{LED} the case $d=2$ is particularly
interesting in the ADD scenario. It corresponds indeed to the
smallest value of $d$ compatible with tests of gravity. Moreover for
$d=2$ the ADD scenario predicts a value of $r$ close to the present
bound given by such experiments and therefore it is a falsifiable
set up. Furthermore models in 6 dimensions have attracted interest
as possible frameworks in which the cosmological constant could be
faced, in both the non supersymmetric
\cite{Sundrum:1998ns,Carroll:2003db,Randjbar-Daemi:2004ni} and the
supersymmetric \cite{Aghababaie:2003wz}-\cite{Nair:2004yu} case. One
of the motivations for that is the numerical equality between $r$
and $\rho^{-1/4}$, where $\rho\sim (10^{-3}eV)^4$ is the measured
vacuum's energy density, in 6D ADD scenario.

In order to implement these ideas usually {\it 3-brane sources} are
introduced in the action, namely terms similar to the second and
third term in the RS action given in (\ref{SRS}). However, branes
whose transverse space is 2D (codimension 2 brane) are qualitative
different from codimension 1 branes, described in Section \ref{RS}.
One of the main features of 6D models is the possibility to find
3-branes solutions of the EOM with a geometry independent of the
value of the brane tensions, at least outside the branes. This
property is not shared by the codimension 1 RS branes because of the
following reason: the 5D cosmological constant $\Lambda$ in that
case depends on the brane tensions because of constraint
(\ref{RSconstraint}), and, on the other hand, $\Lambda$ gives a non
vanishing contribution to the 5D Ricci scalar both outside and on
the branes.

The aim of the present section is to describe codimension 2 branes
in a simple set up because they are usefull to understand the
original results of Chapter \ref{SUGRAspectrum}, in which we will
deal with 6D supersymmetric models having this type of brane
solutions.

We start with the following 6D action
\be S=\frac{1}{\kappa^2}\int d^6 X \sqrt{-G}R-T\int d^4x\sqrt{-g},\label{Sconical} \ee
that is the sum of the standard 6D Einstein-Hilbert action and a single brane source. Moreover we consider the following
simple ansatz for the background metric
\be ds^2=\eta_{\mu \nu}dx^{\mu}dx^{\nu} + h(r)\left(dr^2+r^2 d\varphi^2\right), \label{ansatzconical}\ee
that is we assume 4D Poincar\'e invariance, vanishing warp factor,
axisymmetry of the internal 2D space but we allow a non-trivial
curvature in order to take into account the backreaction of the
geometry due to the brane source. This curvature will be a
functional of $h(r)$. We assume that $r$ and $\varphi$ range from
$0$ to $\infty$ and from $0$ to $2\pi$ respectively. So we interpret
$r$ as a radial coordinate and $\varphi$ as an angular coordinate.

The EOM associated to action (\ref{Sconical}) are the Einstein equations in presence of a brane source:
\be \frac{\sqrt{-G}}{\kappa^2}\left(R^{MN}-\frac{1}{2}R
G^{MN}\right)= -\frac{T}{2}\delta^{(2)}(y)\sqrt{-g}\,g^{\mu
\nu}\delta_{\mu}^{M}\delta_{\nu}^{N},\ee
where $y=(r\cos\varphi, r\sin\varphi)$, $\d^{(2)}(y)$ is the 2D
Dirac $\delta$-function and $G_{MN}$ are now the metric components
in the coordinates $x$ and $y$. By using ansatz
(\ref{ansatzconical}) the $\mu, \nu$ components of the Einstein
equations give
\be \frac{1}{\kappa^2}\sqrt{g_2}R=T\d^{(2)}(y), \label{Rd}\ee
where $g_2$ is the determinant of $g_{mn}$. We observe that the Ricci scalar outside the brane vanishes and therefore
is independent of the brane tension, this is an example of the possible independence of geometry and brane
tensions in codimension 2 brane world. The $m,n$ components are trivially satisfied as they read $R_{mn}-\frac{1}{2}Rg_{mn}=0$
and the 2D metric $g_{mn}$ satisfies $R_{mn}=K(r) g_{mn}$ for some function $K(r)$. On  the other hand the Ricci scalar
is given in terms of $h$ by
\be R=-\frac{1}{h}\nabla^2_E\ln h, \label{hE}\ee
where $\nabla^2_E$ is the covariant Laplacian computed with the
euclidean metric $ds^2_E=dr^2+r^2d \varphi^2$. By putting (\ref{hE})
in (\ref{Rd}) we get a second order differential equation for $h$,
which is satisfied by\footnote{In order to prove that $h\propto
r^{2\zeta}$  is a solution one can use the relation $\nabla^2_E\ln r
=2\pi \delta^{(2)}(y)$.} $h\propto r^{2\zeta}$ where $\zeta$ is
given in terms of the brane tension by
\be \zeta=-\frac{1}{4\pi}\kappa^2 T. \ee
The parameter $\zeta$ has an interesting geometrical meaning that
can be understood by introducing the coordinate $\rho\equiv r^{\zeta
+1}/(\zeta +1)$. Indeed the 2D metric in terms of this coordinate is
$d\rho ^2 + \left(\zeta +1\right)^2\rho^2 d \varphi^2$. Therefore
the effect of a non vanishing value of $\zeta$ is the shift $\varphi
\rightarrow (1+\zeta) \varphi$, that is it produces a {\it deficit
angle} $ \delta$ given by \cite{Chen:2000at}
\be \frac{2\delta}{\kappa^2}=T. \label{Td}\ee
The resulting geometry presents a {\it conical defect} (or {\it conical singularity})  at
$r=0$. Eq. (\ref{Td}) is very important because it establishes a
relation between a geometrical property of the internal space and a
physical property of the 3-brane. We will use this formula in
Chapter \ref{SUGRAspectrum}, where we will study the 6D
supersymmetric and gravitational models compactified on an internal
space that will turn out to have conical singularities.

\addcontentsline{toc}{chapter}{PART I: Heavy Mode Contribution from
Extra Dimensions}
\chapter*{ Part I: Heavy Mode Contribution from Extra Dimensions}

In studying the low energy physics of the light modes of a
(4+d)-dimensional theory the attention is usually paid only to the
spectral aspects. After determining the quantum numbers of the light
modes  the nature and the form of the interaction terms are often
assumed to be dictated by symmetry arguments.  Such arguments fix
the general form of all the renormalilzable terms and if the
effective theory is supersymmetric certain relationship between the
couplings can also be established by supersymmetry. The masses are
derived from the bilinear part of the effective action and the role
of the heavy modes in the actual values of the masses and the
couplings of the effective theory for the light modes are seldom
taken into account. It is, however, well known from the study of the
GUT's in 4-dimensions that the heavy modes have an important role to
play even at low energies \cite{GUT}. This happens through their
contributions to the couplings  entering into the effective
Lagrangians describing the low energy physics of the light modes.
According to Wilsonian approach, in order to obtain an effective
theory applicable in large distances, the heavy modes should be
integrated out \cite{Wilson}. The processes of "integrating out" has
the effect of modifying the couplings of the light modes or
introducing additional terms, which are suppressed by inverse powers
of the heavy masses, as proved\footnote{If the gauge symmetry is not
assumed, the decoupling theorem of \cite{Appelquist:1974tg} in
general does not hold \cite{Collins:1978wz,Chan:1979ce,Li:1980dz}.}
in \cite{Appelquist:1974tg}.

The aim of Part I of the present thesis is to examine the role of
the heavy modes in the low energy description of a higher
dimensional theory. To this end we shall basically perform two
complementary calculations. The first one will start from a solution
of a higher dimensional theory with a 4D Poincar\'e invariance and
develop an action functional for the light modes of the effective 4D
theory. This effective action generally has a local symmetry, which
should be broken by  Higgs mechanism. Our interest is in the
spectrum of the broken theory. The procedure is essentially what is
adopted in the effective description of higher dimensional theories
including superstring and M-theory compactifications. In this
construction the heavy KK modes are generally ignored simply by
reasoning that their masses are of the order of the compactification
mass and this can be as heavy as the Planck mass. Therefore they
cannot affect the low energy physics of the light modes.

 In the second approach, which we shall call the $\it{ geometrical\ approach}$, we shall find a solution of the higher dimensional equations
with the same symmetry group as the one of the broken phase of the
effective 4D theory for the light modes. We shall then study the
physics of the 4D light modes around this solution. The result for
the effective 4D theory will turn out to be {\it different} from the
first approach.  Our aim is to show that the difference is precisely
due to the fact that in constructing the effective theory along the
lines of the first approach the contribution of the heavy KK modes
have been ignored. Indeed it will be argued - and demonstrated by
working out some explicit examples - that taking due care of the
role of the heavy modes a complete equivalence is established
between the two approaches.

Part I contains two chapters and they both present original results
\cite{Randjbar-Daemi:2006gf}. In Chapter \ref{GHBscalar} we motivate
the discussion in a simple context. In Section \ref{simple4D} we
shall work out a simple model of two coupled scalar fields in
4-dimensions, which will be generalized to a multiplet of scalar
fields in arbitrary dimensions in Section \ref{general}. The
examples in Sections \ref{simple4D} and \ref{general} will clarify
the relevance of the heavy modes in the low energy description of
the light modes. In Section \ref{DomainWall} we shall discuss a
simple 5D domain wall model including two bulk scalars, one of which
acquires a kink VEV. In this simple higher dimensional model we will
include also interactions in our study of the broken effective
theory. In Chapter \ref{EMS} we shall study a higher dimensional (in
this case six dimensional) theory of Einstein-Maxwell system
\cite{RSS} coupled to a charged scalar and eventually also to
charged fermions. We will define explicitly this model in Section
\ref{defEMS}. Such a model can arise in the compactification of
string or M-theory to lower dimensions. The system has enough number
of adjustable parameters to allow us to go to various limits in
order to establish the main point of the present part of the thesis.
The result will of course confirm the above mentioned expectation
that in order to obtain a correct 4-dimensional description of the
physics of the light modes the contribution of the heavy modes
should be duly taken into account\footnote{Of course this does not
prove that the heavy mode contribution never vanishes: for instance
\cite{Gibbons:2003gp} proves the decoupling of the heavy modes in
the $(Minkowski)_4\times S^2$ compactification of the 6D chiral
supergravity \cite{Salam:1984cj}, which is basically the
supersymmetric version of our 6D theory.}. The explicit calculations
will be given in Sections \ref{EWSB4D} and \ref{u13} and summarized
in Section \ref{Comments}. This example is particularly interesting
because the first kind of solution will produce an effective 4D
gauge theory with a $SU(2)\times U(1)$ symmetry which will be broken
to $U(1)$ by a complex triplet of Higgs fields. The geometrical
approach, on the other hand, will take us directly to the unbroken
$U(1)$ phase by deforming a round sphere into an ellipsoid\footnote{
This will correspond to the magnetic monopole charge of 2.  A
monopole charge of unity will produce a Higgs doublet of SU(2).}. In
the geometrical approach the W and the Z masses originate from the
deformation of the internal space. In this sense the standard Higgs
mechanism acquires a geometrical origin\footnote{It should be
mentioned that all of our discussion is ( semi-) classical. To
include quantum and renormalization effects is beyond the scope of
the present study.}. We elaborate a little more on this point in
Section \ref{Comments} which summarizes our results.  Some technical
aspects of various derivations have been detailed in Appendix
\ref{AppGHB}.


\chapter{Higher Dimensional Scalar Models.}\label{GHBscalar}

The present chapter will clarify the role of the heavy modes in the
low energy dynamics without introducing any complications due to
gauge and gravitational interactions. We will provide a
generalization to a more sophisticated context in Chapter \ref{EMS}.

\section{A Simple 4D Theory} \label{simple4D}
\setcounter{equation}{0}

 Let us consider a 4D theory,
which contains two real scalar fields $\varphi$ and $\chi$ and with
the lagrangian
$$ \mathcal{L}=
-\frac{1}{2}\partial_{\mu}\varphi\partial^{\mu}\varphi-\frac{1}{2}\partial_{\mu}\chi\partial^{\mu}\chi
-\frac{1}{2}m^2_{\varphi}\varphi^2-\frac{1}{2}m^2\chi^2-\frac{1}{4}\lambda_{\varphi}\varphi^4
-\frac{1}{4}\lambda_{\chi}\chi^4-a\varphi^2\chi^2, $$
where $m^2_{\varphi}$, $m^2$, $\lambda_{\varphi}$, $\lambda_{\chi}$
and $a$ are real parameters\footnote{Of course we consider only the
values of these parameters such that the scalar potential is bounded
from below.}. Here we have the symmetry:
\bea &&Z_2: \varphi\rightarrow \pm \varphi, \nonumber \\
&&Z_2': \chi \rightarrow \pm \chi. \eea
 This is a very particular
example and of course we do not want to present any general result
in this section, we just want to provide a framework in which the
general equivalence that we spoke about in the introduction emerges
in a simple way and is not obscured by technical difficulties.

For $m^2_{\varphi}<0$ we have the following solution of the
EOM:
\be \chi=0, \quad \quad
\varphi=\sqrt{\frac{-m^2_{\varphi}}{\lambda_{\varphi}}}\equiv
\varphi_{eff}, \label{simplefirstbac}\ee
which breaks $Z_2$ but preserves $Z_2'$. We can express the
lagrangian in terms of the fluctuation $\d \varphi $ and $\chi$
around this background:
 \bea \mathcal{L}&=&
-\frac{1}{2}\partial_{\nu}\d\varphi\partial^{\nu}\d\varphi-\frac{1}{2}\partial_{\nu}\chi\partial^{\nu}\chi
+m^2_{\varphi}\left(\d\varphi\right)^2-\frac{1}{2}\mu^2\chi^2
\nonumber
\\
&&-\sqrt{-m^2_{\varphi}\lambda_{\varphi}}\left(\d\varphi\right)^3-\frac{1}{4}\lambda_{\varphi}\left(\d\varphi\right)^4
-\frac{1}{4}\lambda_{\chi}\chi^4-2a\sqrt{\frac{-m^2_{\varphi}}{\lambda_{\varphi}}}\d\varphi\chi^2
\nonumber \\&&
-a\left(\d\varphi\right)^2\chi^2+constants,\label{lagexp} \eea
where
\be \mu^2\equiv m^2-2a\frac{m^2_{\varphi}}{\lambda_{\varphi}}. \ee

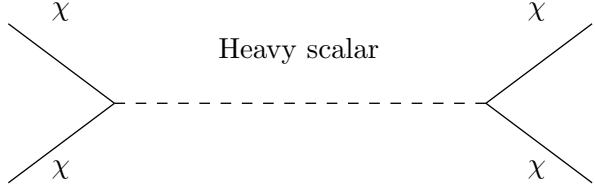
\begin{figure}[t]
\begin{center}
\begin{picture}(300,100)(0,0)

\DashLine(50,80)(190,80){4}\Text(120,100)[]{Heavy scalar}
\Line(190,80)(230,50)\Text(210,55)[]{$\chi$}
\Line(190,80)(230,110)\Text(210,115)[]{$\chi$}
\Line(10,50)(50,80)\Text(30,55)[]{$\chi$}
\Line(10,110)(50,80)\Text(30,115)[]{$\chi$}

\end{picture}
\end{center}
\caption{\footnotesize A tree diagram which describes the scattering
of two light $\chi$, through the exchange of an heavy scalar. This
kind of diagram gives a contribution to the quartic term in the
effective theory potential.} \label{simplediagram}
\end{figure}

If $|\mu^2|\ll |m^2_{\varphi}|$, we expect that the heavy mode
$\d\varphi$ can be integrated out and an effective theory for $\chi$
can be constructed for both the signs of $\mu^2$. However, it's
important to note that $\d\varphi$ cannot be simply neglected
because it gives a contribution, because of the
trilinear\footnote{Also the quartic coupling
$\left(\d\varphi\right)^2\chi^2$ gives a contribution to the
operator $\chi^4$, but this is negligible in the classical limit.}
coupling $\d\varphi \chi^2$ in (\ref{lagexp}), to the operator
$\chi^4$ in the effective theory, through the diagram
\ref{simplediagram}. This is similar to what is usually done in GUT
theories \cite{GUT}, where, for instance, four fermions effective
interactions emerge by integrating out the heavy gauge fields
\cite{Buras:1977yy}.
At the classical level the effective lagrangian for $\chi$ is
\be
\mathcal{L}_{eff}=-\frac{1}{2}\partial_{\nu}\chi\partial^{\nu}\chi-\frac{1}{2}\mu^2\chi^2
-\frac{1}{4}\left(\lambda_{\chi}-\frac{4a^2}{\lambda_{\varphi}}\right)\chi^4+...\,,\label{effsimple}
\ee
where the dots represent higher dimensional operators. The term
$a^2\chi^4/\lambda_{\varphi}$ is the contribution of the heavy mode.
The result (\ref{effsimple}) was originally derived in
\cite{Chan:1979ce}, but here we want also to study the effective
theory with spontaneous symmetry breaking and we want to compare it
with the low energy limit of the fundamental theory.

 For $\mu^2>0$, the minimum of the effective theory potential
is for $\chi=0$. Instead for $\mu^2<0$ we have
\be
\chi=\sqrt{\frac{-\mu^2}{\lambda_{\chi}-\frac{4a^2}{\lambda_{\varphi}}}}\label{chieff}\ee
and the fluctuation $\d\chi$ over this background has the following
mass squared:
\be M^2(\d\chi)=-2\mu^2. \label{Meff}\ee
This results will be not modified by the higher dimensional operator
at the leading order\footnote{The mass $\mu$ is small in the sense
$|\mu|\ll|m_{\varphi}|$.} in $\mu$. The equations (\ref{chieff}) and
(\ref{Meff}) represent the effective theory prediction for the VEV
and the spectrum in the phase where $Z_2'$ is broken.

On the other hand, a solution of the fundamental EOM, namely the EOM
derived from the fundamental lagrangian $\mathcal{L}$, is
\bea
\chi^2&=&\frac{-\mu^2}{\lambda_{\chi}-\frac{4a^2}{\lambda_{\varphi}}}+O(\mu^3),\nonumber
\\
\varphi^2&=&-\frac{m^2_{\varphi}}{\lambda_{\varphi}}+\frac{2a\mu^2}{\lambda_{\varphi}\lambda_{\chi}-4a^2}+O(\mu^3)
\label{correctsimpleVEV}\eea
which is a small deformation of (\ref{simplefirstbac}) at the
leading non-trivial order in $\mu$ and breaks the $Z_2'$ symmetry.
Moreover the light mode which corresponds to this solution has a
mass squared $-2\mu^2$.

Therefore the effective theory prediction for the light mode VEV and
spectrum is correct, at the order $\mu$, in this simple framework,
but the heavy mode contribution is necessary in order the effective
theory prediction to be correct.

\section{A More General Case}\label{general}
\setcounter{equation}{0}

Now we want to extend the result of Section \ref{simple4D} and ref
\cite{Chan:1979ce} to a more general class of theories. We consider
a set of real D-dimensional scalars $\Phi_i$ with a general
potential $V$: the lagrangian is

\be \mathcal{L}=-\frac{1}{2}\partial_M\Phi_i\partial^M\Phi_i -
V(\Phi), \ee
where $M,N,...$ run over all the space-time dimensions, while
$\mu,\nu,...$ and $m,n,...$ are respectively the 4D and the internal
coordinates indices. The EOM are
\be
\partial_M\partial^M\Phi_i-\frac{\partial V}{\partial\Phi_i}(\Phi)=0.\label{generalEOM}
\ee
We consider now a solution $\Phi_{eff}$ of (\ref{generalEOM}) which
preserves the 4D Poincaré invariance and some internal symmetry
group $\mathcal{G}$; the corresponding mass squared eigenvalue
problem for the 4D states is
\be -\partial_m\partial_m\d
\Phi_i+\frac{\partial^2V}{\partial\Phi_i\partial
\Phi_j}(\Phi_{eff})\d \Phi_j=M^2\d\Phi_i,\label{generaleigen} \ee
where $\d \Phi$ is the fluctuation around $\Phi_{eff}$. We assume
that there are $n$ normalizable solutions $\D_l$ with small
eigenvalues ($M^2\sim \mu^2$), other, in principle infinite,
solutions\footnote{In principle $h$ can be a discrete or a
continuous variable.} $\tilde{\D}_h$with large eigenvalues ($M^2\gg
|\mu^2|$) and nothing else. These hypothesis are needed in order to
define the concept of light KK modes.

We can expand the scalars $\Phi_i$ as follows
\be
\Phi_i=\left(\Phi_{eff}\right)_i+\chi_l(x)\D_{li}(y)+\tilde{\chi}_h(x)\tilde{\D}_{hi}(y),\label{expansion}
\ee
where $\chi_l$ and $\tilde{\chi}_h$ are respectively the light and
heavy KK modes. We choose the $\D_l$ and $\tilde{\D}_h$ in order
that they form an orthonormal basis for the functions over the
internal space:
\bea \left<\D_l|\D_{l'}\right>&\equiv& \int
d^{D-4}y\D_{li}(y)\D_{l'i}(y)=\d_{ll'},\nonumber \\
\left<\tilde{\D}_h|\tilde{\D}_{h'}\right>&\equiv& \int
d^{D-4}y\tilde{\D}_{hi}(y)\tilde{\D}_{h'i}(y)=\d_{hh'},\nonumber
\\
\left<\D_l|\tilde{\D}_{h}\right>&\equiv& \int
d^{D-4}y\D_{li}(y)\tilde{\D}_{hi}(y)=0. \eea
We note that $\chi_l$ and $\tilde{\chi}_h$ could both belong to some
non-trivial representation of the internal symmetry group
$\mathcal{G}$.

\subsection{The Effective Theory Method}\label{EFFGeneral}

We construct now some relevant terms in the effective theory for the
light KK modes $\chi_l$. Here "relevant terms" mean relevant terms
in the classical limit and in case we have a small point of minimum
of the order $\mu$ of the effective theory potential: we want to
compare the results of the effective theory for the light KK modes
with the low energy limit of the fundamental theory expanded around
a vacuum which is a small perturbation of $\Phi_{eff}$. Further we
calculate everything at leading non-trivial order\footnote{The $\mu$
mass scale is small in the sense $|\mu|$ is much smaller than the
heavy masses.} in $\mu$.
 The relevant terms can be
computed by putting just the light KK modes in the action and
performing the integration over the extra dimensions and then by
taking into account the effect of heavy KK modes through the
diagrams like Fig. \ref{simplediagram}. In order to calculate those
diagrams, we give the interactions between two light modes $\chi_l$
and one heavy mode $\tilde{\chi}_h$:
\be -\frac{1}{2}\left(\int d^{D-4}y
V_{ijk}\D_{li}\D_{mj}\tilde{\D}_{hk}\right)\chi_l\chi_m\tilde{\chi}_h,
\ee
where we have used the notation
\be V_{i_1 ... i_N}\equiv \frac{\partial^N V}{\partial\Phi_{i_1} ...
\partial\Phi_{i_N} }\left(\Phi_{eff}\right). \ee
We get the following relevant terms in the effective theory
potential $\U$:
\be \U(\chi)=\frac{1}{2}c_l\mu^2\chi_l\chi_l
+\frac{1}{3}\lambda^{(3)}_{lmp}\chi_l\chi_m\chi_p+\frac{1}{4}\lambda_{lmpq}^{(4)}\chi_l\chi_m\chi_p\chi_q+...
\,,\label{Ugeneral}\ee
where the dots represent non relevant terms, $c_l$ are dimensionless
numbers and
\bea\lambda^{(3)}_{lmp}&\equiv&\frac{1}{2}\int d^{D-4}y
V_{ijk}\D_{li}\D_{mj}\D_{pk},\\
\lambda_{lmpq}^{(4)}&\equiv&\frac{1}{3!}\left(\int d^{D-4}y
V_{ijkk'}\D_{li}\D_{mj}\D_{pk}\D_{qk'}\right)+a_{lmpq},
\label{l4}\eea
where the quantities $a_{lmpq}$ represent the heavy modes
contribution and they are given by
\be a_{lmpq}=c_{lmpq}+c_{lpmq}+c_{lqpm} \ee
and
\be c_{lmpq}\equiv -\frac{1}{6}\int
d^{D-4}yd^{D-4}y'V_{ijk}(y)\D_{li}(y)\D_{mj}(y)
G_{kk'}(y,y')V_{i'j'k'}(y')\D_{pi'}(y')\D_{qj'}(y').\ee
The object $G_{kk'}$ is the Green function for the mass squared
operator at the left hand side of (\ref{generaleigen}) and it's
explicitly given by
\be G_{kk'}(y,y')=\sum_h
\frac{1}{m_h^2}\tilde{\D}_{hk}(y)\tilde{\D}_{hk'}(y'),\ee
where $m^2_h$ is the eigenvalue associated to the eigenfunction
$\tilde{\D}_h$.

In the rest of this section we consider the predictions of the
effective theory with spontaneous symmetry breaking.  The potential
(\ref{Ugeneral}) has to be considered as a generalization of
(\ref{effsimple}), which was originally derived in
\cite{Chan:1979ce}. A non vanishing VEV breaks in general
$\mathcal{G}$ to some subgroup and it must satisfies
\bea \frac{\partial
\U}{\partial\chi_l}&=&c_l\mu^2\chi_l+\lambda^{(3)}_{lmp}\chi_m\chi_p+\lambda_{lmpq}^{(4)}\chi_m\chi_p\chi_q=
0. \label{effvacua}\eea
Since we require that $\chi_l$ goes to zero as $\mu$ goes to zero we
have
\be \chi_l=\chi_{l1}+\chi_{l2}+... \ee
where $\chi_{l1}$ is proportional to $\mu$, $\chi_{l2}$ is
proportional to $\mu^2$ and so on. At the order $\mu^2$ the
equations (\ref{effvacua}) reduce to
\be \lambda^{(3)}_{lmp}\chi_{m1}\chi_{p1}=0 \ee
which implies
\be \lambda^{(3)}_{lmp}\chi_{p1}=0. \label{eff2vacua}\ee
While, at the order $\mu^3$, the equations (\ref{effvacua}) reduce
to
\be
c_l\mu^2\chi_{l1}+\lambda_{lmpq}^{(4)}\chi_{m1}\chi_{p1}\chi_{q1}=0,\label{eff3vacua}
\ee
where we have used the equations (\ref{eff2vacua}).

Finally the mass spectrum corresponding to a solution of
(\ref{effvacua}) is given by the eigenvalues of the hessian matrix
of $\U$ in that solution:
\be \frac{\partial^2 \U}{\partial\chi_l
\partial
\chi_{l'}}=c_l\mu^2\d_{ll'}+2\lambda^{(3)}_{ll'm}\chi_m+
3\lambda_{ll'mq}^{(4)}\chi_m\chi_q. \ee
If we assume, for simplicity, $\lambda^{(3)}_{ll'm}=0$, which
corresponds to the absence of cubic terms in $\U$, the leading order
approximation of the hessian is simply given by
\be \frac{\partial^2 \U}{\partial\chi_l
\partial
\chi_{l'}}=c_l\mu^2\d_{ll'}+
3\lambda_{ll'mq}^{(4)}\chi_{m1}\chi_{q1}+O(\mu^3).\label{hessian}
\ee
In Subsection \ref{correct} we show that this matrix, which
represents the mass spectrum for the light KK modes, and the
equations (\ref{eff2vacua}) and (\ref{eff3vacua}) for the light
modes VEVs are exactly reproduced by a D-dimensional analysis.

\subsection{D-dimensional analysis}\label{correct}

Now we present a D-dimensional (or geometrical) approach to compute
low energy quantities: we want to find a solution of
(\ref{generalEOM}) which is a small perturbation, of the order
$\mu$, of $\Phi_{eff}$ and then we want to find the low energy mass
spectrum of the fluctuations around this solution. In general this
solution will break $\mathcal{G}$ to some subgroup like a solution
of (\ref{effvacua}) does in the effective theory method. The
explicit form of such solution in the simple case of Section
\ref{simple4D} is given by (\ref{correctsimpleVEV}) and the low
energy mass spectrum in that simple case is represented by the
squared mass $-2\mu^2$; now we want to generalize these results.

Let us consider the expansion (\ref{expansion}); we observe that the
statement that the solution is a small perturbation of $\Phi_{eff}$
means
\bea&&\chi_l=\chi_{l1}+\chi_{l2}+...\,, \nonumber\\
&&\tilde{\chi}_h=\tilde{\chi}_{h1}+\tilde{\chi}_{h2}+...\,,\label{analitic}
\eea
that is there are no big $\mu-$independent terms in $\chi_l$ and
$\tilde{\chi}_h$. We consider now a Taylor expansion of the
equations (\ref{generalEOM}) around $\Phi_{eff}$:
\bea
&&\partial_m\partial_m\left(\Phi_i-\left(\Phi_{eff}\right)_i\right)\nonumber
\\
&&-\sum_{k=1}^{N}\frac{1}{k!}V_{ii_1...i_k}
\left(\Phi_{i_1}-\left(\Phi_{eff}\right)_{i_1}\right)\cdot ... \cdot
\left(\Phi_{i_k}-\left(\Phi_{eff}\right)_{i_k}\right)\nonumber\\
&&+O(\mu^{N+1})=0.\label{pertEOM}\eea
At the order $\mu$ the equations (\ref{pertEOM}) reduce to
\be \left(\partial_m
\partial_m\d_{ij}-V_{ij}\right)\left(\Phi_{j}-\left(\Phi_{eff}\right)_{j}\right)+O(\mu^2)=0,
\ee
which simply states
\be \tilde{\chi}_{h1}=0.\label{heavyVEV}\ee
 Moreover at the order $\mu^2$ the
equations (\ref{pertEOM}) imply
\be
\tilde{\chi}_{h2}\left(\partial_m\partial_m\d_{ij}-V_{ij}\right)\tilde{\D}_{hj}=
\frac{1}{2}V_{ijk}\D_{lj}\D_{mk}\chi_{l1}\chi_{m1},
\label{pertEOM2}\ee
which has two consequences: the first one is
\be \lambda^{(3)}_{lmp}\chi_{p1}=0,\label{con1}\ee
which can be derived from (\ref{pertEOM2}) by projecting over $\D_l$
and it exactly reproduces (\ref{eff2vacua}) of the effective theory
method; the second consequence is
\be
\tilde{\chi}_{h2}\tilde{\D}_{hi'}(y)=-\frac{1}{2}\chi_{l1}\chi_{m1}\int
d^{D-4}y'G_{i'i}(y,y')V_{ijk}(y')\D_{lj}(y')\D_{mk}(y'),\label{con2}\ee
where $G$ still represents the Green function for the operator at
the left hand side of (\ref{generaleigen}). Now we can write the
$\mu^3$ part of the Eq. (\ref{pertEOM}) as follows
\bea
&&-c_l\mu^2\chi_{l1}\D_{li}-m^2_h\tilde{\chi}_{h3}\tilde{D}_{hi}\nonumber
\\
&&-\frac{1}{2}V_{ijk}\chi_{l1}\D_{lj}\left(\tilde{\chi}_{h2}\tilde{\D}_{hk}+
\chi_{m2}\D_{mk}\right)\nonumber\\
&&-\frac{1}{2}V_{ijk}\left(\chi_{l2}\D_{lj}+
\tilde{\chi}_{h2}\tilde{\D}_{hj}\right)\chi_{m1}\D_{mk}\nonumber\\
&&-\frac{1}{3!}V_{ijkk'}\D_{lj}\D_{mk}\D_{pk'}\chi_{l1}\chi_{m1}\chi_{p1}=0.\eea
If one projects this equation over $\D_l$ and uses the equations
(\ref{con1}) and (\ref{con2}) one gets exactly the equations
(\ref{eff3vacua}). Therefore, at the order $\mu$, all the solutions
of (\ref{effvacua}) are reproduced by the D-dimensional analysis and
viceversa. Moreover we observe that these light KK modes VEVs,
predicted by the effective theory, constitute approximate solutions
of the fundamental D-dimensional EOM at leading non-trivial order
because of the Eq. (\ref{heavyVEV}), which states that the heavy KK
modes VEVs are higher order quantity with respect to the light KK
modes VEVs.

Now we consider the mass squared eigenvalue problem which
corresponds to a solution $\Phi$; moreover we assume for simplicity
$\lambda^{(3)}_{lmp}=0$, like in the effective theory method. This
eigenvalue problem is

\be \mathcal{O}_{ij}\d\Phi_j\equiv -\partial_m\partial_m\d
\Phi_i+\frac{\partial^2V}{\partial\Phi_i\partial \Phi_j}(\Phi)\d
\Phi_j=M^2\d\Phi_i,\label{perteigen} \ee
where $\d \Phi_i$ represents the fluctuations of the scalars around
the solution $\Phi$. We observe now that the equation
(\ref{perteigen}) can be considered a time-independent Schrodinger
equation: $\mathcal{O}$ is the hamiltonian and $M^2$ the generic
energy level. Moreover we can perform a Taylor expansion of
$\mathcal{O}$ around $\mu=0$:
\be \mathcal{O}=\mathcal{O}_0+\mathcal{O}_1+\mathcal{O}_2+... \,.\ee
The operators $\mathcal{O}_1$ and $\mathcal{O}_2$ can be easily
expressed just in terms of $\chi_{l1}$ and $\chi_{l2}$ by using
(\ref{expansion}), (\ref{analitic}) and the constraints (\ref{con2})
and (\ref{heavyVEV}) which come from the EOM. From the perturbation
theory of quantum mechanics we know that the leading value of the
low energy mass spectrum is given by the eigenvalues of the
following mass squared matrix:
\be M^2_{ll'}\equiv A_{ll'}+B_{ll'},\ee
where
\be A_{ll'}\equiv <\D_l|\mathcal{O}_2|\D_{l'}>\ee
and
\be B_{ll'}\equiv
-\sum_{h}\frac{1}{m^2_h}<\D_l|\mathcal{O}_1|\tilde{\D}_h><\tilde{\D}_h|\mathcal{O}_1|\D_{l'}>.\ee
If one express the matrices $A$ and $B$ in terms\footnote{The
dependence on $\chi_{l2}$ disappears because we assume
$\lambda^{(3)}_{lmp}=0$, as one can easily check.} of $\chi_{l1}$ ,
one finds exactly the corresponding result (\ref{hessian}) predicted
by the effective theory.

So we have two equivalent (at least at the leading non-trivial order
in $\mu$) approaches to study the breaking of $\mathcal{G}$: the
spontaneous symmetry breaking in the 4D effective theory and the
D-dimensional analysis. We stress that, like in the simple model of
Section \ref{simple4D}, also in this more general case the heavy KK
mode contribution in the effective theory can't be neglected if one
wants to reproduce the D-dimensional result, even at the classical
level. In general this is true not only in scalar theories but also
in theories which involve gauge and gravitational interactions, as
we illustrate in Chapter \ref{EMS}.

\section{A Domain Wall Example}\label{DomainWall}\setcounter{equation}{0}

Here we provide an explicit application of previous results in the
case of brane world models, where we have a non-trivial heavy KK
mode contribution to the 4D effective theory. Moreover we study the
role of heavy modes in the interactions of the 4D effective theory:
we analyze the cubic interaction of the {\it Higgs field} after SSB,
which is reproduced by the geometrical approach as well as masses
and VEV.

\subsection{The Model}

 We consider a
5D model with two scalar fields $\varphi$ and $\phi$.
The lagrangian is
\bea \mathcal{L}&=&
-\frac{1}{2}\partial_{M}\varphi\partial^{M}\varphi-\frac{1}{2}\partial_{M}\phi\partial^{M}\phi
-\frac{1}{2}m^2\phi^2 \nonumber
\\
&&-\frac{1}{4}\xi\phi^4
-\lambda(\varphi^2-v^2)^2-\frac{\alpha}{2}\varphi^2\phi^2, \eea
where $m^2$, $\xi$, $\lambda$, $v$ and $\a$ are real
parameters\footnote{We consider only the values of these parameters
which correspond a potential bounded from below.}. This model
reduces to the one of Section \ref{DW} for $\phi=0$. The internal
symmetry of this theory is $Z_2\times Z_2'$, where
\bea &&Z_2:\varphi\rightarrow \pm \varphi,\nonumber \\
&&Z_2':\phi\rightarrow \pm \phi. \eea
The EOM are
\bea
&&\partial_M\partial^M\varphi-4\lambda(\varphi^2-v^2)\varphi-\a\phi^2\varphi=0,\nonumber
\\
&&\partial_M\partial^M\phi-m^2\phi-\xi\phi^3-\a\varphi^2\phi=0. \eea
Like in Section \ref{DW} we consider the domain wall solution for $\varphi$, whereas $\phi$ is assumed to vanish
at the background level:
\be \phi=0,\quad \quad \varphi=v\tanh(\sqrt{2\lambda} v y)\equiv
\varphi_c(y). \label{domain}\ee
The VEV in (\ref{domain}) preserves the internal symmetry $Z_2'$.
The mass squared eigenvalue problem which corresponds to solution
(\ref{domain}), is
\bea \mathcal{O}^{(1)}\d\varphi\equiv&&-\partial_y^2\d
\varphi+4\lambda\left(3\varphi_c^2-v^2\right)\d\varphi=M^2_{\d\varphi}\d\varphi,\label{fluctKink}\\
\mathcal{O}^{(2)}\phi\equiv&&-\partial_y^2\phi+m^2\phi+\a\varphi_c^2\phi=M^2_{\phi}\phi,\label{fluctphi}\eea
where $\d\varphi$ is the fluctuation of $\varphi$ around
$\varphi_c$. Equations (\ref{fluctKink}) and (\ref{fluctphi}) are
studied in the literature \cite{Dashen,DomainWall,Morse,Kink}. They
are Schroedinger equations with a potential\footnote{We require
$\a>0$ in order to get a localized wave function from equation
(\ref{fluctphi}).} $V(y)=a\tanh^2(\sqrt{2\lambda} v y)+b$, where $a$
and $b$ are constants. Like in Section \ref{DW} we can derive boundary conditions of the form (\ref{boundarydomain})
for both $\varphi$ and $\phi$. Therefore we project out exponentially
growing solutions of (\ref{fluctKink}) and (\ref{fluctphi}). There
is a wave function $\D_1$ for $\d\varphi$ with $M^2_{\d\varphi}=0$:
\be \D_1(y)=\frac{N_1}{\cosh^2(\sqrt{2\lambda}vy)}, \ee
where $N_1$ is a normalization constant such that $\int dy
(\D_1(y))^2=1.$
Concerning $\phi$, we find an eigenfunction
\be \D_2(y)=\frac{N_2}{\cosh^{\sigma}(\sqrt{2\lambda}vy)},\ee
where $N_2$ ensures $\int dy (\D_2(y))^2=1.$
and
\be \sigma\equiv
\frac{1}{2}\left(\sqrt{1+2\frac{\a}{\lambda}}-1\right)\ee
The mass which corresponds to $\D_2$ is very small if we choose
\be m^2=-(1+\eta)M_0^2\label{m2}\ee
where
\be M_0^2\equiv \frac{2\a v^2}{\sqrt{1+2\frac{\a}{\lambda}}+1}\ee
and $\eta$ is a very small dimensionless parameter. Henceforth we
assume (\ref{m2}) so the mass squared
\be M_{\phi}^2= -\eta M^2_0\equiv \mu^2, \ee
associated to $\D_2$, is small. The wave functions $\D_1$ and $\D_2$
are the ground states of the Schroedinger equations
(\ref{fluctKink}) and (\ref{fluctphi}) respectively, because they
have no nodes. All the remaining solutions of (\ref{fluctKink}) and
(\ref{fluctphi}) have $M^2_{\d\varphi},\,M^2_{\phi}\gg |\mu^2|$. We
can perform the following expansion
\bea
\varphi(x,y)&=&\varphi_c(y)+\chi_1(x)\D_1(y)+\sum_h\tilde{\chi}_{h1}(x)\tilde{\D}_{h1}(y),\nonumber
\\
\phi(x,y)&=&\chi_2(x)\D_2(y)+\sum_h\tilde{\chi}_{h2}(x)\tilde{\D}_{h2}(y),
\eea
where $\tilde{\D}_{h1}$ and $\tilde{\D}_{h2}$ are the solutions of
(\ref{fluctKink}) and (\ref{fluctphi}) with
$M^2_{\d\varphi},\,M^2_{\phi}\gg |\mu^2|$ and $\chi_{i}$ and
$\tilde{\chi}_{hi}$ are respectively the light modes and the heavy
modes.

\subsection{The 4D Effective Theory for the Light Modes}

The relevant terms in the effective theory for the light modes can
be computed with the general argument given in Section
\ref{EFFGeneral}. By using the residual $Z_2'$ symmetry and the fact
that $\D_i$ are even functions of $y$ while $\varphi_c$ is odd we
find that the effective theory potential has the following form:
\be \U(\chi)=\frac{1}{2}\mu^2\chi_2^2+\frac{1}{4}\lambda_1\chi_1^4
+\frac{1}{4}\lambda_2\chi_2^4+a\chi_1^2\chi_2^2+...\,,\label{5DeffPot}\ee
where the dots represent higher order operators. The semiclassical
approximation of $\lambda_1$, $\lambda_2$ and $a$ including the
heavy mode contribution is given by
\bea \lambda_1&=&\lambda_{1L}+\lambda_{1H},\\
\lambda_2&=&\lambda_{2L}+\lambda_{2H},\\
a &=&a_L+a_{H},\eea
where $\lambda_{1L}$, $\lambda_{2L}$ and $a_L$ represent the light
mode contribution and they are explicitly given by
\bea \lambda_{1L}=4\lambda\int dy (\D_1(y))^4, \quad \lambda_{2L}=\xi\int dy (\D_2(y))^4,\nonumber \\
a_L=\frac{\a}{2}\int dy (\D_1(y))^2(\D_2(y))^2, \label{light}\eea
whereas $\lambda_{1H}$ $\lambda_{2H}$ and $a_H$ represent the heavy
mode contribution:
\bea \lambda_{1H}&=&-\frac{1}{2}(4!\lambda)^2\int\,dy\, dy'
(\D_1(y))^2\varphi_c(y)G_1(y,y')(\D_1(y'))^2\varphi_c(y'),
\nonumber \nonumber \\
\lambda_{2H}&=&-2\a^2\int\,dy\, dy' (\D_2(y))^2\varphi_c(y)G_1(y,y')(\D_2(y'))^2\varphi_c(y'),\nonumber \\
a_H &=& a_{H1}+a_{H2}=-12\a\lambda\int dy dy' (\D_2(y))^2\varphi_c(y)G_1(y,y')(\D_1(y'))^2\varphi_c(y')\nonumber \\
&&-2\a^2\int dy dy'
\D_1(y)\D_2(y)\varphi_c(y)G_2(y,y')\D_1(y')\D_2(y')\varphi_c(y'),\label{heavy}\eea
where the integrals over $y$ are from $-\infty$ to $+\infty$, $G_i$
are defined by
\be G_i(y,y')\equiv \sum_h
\frac{1}{m^2_{hi}}\tilde{\D}_{hi}(y)\tilde{\D}_{hi}(y')\label{Greendef}\ee
and $m_{hi}$ is the mass of $\tilde{\chi}_{hi}$. The functions in
(\ref{Greendef}) are the Green functions for the operators
$\mathcal{O}^{(i)}$ defined in equations\footnote{More precisely
$G_2$ is the Green function of $\mathcal{O}^{(2)}$ with
$m^2=-M_0^2$.} (\ref{fluctKink}) and (\ref{fluctphi}):
\be \mathcal{O}^{(i)}G_i(y,y')=\d(y-y')-\D_i(y)\D_i(y').
\label{Greenproperty}\ee
The general solution of (\ref{Greenproperty}) is
\bea
&&G_i(y,y')=c_i^{(1)}(y')\D_i(y)+c^{(2)}_i(y')\D_i^{\bot}(y)\nonumber\\
&&+\theta(y-y')\left(\D_i(y')\D_i^{\bot}(y)-\D_i(y)\D_i^{\bot}(y')\right)
\nonumber\\
&&+\D_i(y')\left[\D_i(y)\int_0^ydy''\D_i(y'')\Do_i(y'')-\Do_i(y)\int_0^y
dy'' (\D_i(y''))^2\right],\label{G}\eea
where $c_i^{(1)}(y')$ and $c^{(2)}_i(y')$ are generic functions of
$y'$ and
$$\Do_i(y)\equiv -\D_i(y)\int_0^y \frac{dy'}{(\D_i(y'))^2}; $$
$\D_i$ and $\Do_i$ are two independent eigenfunctions of
$\mathcal{O}^{(i)}$ with vanishing eigenvalue. In our case we can
compute $c^{(2)}_i(y')$ by requiring $G_i(y,y')$, as function of
$y$, to have no exponentially growing part: we get
\be c^{(2)}_i(y')=-\frac{1}{2}\D_i(y').\ee
Moreover, from (\ref{Greendef}), we are forced to project
$G_i(y,y')$ in the subspace orthogonal to $\D_i(y)$, this fixes
$c_i^{(1)}(y')$; however, here we do not need the explicit
expression for $c_i^{(1)}(y')$.

\subsubsection{Computation of the Effective Coupling
Constants}\label{eff-coupling}

By using the explicit form of $\D_{i}$ and $G_{i}$ we can compute
$\lambda_1$, $\lambda_2$ and $a$ by means of equations (\ref{light})
and (\ref{heavy}). The expressions of such coupling constants
simplify because $\D_i$ is even and $\varphi_c$ is odd under
$y\rightarrow -y$; in particular first and third line in (\ref{G})
do not give any contribution to $\lambda_1$, $\lambda_2$ and $a$
because of those parities. Concerning $\lambda_1$, a non-trivial
balancing between the heavy mode contribution and the light modes
one gives
\be \lambda_1=0. \ee
The explicit expression for $\lambda_2$ is
\be \lambda_2=\frac{N_2^4}{v\sqrt{2\lambda}}\xi
J_L(\sigma)\left(1+\frac{2\a^2}{\lambda
\xi}\frac{J_H(\sigma)}{J_L(\sigma)}\right), \label{lambda2}\ee
where
\be J_L(\sigma)\equiv
\int_{-\infty}^{+\infty}\frac{dx}{\cosh^{4\sigma}(x)}=
\frac{\sqrt{\pi} \Gamma(2\sigma)}{\Gamma(\frac{1}{2}+2\sigma)}, \ee
where $\Gamma$ is the Euler gamma function, and \be
J_H(\sigma)=-\frac{1}{2+2\sigma}\int_{-\infty}^{+\infty}dx\frac{\tanh(x)}{\cosh^{4+4\sigma}(x)}
\left(\frac{3}{8}x+\frac{1}{4}\sinh(2x)+\frac{1}{32}\sinh(4x)\right).
\ee
The ratio
\be \rho\equiv\frac{2\a^2}{\lambda
\xi}\frac{J_H(\sigma)}{J_L(\sigma)} \label{rhogeneral}\ee
in (\ref{lambda2}) represents the heavy mode contribution to this
coupling constant. In general $\rho$ is not negligible: for
instance, by choosing $\sigma=2$ that is $\alpha=12\lambda$, we get
\be \rho = -\frac{1}{3}\frac{\alpha}{\xi}, \label{rho}\ee
which is not small for $\alpha$, $\xi$ $\sim 1$. Finally we note
that $a$ depends on $\mu^2$, because of $G_2$ in $a_{H2}$. We
perform a Taylor expansion of this coupling around $\mu^2=0$:
\be a=a_0+O(\mu^2), \ee
where
\be
a_0=N_1^2N_2^2\frac{\a}{v\sqrt{2\lambda}}\left(\frac{1}{2}I_L(\sigma)+6I_{H1}(\sigma)+\frac{\a}{\lambda}I_{H2}(\sigma)\right),
\ee
and
\bea I_L(\sigma)&\equiv& \int_{-\infty}^{+\infty}\frac{dx}{\cosh^{2\sigma+4}(x)},\nonumber \\
I_{H1}(\sigma)&\equiv &\int_{-\infty}^{+\infty}dx
\int_{-\infty}^{+\infty}dx'\frac{\tanh(x)}{\cosh^{2+2\sigma}(x)}\theta(x-x')
\int_{x'}^{x}ds\cosh^4(s)\frac{\tanh(x')}{\cosh^{6}(x')},\nonumber\\
I_{H2}(\sigma)&\equiv &\int_{-\infty}^{+\infty}dx
\int_{-\infty}^{+\infty}dx'\frac{\tanh(x)}{\cosh^{2+2\sigma}(x)}\theta(x-x')
\int_{x'}^{x}ds\cosh^{2\sigma}(s)\frac{\tanh(x')}{\cosh^{2+2\sigma}(x')},\nonumber\eea
If we set for simplicity $\sigma=2$, we get $a_0=0$ again by means
of a non-trivial balance between light and heavy mode contribution.

\subsubsection{Broken Effective Theory and the Role of Heavy Modes}

Here we show that the heavy modes have a non-trivial role in the low
energy physics. The potential now looks like
\be
\U(\chi)=\frac{1}{2}\mu^2\chi_2^2+\frac{1}{4}\lambda_2\chi_2^4+a\chi_1^2\chi_2^2+...\,,\ee
where the dots represent higher order terms (powers of $\chi_{1,2}$
greater than $4$) and $a=O(\mu^2)$.

A consistent vacuum at the leading order in $\mu$ is (for
$\mu^2<0$):
\be <\chi_1>=0,\quad
<\chi_2>=\sqrt{\frac{-\mu^2}{\lambda_2}}\label{vacuum} \ee
Vacuum (\ref{vacuum}) spontaneously breaks $Z_2'$.
The corresponding leading order mass spectrum of the fluctuations around (\ref{vacuum}) is
\be M_1^2=0,\qquad M_2^2=-2\mu^2. \ee
 We observe
that the contribution of the heavy modes to $<\chi_2>$ is not
trivial because the quantity $\rho$ in (\ref{rho}) is not
negligible. This contribution is a modification of the cubic
self-interaction of $H\equiv\chi_2-<\chi_2>$:
\be \mathcal{U}(\chi_1,H)=\lambda_2<\chi_2>H^3+O(\mu^2).\ee
We observe that this is the only cubic interaction at the order
$\mu$. Such an interaction is reproduced by the 5D analysis
explained in Subsection \ref{correct}, as one can expect, only if
the heavy mode contributions are taken into account, that is only if
the quantity $\rho$ in (\ref{rhogeneral}) is not neglected.

So in this section we have extended the results of Sections
\ref{simple4D} and \ref{general}, concerning only the spectral
aspects of the broken effective theory, by including the description
of the heavy mode contribution to interactions in the broken
effective theory. Also at the interaction level such contribution
turns out to be needed to reproduce the D-dimensional, in this case
5-dimensional, approach.

However, we observe that the models presented in this chapter do not
include gauge and gravitational interactions. Therefore in the next
chapter we will introduce a gauge and gravitational model and we
will discuss again the role of the heavy modes in the effective
theory taking into account the extra terms in the lagrangian that
are implied by gravity and gauge invariance.


\chapter{6D Einstein-Maxwell-Scalar Model.}\label{EMS}

In Chapter \ref{GHBscalar} we have proved that the heavy mode
contribution is necessary to reproduce the correct low energy
dynamics because, without this contribution, the 4D effective theory
approach cannot reproduce in general the D-dimensional (or
geometrical) approach to spontaneous symmetry breaking. The aim of
the present chapter is to prove a similar statement in a more
interesting context which can be extended to a semi-realistic
theory. Our model will be an Einstein-Maxwell-Scalar model in 6D,
which will be compactified over an internal space with the $S^2$
topology. This is an ordinary KK theory in which the KK mass scale
is naturally of the order of the Planck mass. We will be able to
prove that the heavy mode contribution is not negligible even if
they have such a large mass.


\section{Definition of the Model and \\ 6D Equations of
Motion}\label{defEMS}

We consider a 6D field theory of gravity with a $U(1)$ gauge
invariance, including a charged scalar field $\phi$ and eventually
fermions. The bosonic action is\footnote{Our conventions are fixed
in Appendix \ref{GR}.}
\be S_B=\int d^6 X\sqrt{-G}
\left[\frac{1}{\kappa^2}R-\frac{1}{4}F_{MN}F^{MN} -(\nabla_M
\phi)^*\nabla^M \phi -V(\phi)  \right], \label{action} \ee
where $R$ is the Ricci scalar, $\kappa$ represents the 6D Planck
scale, $F_{MN}$ is the field strength of the $U(1)$ gauge field
$A_M$, defined by
\be F_{MN}=\partial_M A_N- \partial_N A_M \ee
and
\be \nabla_M \phi = \partial_M \phi +ie A_M \phi,\ee
where $e$ is the $U(1)$ gauge coupling. Moreover $V$ is a scalar
potential and we choose
\be V(\phi)=m^2\phi^*\phi +\xi (\phi^* \phi)^2 + \lambda,  \label{V}
\ee
where $m^2$ and $\xi$ are generical real constants, with the
constraint $\xi > 0$ and $\lambda $ represents the 6D cosmological
constant.

From the action (\ref{action}) we can derive the general bosonic
EOM. However, we focus on the following class of backgrounds, which
are invariant under the 4D Poincar\'e group:
\bea ds^2 &=&\eta_{\mu \nu}dx^{\mu}dx^{\nu}
+g_{mn}(y)dy^m dy^n.\label{Bmetric}\\
A&=&A_m(y)dy^m, \label{BA}\\
\phi &=&\phi(y), \label{Bphi} \eea
where $g_{mn}$ is the metric of a  2-dimensional compact internal
manifold $K_2$; so the 6D space-time manifold is $(Minkowski)_4
\times K_2$. By using (\ref{Bmetric}), (\ref{BA}) and (\ref{Bphi}),
we can write the bosonic EOM in the following form:
\bea &&\nabla^2 \phi -m^2\phi-2\xi(\phi^* \phi)\phi=0,\nonumber \\
&& \nabla_mF^{mn}+ie\left[\phi^*\nabla^n\phi
-(\nabla^n \phi)^*\phi \right]=0,\nonumber \\
&&\frac{1}{\kappa^2}R_{mn}-\frac{1}{2}F_{mp}F_n^{\,\,p}
-\frac{1}{2}(\nabla_m\phi)^*\nabla_n\phi-
\frac{1}{2}(\nabla_n\phi)^*\nabla_m\phi=0,\nonumber \\
&&\frac{1}{4}F^2-\lambda-m^2\phi^*\phi-\xi(\phi^*\phi)^2=0,
\label{EOM}
 \eea
 where $\nabla^2\equiv\nabla_m\nabla^m$ is the covariant Laplacian over
the internal manifold. The equations (\ref{EOM}) must be satisfied
by the bosonic VEV.

We introduce also fermions and gauge invariant coupling with the
scalar $\phi$. In order to do that it is necessary to introduce at
least a pair of 6D {\it Weyl spinors} $\psi_+$ and $\psi_-$,
where $\psi_+$ and $\psi_-$ are eigenvectors of $\Gamma^7$ with eigenvalues $+1$ and $-1$ respectively\footnote{Our conventions for the
6D gamma matrices are given in Appendix \ref{GR}.}. We
consider the following fermionic action:
\be S_F= \int d^6 X\sqrt{-G} \left(\overline{\psi_+}\Gamma^M
\nabla_M \psi_+ +\overline{\psi_-}\Gamma^M \nabla_M \psi_-
+g_Y\phi^* \overline{\psi_+}\psi_- + g_Y\phi
\overline{\psi_-}\psi_+\right), \label{fermionL}\ee
where $g_Y$ is a real Yukawa coupling constant. In (\ref{fermionL})
$\nabla_M$ represents the covariant derivative acting on spinor,
which includes the gauge and the spin connection. The
$U(1)$ charge $e_+$ and $e_-$ of $\psi_+$ and $\psi_-$ have to
satisfy the condition $e_-=e_+ +e$ coming from the gauge invariance
of the Yukawa terms. In the following we consider the choice
$e_+=e/2$ and $e_-=3e/2$, corresponding to a simple harmonic
expansion for the compactification over $(Minkowski)_4\times S^2$.
From (\ref{fermionL}) we get the following EOM:
\be \Gamma^M \nabla_M \psi_+ +g_Y\phi^* \psi_-=0, \quad \Gamma^M
\nabla_M \psi_- +g_Y\phi \psi_+=0. \ee
Now we define the following 4D Weyl spinors:
\be \psi_{\pm L}=\frac{1-\gamma^5}{2}\psi_{\pm }, \quad \psi_{\pm
R}=\frac{1+\gamma^5}{2}\psi_{\pm}, \ee
where $\gamma^5$ is the 4D chirality matrix. In terms of $\psi_{\pm
L}$ and $\psi_{\pm R}$ the EOM, for a $(Minkowski)_4\times K_2$
background space-time, are\footnote{We rearrange the equations in a
way that the left handed and right handed sector are split.}

\bea \left(\partial^2 + 2 \nabla_+ \nabla_-
-g_Y^2|\phi|^2\right)\psi_{+L}
-\sqrt2 g_Y\left(\nabla_+\phi^* \right)\psi_{-L}=0, \nonumber \\
\left(\partial^2 + 2 \nabla_- \nabla_+ -g_Y^2|\phi|^2\right)\psi_{-L}-\sqrt2 g_Y\left(\nabla_-\phi \right)\psi_{+L}=0, \nonumber \\
\left(\partial^2 + 2 \nabla_- \nabla_+
-g_Y^2|\phi|^2\right)\psi_{+R}
+\sqrt2 g_{Y}\left(\nabla_-\phi^* \right)\psi_{-R}=0, \nonumber \\
\left(\partial^2 + 2 \nabla_+ \nabla_-
-g_Y^2|\phi|^2\right)\psi_{-R} +\sqrt2 g_{Y}\left(\nabla_+\phi
\right)\psi_{+R}=0, \label{fermioneq} \eea
where $\partial^2\equiv \eta^{\mu \nu}\partial_{\mu}\partial_{\nu}$,
\be \nabla_{\pm}= \frac{1}{\sqrt2}(\nabla_5 \pm i\nabla_6) \ee
and $\nabla_{5,6}$ are the covariant derivative components in an
orthonormal basis. The equations (\ref{fermioneq}) will be used in
order to compute the fermionic spectrum.

\section{4D Electroweak Symmetry Breaking}\label{EWSB4D}\setcounter{equation}{0}

\subsection{The $SU(2)\times U(1)$ Background Solution} \label{su2 x
u1}

An $SU(2)\times U(1)$-invariant  solution of (\ref{EOM}) is
\cite{RSS}
\bea ds^2 &=&\eta_{\mu \nu}dx^{\mu}dx^{\nu}+
a^2\left(d\theta^2+\sin^2\theta d\varphi^2\right), \label{sphere}\\
A&=&\frac{n}{2e}(\cos\theta -1)d\varphi
\equiv -\frac{n}{2e}e^3(y), \label{monopole} \\
\phi &=&0, \label{phi0} \eea
subject to the constraints
\be \lambda=\frac{n^2}{8e^2a^4}=\frac{1}{\kappa^2a^2},
\label{constraint}\ee
where $n$ is the monopole number. The metric (\ref{sphere}) is the
sum of the 4D Minkowski metric and the metric of the 2D sphere
$S^2$, with radius $a$. So we have $K_2=S^2$ and our internal space
is maximally symmetric. We use the spherical coordinates $\theta$
and $\varphi$, so $dy^5=a\,d\theta$, $dy^6=a\,d\phi$ . The 1-form
(\ref{monopole}) is a monopole configuration for the $U(1)$ gauge
field. In (\ref{monopole}) $A$ is expressed in the chart $0\leq
\theta <\pi$, $0\leq \varphi <2\pi$. Instead in the chart
$0<\theta\leq\pi$, $0\leq \varphi <2\pi$, $A$ has the form
\be A =\frac{n}{2e}(\cos\theta +1)d\varphi. \label{monopole2} \ee
The two 1-forms (\ref{monopole}) and (\ref{monopole2}) must differ
by a single valued gauge transformation and so we have that $n$ is an
integer. This rule is called Dirac quantization condition. We note that the solution in (\ref{sphere}),
(\ref{monopole}) and (\ref{phi0}) has an $SU(2)\times U(1)$
symmetry. It's useful to introduce an orthonormal basis in the
internal cotangent space \cite{RSS}. We choose the following 1-forms
basis
\be e^{\pm}(y)=\pm \frac{i}{\sqrt{2}}e^{\pm i\varphi}\left(d\theta
\pm i\sin\theta d\varphi \right). \label{epm}\ee
In this basis the metric (\ref{sphere}) has the form
\be ds^2 =\eta_{\mu \nu}dx^{\mu}dx^{\nu}+
a^2\left(e^+e^-+e^-e^+\right). \label{sphere2} \ee
Under a rotation on the sphere we have \cite{RSS}
\bea e^{\pm}&\rightarrow& e^{\mp i\zeta}e^{\pm}, \label{pullbacke12}\\
e^3&\rightarrow& e^3-d\zeta, \label{pullbacke3} \eea
where $\zeta$ depends on the internal coordinates $\theta$ and
$\varphi$, and the group element of $SU(2)$, associated to the
rotation, but it does not depend on the 4D coordinates\footnote{The
transformation laws (\ref{pullbacke12}) and (\ref{pullbacke3}) can
be extended to an $x$-dependent rotation \cite{RSS}.} $x^{\mu}$. So
the 1-form (\ref{monopole}) has the following transformation
property
\be A\rightarrow A+\frac{n}{2e}d\zeta. \ee
We can now introduce the iso-helicity by saying that the
iso-helicity of $e^{\pm}$ is $\pm 1$. Further more, if we consider
the background covariant derivative of $\phi$ and we remember this
object must have the same iso-helicity of $\phi$, we obtain that
$\phi$ has iso-helicity $n/2$. Generally rotations act on tensors
like an $SO(2)$ group, so we can group the components of tensors in
$SO(2)$ irreducible pieces \cite{RSS}: the iso-helicity of a field
is nothing but its $SO(2)$ charge.

Generally if $\Phi_{\lambda}$ is a field with an integer or
half-integer iso-helicity $\lambda$, we can perform an harmonic
expansion \cite{RSS}:
\be \Phi_{\lambda}(x,\theta, \phi)=\sum_{l\geq |\lambda|}
\sum_{|m|\leq l}\Phi_m^l(x)
\sqrt{\frac{2l+1}{4\pi}}\mathcal{D}^{(l)\lambda}_{m}(\theta ,
\varphi), \label{lexpansion}\ee
where, for a given $l$, $\mathcal{D}^{(l)\lambda}_{m}$ is a
$(2l+1)\times (2l+1)$ unitary matrix. For example $\phi$ has an
expansion like (\ref{lexpansion}) with $\lambda=n/2$. The
$\mathcal{D}^{(l)\lambda}_{m}$ were originally introduced in
\cite{Wigner} and in the following we give our conventions.
 We define the harmonics $\mathcal{D}^{(l)\lambda}_m$
as proportional to the matrix element
\be \left<l,\lambda \right|e^{i\varphi
Q_3}e^{i(\pi-\theta)Q_2}e^{i\varphi Q_3}\left|l,m\right>, \ee
where the $Q_j$, $j=1,2,3$, are the generators of $SU(2)$:
\be \left[Q_j,Q_k\right]=i\epsilon_{jkl}Q_l, \ee
where $\epsilon_{jkl}$ is the totally antisymmetric Levi-Civita
symbol with $\epsilon_{123}=1$. Moreover $\left|l,m\right>$ is the
eigenvector of $\sum_j Q_j^2$ with eigenvalue $l(l+1)$ and the
eigenvector of $Q_3$ with eigenvalue $m$. We introduce also
$\mathcal{D}^{(l)}_{\lambda, m}\equiv \mathcal{D}^{(l)-\lambda}_m $.
The explicit harmonic expansions for $\phi$ and the fluctuations
$h_{MN}$ and $\mathcal{V}_M$ of the metric and the gauge field are
\be \phi=\sum_{l\geq |n|/2} \sum_{|m|\leq l}\phi_{\,\,\,m}^{l}(x)
\sqrt{\frac{2l+1}{4\pi}}\mathcal{D}^{(l)}_{-n/2,m}(\theta ,
\varphi),\ee
\be \mathcal{V}_{\mu}=\sum_{l\geq 0} \sum_{|m|\leq
l}\mathcal{V}_{\mu\,\,\,m}^{\,l}(x)
\sqrt{\frac{2l+1}{4\pi}}\mathcal{D}^{(l)}_{0,m}(\theta ,
\varphi),\label{Vexp} \ee
\be h_{\mu +}=\sum_{l\geq 1} \sum_{|m|\leq l}h_{\mu +
\,\,m}^{\,l}(x) \sqrt{\frac{2l+1}{4\pi}}\mathcal{D}^{(l)}_{+
,m}(\theta , \varphi),\label{hexp}\ee
\be h_{\mu \nu}=\sum_{l\geq 0} \sum_{|m|\leq l}h_{\mu \nu
\,\,m}^{\,l}(x) \sqrt{\frac{2l+1}{4\pi}}\mathcal{D}^{(l)}_{0
,m}(\theta , \varphi), \ee
\bea \mathcal{V}_{+}&=&\sum_{l\geq 1} \sum_{|m|\leq l}\mathcal{V}_{+
\,\,m}^{\,l}(x)
\sqrt{\frac{2l+1}{4\pi}}\mathcal{D}^{(l)}_{+ ,m}(\theta , \varphi),\nonumber \\
 h_{ ++}&=&\sum_{l\geq 2}
\sum_{|m|\leq l}h_{++ \,\,m}^{\,l}(x)
\sqrt{\frac{2l+1}{4\pi}}\mathcal{D}^{(l)}_{2 ,m}(\theta , \varphi)\nonumber \\
h_{ +-}&=&\sum_{l\geq 0} \sum_{|m|\leq l}h_{+- \,\,m}^{\,l}(x)
\sqrt{\frac{2l+1}{4\pi}}\mathcal{D}^{(l)}_{0 ,m}(\theta ,
\varphi),\label{bosonicexpansion} \eea
where the subscripts $+$ and $-$ refer to the basis (\ref{epm}). For
$l=1$ our choice is
\be \mathcal{D}_{\hat{\alpha},\hat{\beta}}(\theta,\varphi)=
\left(\ba {ccc} \frac{1}{2}(\cos\theta+1) &
\frac{1}{2}(\cos\theta-1)e^{-2i\varphi} &
-\frac{1}{\sqrt2}\sin\theta
e^{-i\varphi} \\
 \frac{1}{2}(\cos\theta-1)e^{2i\varphi}
 & \frac{1}{2}(\cos\theta+1)
 & -\frac{1}{\sqrt2}\sin\theta
e^{i\varphi} \\
 \frac{1}{\sqrt2}\sin\theta e^{i\varphi} &
\frac{1}{\sqrt2}\sin\theta e^{-i\varphi} & \cos\theta
\ea\right),\label{harmonics} \ee
where we have introduced $\mathcal{D}_{\lambda, m}\equiv
\mathcal{D}^{(1)}_{\lambda, m}$. In (\ref{harmonics}) the first,
second and third rows correspond to $\hat{\alpha}=+,-,3$, the first,
second and third columns to $\hat{\beta}=+,-,3$. While our choice
for $\mathcal{D}^{(2)}_{\lambda,m}$ is
 $$\mathcal{D}^{(2)}_{\lambda,2}(\theta,\varphi)=\left(\ba {c} \frac{1}{4}\left(1+\cos\theta \right)^2 \\
  -\frac{1}{2}\sin\theta (1+\cos\theta)e^{i\varphi}
 \\
  \sqrt{\frac{3}{8}}\sin^2\theta e^{2 i\varphi} \\ -\frac{1}{2}\sin\theta (1-\cos\theta)e^{3 i\varphi}  \\
\frac{1}{4}\left( 1-\cos\theta\right)^2 e^{4i\varphi}
\ea\right),\, \mathcal{D}^{(2)}_{\lambda,1}(\theta,\varphi)=\left(\ba {c} -\frac{1}{2}\sin\theta (1+\cos\theta)e^{-i\varphi}  \\

  \frac{1}{2}(1-\cos\theta-2\cos^2\theta)  \\

 \sqrt{\frac{3}{2}}\sin\theta \cos\theta e^{i\varphi} \\

\frac{1}{4}(4\cos^2\theta-2\cos\theta -2) e^{2i\varphi}   \\

\frac{1}{2}\sin\theta (1-\cos\theta)e^{3i\varphi} \ea\right), $$
 $$\mathcal{D}^{(2)}_{\lambda,0}(\theta,\varphi)=\left(\ba {c} \sqrt{\frac{3}{8}}\sin^2\theta e^{-2 i\varphi} \\
  \sqrt{\frac{3}{2}}\sin\theta \cos\theta e^{-i\varphi}
 \\
  \frac{1}{2}(3\cos^2\theta -1)\\-\sqrt{\frac{3}{2}}\sin\theta \cos\theta e^{i\varphi}  \\
\sqrt{\frac{3}{8}}\sin^2\theta e^{2 i\varphi}
\ea\right),\, \mathcal{D}^{(2)}_{\lambda,-1}(\theta,\varphi)=\left(\ba {c} -\frac{1}{2}\sin\theta (1-\cos\theta)e^{-3i\varphi}  \\

  \frac{1}{4}(4\cos^2\theta-2\cos\theta -2) e^{-2i\varphi}  \\

 -\sqrt{\frac{3}{2}}\sin\theta \cos\theta e^{-i\varphi} \\

\frac{1}{2}(1-\cos\theta-2\cos^2\theta)   \\

\frac{1}{2}\sin\theta (1+\cos\theta)e^{i\varphi} \ea\right), $$
$$\mathcal{D}^{(2)}_{\lambda,-2}(\theta,\varphi)=\left(\ba {c} \frac{1}{4}\left( 1-\cos\theta\right)^2 e^{-4i\varphi} \\
  \frac{1}{2}\sin\theta (1-\cos\theta) e^{-3i\varphi}
 \\
  \sqrt{\frac{3}{8}}\sin^2\theta e^{-2 i\varphi}\\\frac{1}{2}\sin\theta (1+\cos\theta)e^{-i\varphi}   \\
\frac{1}{4}\left(1+\cos\theta \right)^2 \ea\right),$$ where
$\lambda$ is a row index. We could continue and compute the
harmonics for every value of $l$ but we do not do that as we do not
need their explicit expression for $l>2$.

It is useful to compute the effect of the background covariant
derivatives on the harmonics. We have
\be
\nabla_{\alpha}\mathcal{D}^{(l)}_{\lambda,m}=e^n_{\alpha}\left(\partial_n-\lambda\omega_n
\right) \mathcal{D}^{(l)}_{\lambda,m},
 \ee
where $\nabla_{\alpha}$ is the background covariant derivative,
$e^n_{\alpha}$ is the inverse of $e^{\alpha}_n$, which can be
calculated from (\ref{epm}), and $\omega_{n}$ represents the
background spin connection: we have
\be \omega_{\varphi} =\frac{i}{a}(\cos\theta-1),\quad
\omega_{\theta}=0. \ee
It can be proved the following effects of
\be \nabla^2=\nabla_{\a}\nabla^{\a}=\nabla_+\nabla_- +\nabla_-
\nabla_+ \ee
 over the harmonics:
\be \nabla^2 \mathcal{D}^{(l)}_{\lambda
,m}=-\frac{1}{a^2}\left[l(l+1)- \lambda^2
\right]\mathcal{D}^{(l)}_{\lambda ,m}.\label{nabla2} \ee

In the following we consider, just for simplicity, the case
\be n=2. \ee
In fact for this value of the monopole charge we can find a very
simple solution of the fundamental 6D EOMs (\ref{EOM}) which is
invariant under a $U(1)$ subgroup of $SU(2)\times U(1)$; this
solution is discussed in Section \ref{u13}. Like in Section
\ref{general} our purpose is in fact to construct the 4D
$SU(2)\times U(1)$-invariant effective theory, study the spontaneous
symmetry breaking $SU(2)\times U(1)\rightarrow U(1)$ and the Higgs
mechanism in the effective theory and then compare the results with
the corresponding quantities predicted by the 6D theory; therefore,
in order to do that, one has to find a 6D U(1)-invariant solution of
the EOMs. We observe that for $n=2$ the iso-helicity of $\phi$ is
$1$.

The low energy 4D spectrum coming from the background
(\ref{sphere}), (\ref{monopole}) and (\ref{phi0}) is given in Ref.
\cite{RSS} for the spin-1 and spin-2 sectors. The massless sector is
the following: there are a graviton (helicities $\pm2$, $l=0$), a
$U(1)$ gauge field (helicities $\pm1$, $l=0$) coming from
$\mathcal{V}_{\mu}$ and a Yang-Mills $SU(2)$ triplet (helicities
$\pm 1$, $l=1$) coming from $h_{\mu \a}$ and $\mathcal{V}_{\mu}$,
where $\mathcal{V}_M$ and $h_{MN}$ are the fluctuations of the gauge
field and the metric around solution (\ref{sphere}),
(\ref{monopole}) and (\ref{phi0}). Regarding the scalar spectrum all
the scalars from $G_{MN}$ and $A_M$ have very large masses, of the
order $1/a$, and we can get only an $SU(2)$-triplet with mass
squared $\mu^2$ from $\phi$ in the low energy spectrum if we choose
$m^2$ such that
\be |\mu^2| \ll \frac{1}{a^2}, \label{assumption1}\ee
where
\be \mu^2 \equiv -\frac{1}{a^2}\eta \equiv m^2+\frac{1}{a^2}.
\label{edefin}\ee

In fact $-1/a^2$ is the eigenvalue of the Laplacian operator acting
on the harmonic with $l=1$ and $\lambda=1$ , as one can check using
the related formula of \cite{RSS}. The parameter $\mu^2$ is in fact
the squared mass of the triplet from $\phi$, and it can be in
principle either positive or negative. If (\ref{assumption1}) holds
all the remaining scalars have masses at least of the order $1/a$
and they do not appear in the low energy theory. So we assume that
(\ref{assumption1}) holds. Finally in order to find the low energy
fermionic spectrum we have to calculate the associated
iso-helicities by using the explicit expression for the background
covariant derivative of $\psi_{\pm}$ along the internal space:
\be \nabla_m \psi_{\pm} = \left( \partial_m \pm\omega_m \frac{1}{2}
\gamma^5 +ie_{\pm} A_m \right)\psi_{\pm}, \ee
where $\omega_\theta = 0$, $\omega_{\varphi}=\frac{i}{a}( \cos\theta
- 1)$, $e_+ = e/2$ and $e_- = 3e/2$. We get
\be \lambda_{+L}=0, \quad \lambda_{+R}=1, \quad \lambda_{-L}=2,
\quad \lambda_{-R}=1 \ee
and the corresponding expansions are given by (\ref{lexpansion}). So
the equations (\ref{fermioneq}) tell us that there are 4 zero-modes:
the $l=0$, $m=0$ mode in $\psi_{+L}$ and the $l=1$, $m=+1,-1,0$ in
$\psi_{-R}$. So we have a massless $SU(2)$ singlet from $\psi_{+L}$
and a massless $SU(2)$ triplet from $\psi_{-R}$.

\subsection{The 4D $SU(2)\times U(1)$ Effective Lagrangian \\ and the
Higgs Mechanism} \label{effective}

Now we want to study the 4D effective theory: which is the 4D
 theory obtained
from the background (\ref{sphere}), (\ref{monopole}) and
(\ref{phi0}) retaining only
 the low energy spectrum we discussed at the end of Subsection \ref{su2 x u1}, that is the particles with
masses much smaller than $1/a$, and integrating out all the heavy
modes, namely those with mass at least of the order $1/a$. This is
an $SU(2)\times U(1)$-invariant theory, which includes a charged
scalar, that we call $\chi$, in the $3$-dimensional representation
of $SU(2)$, and, if we want, two Weyl spinors in the $1_{1/2}$ and
$3_{3/2}$ of $SU(2)\times U(1)$.
 The background (\ref{sphere}), (\ref{monopole}) and (\ref{phi0}) is the analogous of what we
 called $\Phi_{eff}$ in Section \ref{general}. In this section we give
only some relevant terms\footnote{Here ``relevant terms'' has the
same meaning as in the Subsection \ref{EFFGeneral}.} appearing in
the lagrangian of this theory. In particular we calculate the scalar
potential, we study the Higgs mechanism, which is active only for
$\mu^2 <0$, and we give in this case the masses of the spin-1,
spin-0 and spin-1/2 particles.

Like in the general scalar theory of Section \ref{general}, in the
following we perform all the calculations at the order $\eta$. If we
use the information regarding the low energy spectrum which we
discussed at the end of Subsection \ref{su2 x u1}, we can construct
some relevant terms of the 4D effective theory through the following
ansatz\footnote{The ansatz (\ref{0ansatz}) is a generalization of
the zero-mode ansatz of \cite{RSS}, which does not include scalar
fields.}
\bea E^a(x)&=& E^a_{\mu}(x)dx^{\mu}, \nonumber \\
E^{\alpha}(x,y)&=&e^{\alpha}(y)-\frac{\kappa}{a\sqrt{4\pi}}W_{\mu}^{\hat{\alpha}}(x)dx^{\mu}
\mathcal{D}^{\alpha}_{\hat{\alpha}}(y),
\nonumber  \\
A(x,y)&=& -\frac{n}{2ea}e^3(y) \nonumber\\
&&+\frac{1}{a\sqrt{4\pi}}V_{\mu}(x)dx^{\mu}
-\frac{n\kappa}{2ea^2\sqrt{4\pi}}U_{\mu}^{\hat{\alpha}}(x)dx^{\mu}
\mathcal{D}_{\hat{\alpha}}^3(y), \nonumber \\
 \phi(x,y)&=&\frac{1}{a}\sqrt{\frac{3}{4\pi}}\chi^{\hat{\a}}(x)\mathcal{D}_{-,\hat{\a}}(y),\nonumber \\
\psi_{+R}&=&\psi_{-L}=0, \nonumber \\
\psi_{-R}&=&\frac{1}{a}\sqrt{\frac{3}{4\pi}}\psi_R^{\hat{\a}}(x)\mathcal{D}_{-,\hat{\a}}(y),\nonumber \\
\psi_{+L}&=&\frac{1}{a\sqrt{4\pi}}\psi_L(x), \label{0ansatz} \eea
where $E^A$, $A=0,1,2,3,+,-$, are the 6D $(x,y)$-dependent
orthonormal 1-form basis, $E^a_{\mu}$ is the 4D $x$-dependent
vielbein, $V_{\mu}$ is the 4D $U(1)$ gauge field coming from
$\mathcal{V}_{\mu}$, a linear combination\footnote{The orthogonal
linear combination has a large mass; we show this in Appendix
\ref{S^2simm}.} of $W_{\mu}$ and $U_{\mu}$ is the Yang-Mills $SU(2)$
triplet \cite{RSS} coming from $h_{\mu \a}$ and $\mathcal{V}_{\mu}$;
finally $\psi_L$ and $\psi_R$ are the $SU(2)$ fermion singlet and
fermion triplet, respectively.
 Actually the ansatz (\ref{0ansatz}) is the
background (\ref{sphere}), (\ref{monopole}) and (\ref{phi0}) plus
some fluctuations, which include all the light KK states.

Now we want to write some relevant terms of the effective lagrangian
for $\chi$ by using the light-mode ansatz (\ref{0ansatz}) and by
taking into account the heavy mode contribution. Concerning the
scalar potential in the 4D effective theory, we already know that
the bilinear part is simply $\mu^2\chi^{\dagger}\chi$. Whereas the
quartic terms are non-trivial and they have two different
contributions: the quartic term in the 6D potential $V$ in (\ref{V})
computed with $\phi(x,y)$ in Eq. (\ref{0ansatz}) and the heavy
scalar modes ($h_{\a\b}$ and $\mv_{\a}$) contribution through
diagrams like Fig. \ref{simplediagram}, evaluated at transferred
momentum equal to zero. The first contribution is
\be
\xi \int a^2 \sin \theta d\theta
d\varphi\left|\phi(x,\theta,\phi)\right|^4. \ee
This object is equal to
\be \frac{\xi}{a^2}\left(\chi_{m_1}\right)^*\chi_{m_2}
\left(\chi_{m_3}\right)^*\chi_{m_4}J_{m_1 m_2 m_3 m_4}, \ee
where $J_{m_1 m_2 m_3 m_4}$ is an invariant tensor in the $3\times
3\times 3\times 3$ representation  of $SU(2)$. We obtain
\be J_{\hat{\alpha}_1
\hat{\alpha}_2\hat{\alpha}_3\hat{\alpha}_4}=j_1\delta_{\hat{\alpha}_1
\hat{\alpha}_2}\delta_{\hat{\alpha}_3
\hat{\alpha}_4}+j_2g_{\hat{\alpha}_1
\hat{\alpha}_3}g_{\hat{\alpha}_2
\hat{\alpha}_4}+j_3\delta_{\hat{\alpha}_1
\hat{\alpha}_4}\delta_{\hat{\alpha}_2 \hat{\alpha}_3}, \label{J}\ee
where $j_1$, $j_2$ and $j_3$ are some constants. By explicit
calculations we get
\be j_1+j_3=\frac{9}{20\pi}, \,\,\,\,j_2=-\frac{3}{20\pi}. \ee
The final expression for the scalar potential $\mathcal{U}$ in the
4D effective theory, including the bilinear and the quartic
interactions and the light and heavy mode contributions, is
\be \mathcal{U}(\chi)=\mu^2 \chi^{\dag} \chi
+\left(\lambda_H+c_1\lambda_G\right)\left(\chi^{\dag}\chi\right)^2
-\frac{\lambda_H+c_2\lambda_{G}}{3}\left|\chi^{\hat{\alpha}}g_{\hat{\alpha}\hat{\beta}}
\chi^{\hat{\beta}}\right|^2+...,\label{U} \ee
where $c_1$ and $c_2$ are dimensionless parameters,
\be \lambda_H\equiv \frac{9}{20 \pi a^2}\xi,\quad \quad
\lambda_G\equiv \frac{9\kappa^2}{80\pi a^4} \label{lGlH}\ee
and the dots represent higher order non relevant terms, for example
terms with a product of 6 $\chi$ or 8 $\chi$. These terms do not
contribute to the VEV of $\chi$ as we want this VEV to be of the
order\footnote{The order $\eta^{1/2}$ corresponds to the order $\mu$
because of Eq. (\ref{edefin}).} $\eta^{1/2}$. In (\ref{U}) the
contribution of the heavy scalars, namely $h_{\a\b}$ and $\mv_{\a}$,
is represented by $c_1\lambda_G$ and $c_2\lambda_G$, the analogous
of $a_{lmpq}$ in the Eq. (\ref{l4}). Moreover we give also the
expression for the gauge covariant derivative of $\chi$:
\be D_{\mu} \chi ^{\hat{\alpha}}=\partial_{\mu}\chi^{\hat{\alpha}}
+ig_1V_{\mu}\chi^{\hat{\alpha}} +g_2\mathcal{A}_{\mu}^{\hat{\beta}}
\epsilon_{\hat{\beta}\hat{\gamma}}^{\,\,\,\,\,\,\hat{\alpha}}\chi^{\hat{\gamma}},\label{covphi}
\ee
where $\mathcal{A}_{\mu}$ is defined in Appendix \ref{S^2simm} and
it represents the $SU(2)$ Yang-Mills field,
$\epsilon_{\hat{\gamma}\hat{\beta}\hat{\alpha}}$ is a totally
antisymmetric symbol with $\epsilon_{+-3}=i$, and
\be g_1=\frac{e}{\sqrt{4\pi}a}, \,\,\,\,
g_2=\sqrt{\frac{3}{16\pi}}\frac{\kappa}{a^2},\label{g12}\ee
are the 4D $U(1)$ and $SU(2)$ gauge couplings. Therefore the
complete lagrangian for $\chi$ is
\be \mathcal{L}_{\chi
eff}=-\left(D_{\mu}\chi\right)^{\dagger}D^{\mu}\chi-\mathcal{U}(\chi).
\label{actionchi} \ee

Let us look for the points of minimum of the order $\eta^{1/2}$ of
the potential $\mathcal{U}$ in (\ref{U}). We have a minimum, in the
case $\mu^2<0$, for
\be \chi_1=\chi_2=0, \,\,\,\,
\chi_3=v\equiv\sqrt{\frac{-3\mu^2}{4\left[\lambda_H+\frac{1}{2}\left(3c_1-c_2\right)\lambda_G\right]}},\label{minimum}\ee
which corresponds to the global minimum
\be \mathcal{U}_0=0 \label{Umin}\ee
at the order $\eta$. This fact states that, at leading order, the 4D
flatness condition in the background is compatible with the
procedure of the 4D effective theory. In fact $\mathcal{U}_0$ can be
interpreted as a 4D cosmological constant and the flatness implies
$\mathcal{U}_0=0$. Instead for $\mu^2>0$ we do not have any order
parameter because the global minimum $\mathcal{U}_0=0$ corresponds
to $\chi=0$.

If we take, for $\mu^2<0$, the vacuum (\ref{minimum}),
 $SU(2)\times U(1)$ breaks to $U(1)_3$, where $U(1)_3$ is the $U(1)$-subgroup
of $SU(2)$ generated by its third generator. The gauge field of
$U(1)$ and $SU(2)$ are respectively $V_{\mu}$ and
$\mathcal{A}_{\mu}$; before Higgs mechanism these gauge fields are
of course massless as one can see by looking at their bilinear
lagrangian given in Appendix \ref{S^2simm}. From (\ref{actionchi})
and (\ref{covphi}) we can calculate the masses of these vector
fields in the 4D effective theory after the Higgs mechanism. We get
a massless vector field $\mathcal{A}_{\mu}^3$, which corresponds to
the unbroken $U(1)_3$ gauge symmetry. Instead $V_{\mu}$ and
$\mathcal{A}_{\mu}^{\pm}$ acquire respectively the following squared
masses
\be M^2_{V}  = \frac{3e^2}{8\pi
a^2}\frac{-\mu^2}{\lambda_H+\frac{1}{2}\left(3c_1-c_2\right)\lambda_G},\label{MV}
\ee
\be M^2_{V\pm}=\frac{9e^2}{16\pi
a^2}\frac{-\mu^2}{\lambda_H+\frac{1}{2}\left(3c_1-c_2\right)\lambda_G},
\label{MA} \ee
where the subscript $V$ indicates that we are dealing with vector
particles. Moreover, in the spin-0 sector, we have two physical
scalar fields: a real scalar and a complex one, which is charged
under the residual $U(1)_3$ symmetry. Their squared masses are
respectively
\be M^2_S=-2\mu^2,\ee
\be
M^2_{S\pm}=-\mu^2\frac{\lambda_H+c_2\lambda_G}{\lambda_H+\frac{1}{2}\left(3c_1-c_2\right)\lambda_G}.
\label{MS+-}\ee
Finally we can determine the fermionic spectrum by examining the
fermionic lagrangian in the effective theory:
\be \mathcal{L}_{F eff}=\overline{\psi_L}\gamma^{\mu} D_{\mu}
\psi_L+\overline{\psi_R}\gamma^{\mu} D_{\mu} \psi_R +
g_4\overline{\psi_L}\chi^{\dag}\psi_R +
g_4\overline{\psi_R}\chi\psi_L,\ee
where
\be g_4=\frac{g_Y}{a\sqrt{4\pi}}. \ee
The result is a neutral Dirac fermion, with squared mass
\be M_F^2=\frac{3g_Y^2}{16\pi
a^2}\frac{-\mu^2}{\lambda_H+\frac{1}{2}\left(3c_1-c_2\right)\lambda_G},
\label{efffermion}\ee
and a pair of massless right-handed Weyl fermions. We observe that
the mass spectrum that we gave here is parametrized by the $c_i$. Of
course these constants are not free parameters but they can be in
principle computed by evaluating explicitly the heavy modes
contribution. In the rest of the present chapter we do not compute
the $c_i$ but we prove that the 4D effective theory without heavy
mode contribution, that is $c_i=0$, is not correct because it
predicts a wrong VEV of the light KK scalars and a wrong mass
spectrum.

\section{6D Electroweak Symmetry Breaking} \label{u13}\setcounter{equation}{0}

Now we perform a 6D (or geometrical) analysis of spontaneous
symmetry breaking: this method corresponds to the contents of
Section \ref{general} for scalar theories. Of course we perform all
the calculations at the order $\eta$, as in the effective theory
method. So our first purpose is finding a solution of the 6D EOM
which breaks the $SU(2)\times U(1)$ symmetry at the 6D level and
which is a small perturbation of the order $\eta^{1/2}$ of the
sphere solution (\ref{sphere}), (\ref{monopole}) and (\ref{phi0}).

In order to find such a solution we consider an expansion of all
background tensors in powers of $\eta^{1/2}$. For the ansatz
(\ref{Bmetric}), (\ref{BA}) and (\ref{Bphi}) our tensors are
$g_{mn}$, $A_m$ and $\phi$; the expansion of the latter is
\be \phi=\sum_{k=1}^{\infty}\phi_k \eta^{k/2}. \label{phiexp}\ee
We have omitted the $k=0$ term because we want that $\phi$ goes to
zero as $\eta$ goes to zero. Now we are interested in the EOM for
$\phi$, namely the first equation of (\ref{EOM}). Since that
equation involves also the Laplacian $\nabla^2$ acting on charged
scalar, we expand also this operator in powers of $\eta^{1/2}$:
\be \nabla^2= \nabla^2_0 +\sum_{k=1}^{\infty}L_k,\label{nablaexp}
\ee
where $\nabla^2_0$ is the Laplacian corresponding to the
$SU(2)\times U(1)$-invariant solution (\ref{sphere}) and
(\ref{monopole}) and $L_k$ is an operator proportional to
$\eta^{k/2}$. By putting (\ref{phiexp}) and (\ref{nablaexp}) in the
first equation of (\ref{EOM}) we get one equation for every power of
$\eta^{1/2}$. The first one is
\be \eta^{1/2}\left(\nabla^2_0 +\frac{1}{a^2} \right) \phi_1 =0,\ee
which implies that $\phi_1$ must be proportional to the harmonic
with $l=1$ and $\lambda=1$. Further we impose $m=0$, otherwise we do
not have an $U(1)_3$-invariant background. So we have
\be \phi_1 \propto \D^{(1)1}_0\equiv D. \ee
Moreover the equation proportional to $\eta$ is
\be \eta  \left(\nabla^2_0 +\frac{1}{a^2} \right) \phi_2 +\eta^{1/2}
L_1\phi_1=0. \label{phi2eq} \ee
On the other hand the operator $L_1$ must vanish because from
(\ref{EOM}) follows
\be \frac{1}{k^2}R_{mn}-g_{mn}\left(\lambda +m^2|\phi|^2 +\xi
|\phi|^4 \right) -\frac{1}{2}(\nabla_m\phi)^*\nabla_n\phi-
\frac{1}{2}(\nabla_n\phi)^*\nabla_m\phi=0,\ee
which, up to O($\eta$), reduces to
\be \frac{1}{k^2}R_{mn}-g_{mn}\lambda=0. \label{zero1/2} \ee
Since the only solution of (\ref{zero1/2}) is the round $S^2$, there
is no $\eta^{1/2}$ terms in $g_{mn}$. By putting this result in the
last equation of (\ref{EOM}) we get that also the gauge field $A_m$
cannot have $\eta^{1/2}$ terms. So we have $\mathcal{O}_1=0$ and
(\ref{phi2eq}) becomes
\be \left(\nabla^2_0 +\frac{1}{a^2} \right) \phi_2 =0,\ee
which, taking into account also the $U(1)_3$-invariance, implies
\be \phi_2 \propto D. \ee
For simplicity we take $\phi_2=0$ because in any case $\phi_2$ must
be proportional to the same harmonic of $\phi_1$. So
$\phi=\eta^{1/2} \phi_1$ up to O($\eta^{3/2}$). The equation
proportional to $\eta^{3/2}$ is then
\be   \eta^{3/2}  \left(\nabla^2_0 +\frac{1}{a^2} \right) \phi_3
+\eta^{1/2} L_2\phi_1 +\eta^{3/2}\frac{1}{a^2}\phi_1=2\xi
\eta^{1/2}|\eta||\phi_1|^2\phi_1. \ee
By projecting this equation over the harmonic $D$, the first term
disappears and we get
\be \frac{1}{a^2}\eta^{3/2} \int \,D^* \phi_1 + \eta^{1/2}\int
\,D^*L_2\phi_1= \eta^{1/2}|\eta| 2\xi\int \,
D^*|\phi_1|^2\phi_1.\label{alpha0}\ee

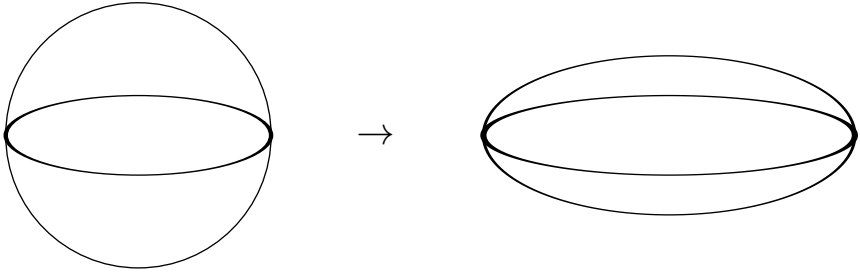
\begin{figure}[t]
\begin{center}
\begin{picture}(450,160)(0,120)

\Text(100,280)[]{{\bf Sphere ($SU(2)$ symmetry)}}
\BCirc(100,200){50} \Oval(100,200)(15,50)(0)

\Text(190,200)[]{{\Large $\rightarrow$}} \Text(300,280)[]{{\bf
Ellipsoid ($U(1)$ symmetry)}} \Oval(300,200)(30,70)(0)
\Oval(300,200)(15,70)(0)

\end{picture}
\end{center}\caption{\footnotesize We show the deformation of the internal space in the 6D approach
to the electroweak symmetry breaking. The elecroweak gauge symmetry
is broken to $U(1)$ through the ellipsoid background.}
\label{distortion}
\end{figure}

 The most simple solution of this kind, up to
higher order terms in $\eta$, that we find is similar to the
background which appears in Ref. \cite{S} \footnote{This solution
was discussed in Ref. \cite{S}, but incorrectly.} :
\bea ds^2 &=&\eta_{\mu \nu}dx^{\mu}dx^{\nu}+ a^2\left[(1+|\eta
|\beta \sin^{2}\theta)d\theta^2+
\sin^2\theta d\varphi^2\right], \nonumber \\
A&=& -\frac{1}{e}e^3, \nonumber \\
\phi &=&\eta^{1/2}\alpha\exp\left(i\varphi \right) \sin\theta,
\label{solution2} \eea
where $\beta\equiv \kappa^2 |\alpha|^2$. As required, for $\eta=0$
this background reduces to the background of Subsection \ref{su2 x
u1}. The absolute value of $\alpha$ can be computed by using Eq.
(\ref{alpha0}), which, through the redefinition
$\eta^{1/2}\phi_1\rightarrow \phi$ reads
\be \frac{1}{a^2}\eta \int \,D^* \phi + \int \,D^*L_2\phi= 2\xi\int
\, D^*|\phi|^2\phi. \label{alpha}\ee
The metric appearing in solution (\ref{solution2}) is the metric of
an ellipsoid. From the geometrical point of view we have deformed
our internal space as shown in Fig. \ref{distortion}.

 For $\mu^2<0$ the equation
(\ref{alpha}) has a solution for
\be \lambda_H>\lambda _G, \label{geometricalstability}\ee
where $\lambda_H$ and $\lambda _G$ are defined by (\ref{lGlH}),
while, for $\mu^2>0$, we have a solution for
\be \lambda_H<\lambda _G. \ee
Whether $\mu^2>0$ or $\mu^2<0$, the solution of (\ref{alpha}) is
\be |\alpha|^2=\frac{5}{|8\xi a^2-2\kappa^2|}=\frac{9}{32 \pi
a^4}\frac{1}{|\lambda_H-\lambda_G|}.\label{malpha} \ee
Note that here we have symmetry breaking for both signs of $\mu^2$.
 This is not so interesting because the solution with
$\mu^2>0$ is unstable, as it is discussed in Subsection
\ref{spin-0}. We want to stress that the value of $|\alpha|^2$
predicted by the 4D effective theory is not equal to (\ref{malpha})
if we neglect the heavy mode contribution to the effective theory,
namely for $c_i=0$: indeed in this case the effective theory
predicts a value of $|\a|^2$ equal to
\be |\alpha|_{eff}^2=\frac{9}{32 \pi
a^4}\frac{1}{\lambda_H},\label{malpha2} \ee
which is equal to (\ref{malpha}) only for $\lambda_G=0$. However,
from (\ref{lGlH}) it's clear that $\lambda_G$ cannot be taken equal
to zero. Therefore we have already proved that the heavy modes
contribution is needed at least for the light mode VEV. We shall
prove that this is the case also for the mass spectrum.

 As required
the background (\ref{solution2}) has the symmetry
\be U(1)_3 \subset SU(2).\label{symmetrybr}\ee
  So the 4D effective low energy
theory, which follows from this background, is $U(1)_3$-invariant
and comparing these results with the effective theory predictions
makes sense.

We note that the symmetry breaking (\ref{symmetrybr}) is associated,
in the 6D theory, to a geometrical deformation of the internal
space. Further we observe that (\ref{solution2}) tell us the heavy
modes VEVs are higher order corrections with respect to the light
modes VEVs like in the scalar theories of section \ref{general}.

Now we calculate the low energy vector, scalar and fermion spectrum
by analyzing the 4D bilinear lagrangian for the fluctuations around
the solution (\ref{solution2}).

\subsection{Spin-1 Spectrum}

The spin-1 spectrum can be calculated in a way similar to the light
mode ansatz (\ref{0ansatz}). However, it must be noted that the
sectors with different $l$ no longer decouple for $\eta\neq 0$, but
the mixing terms are of the order $\eta$ and they give negligible
corrections of the order $\eta^2$ to the vector boson masses. These
facts are evident from the general formula of \cite{RS}. So we can
neglect the modes with $l>1$ in the calculation of spin-1 spectrum.
 Therefore we can compute the vector boson masses
by putting the following ansatz in the action and integrating over
the extra dimensions:

\bea E^a(x)&=& E^a_{\mu}(x)dx^{\mu}, \nonumber \\
E^{\alpha}(x,y)&=&e^{\alpha}(y,\eta)-\frac{\kappa}{a\sqrt{4\pi}}W_{\mu}^{\hat{\alpha}}(x)dx^{\mu}
\mathcal{D}^{\alpha}_{\hat{\alpha}}(y) ,
\nonumber  \\
A(x,y)&=& -\frac{1}{ea}e^3(y) \nonumber\\
&&+\frac{1}{a\sqrt{4\pi}}V_{\mu}(x)dx^{\mu}
-\frac{\kappa}{ea^2\sqrt{4\pi}}U_{\mu}^{\hat{\alpha}}(x)dx^{\mu}
\mathcal{D}_{\hat{\alpha}}^3(y), \nonumber \\
 \phi(x,y)&=&\eta^{1/2}\alpha\exp\left(i\varphi \right)
\sin\theta, \label{ansatz} \eea
where $e^{\a}(y,\eta)$ is the orthonormal basis for the
2-dimensional metric in (\ref{solution2}): :
\be e^{\pm}(y,\eta)=\pm \frac{i}{\sqrt{2}}e^{\pm i\varphi}
\left[\left(1+|\eta|\frac{\beta}{2}\sin^{2}\theta\right)d\theta \pm
i\sin\theta d\varphi \right]. \label{epmeps}\ee
In (\ref{ansatz}) we consider the spin-1 fluctuations but we do not
consider the spin-0 fluctuations, because they are not necessary for
the calculation of vector boson masses. It's important to note that
in (\ref{ansatz}) the VEV of $E^{\alpha}$ is $e^{\alpha}(y,\eta)$,
it's not $e^{\alpha}(y)$ as in (\ref{0ansatz}).

From (\ref{ansatz}) it follows that some of the previous ($\eta=0$)
massless states acquire masses for $\eta\neq 0$. Up to
$O(\eta^{3/2})$, the $U(1)$ gauge boson ($l=0$) has the mass squared
\be M^2_V=\eta \frac{20}{3}\frac{e^2}{8\xi
a^2-2\kappa^2}=\frac{3e^2}{8\pi
a^2}\frac{-\mu^2}{\lambda_H-\lambda_G},\label{M0} \ee
while the Yang-Mills triplet $\mathcal{A}$ ($l=1$) is separated in a
massless gauge boson, which is associated to $U(1)_3$ gauge
invariance, and a couple of massive vector fields with the same mass
squared
\be M^2_{V\pm}=\eta \frac{10e^2}{8\xi
a^2-2\kappa^2}=\frac{9e^2}{16\pi
a^2}\frac{-\mu^2}{\lambda_H-\lambda_G}. \label{M1} \ee
By comparing (\ref{M0}) and (\ref{M1}) with (\ref{MV}) and
(\ref{MA}), we get that the heavy mode contribution is needed in the
effective theory. However, we observe that the ratio
$M^2_V/M^2_{V\pm}$ is correctly predicted by the 4D effective theory
for every $c_i$.

 Since the computation of vector bosons masses is complicated we
present it explicitly. In order to prove (\ref{M0}) and (\ref{M1})
it's useful to split the action in four terms:
\be S_B=S_R+S_F+S_{\lambda}+S_{\phi}, \ee
where
\bea S_R&=&\int d^6X \sqrt{-G}\frac{1}{\kappa^2}R, \\
 S_F&=&-\frac{1}{4}\int d^6X \sqrt{-G}F^2, \\
S_{\lambda}&=&\int d^6X\sqrt{-G}\left(-\lambda \right), \\
S_{\phi}&=&\int d^6X
\sqrt{-G}\left[-\left(\nabla_M\phi\right)^*\nabla^M\phi-V(\phi)
\right].\label{actiondecomposition} \eea
In Appendix \ref{SFSR} we prove that the contributions coming from
$S_R$ and $S_F$ vanish, so only $S_{\phi}$ contributes to the spin-1
masses up to $O(\eta^{3/2})$. The same low energy spin-1 masses in
(\ref{M0}) and (\ref{M1}) can be obtained also by using the general
formula of \cite{RS}, which contains all the bilinear terms in the
light cone gauge. The light cone gauge advantage is that the sectors
with different spin decouple. However, the derivation that we
presented here shows that the unique contribution (at the leading
order) to the spin-1 masses comes from $S_{\phi}$, like in the
effective theory approach. This explains why the ratio
$M^2_V/M^2_{V\pm}$ is correctly predicted by the 4D effective theory
for every values of $c_i$.

\subsection{Spin-0 Spectrum} \label{spin-0}

We choose the light cone gauge \cite{RS, RSS2, RS2} in order to
evaluate the spin-0 spectrum. In this gauge we have just two
independent values for the indexes $\mu,\nu,...$ which label the 4D
coordinates.
  The bilinears for the
fluctuations over the solution (\ref{solution2}) can be simply
computed with the general formula of \cite{RS}. For our model the
helicity-0 $\mathcal{L}_0$ part is given by
\be
\mathcal{L}_0=\mathcal{L}_0(\phi,\phi)+\mathcal{L}_0(h,h)+\mathcal{L}_0(\mathcal{V},\mathcal{V})+\mathcal{L}_0(\phi,h)
+\mathcal{L}_0(\phi,\mathcal{V})+\mathcal{L}_0(h,\mathcal{V}),\ee
where
\bea
\mathcal{L}_0(\phi,\phi)&=&\phi^*\partial^2\phi+\phi^*\nabla^2\phi-\left[m^2+(4\xi+e^2)|\Phi|^2
+\kappa^2\left(\nabla_m\Phi\right)^*\nabla^m\Phi \right]|\phi|^2 \nonumber \\
&&-\frac{1}{2}\left\{\left[(2\xi-e^2)\left(\Phi^*\right)^2+\kappa^2
\left(\nabla_m\Phi \nabla^m\Phi \right)^*\right]\phi^2
+c.\,c.\right\},\label{L01} \\
\mathcal{L}_0(h,h)&=& \frac{1}{4\kappa^2}\left\{h_{mn}\partial^2
h^{mn}+h_{mn}\nabla^2 h^{mn}
+2R_{mn}^{\,\quad kl}h_l^m h_k^n \right. \nonumber \\
 &&\left.+\kappa^2h_{ks}h_{mn}F^{km}F^{sn}-2\kappa^2h^l_mh_{ln}\left[\frac{1}{2}F^m_{\,\,\,\,k}F^{nk}+\left(\nabla^m\Phi\right)^*\nabla^n\Phi
 \right]\right. \nonumber \\
&&\left.+\frac{1}{2}
h^i_i\partial^2h_j^j+\frac{1}{2}h_i^i\nabla^2h_j^j
\right\},\label{L02}\\
\mathcal{L}_0(\mv,\mv)&=& \frac{1}{2}\left\{\mv_m\partial^2\mv^m +\mv_m\nabla^2\mv^m-R_{mn}\mv^m\mv^n\right. \nonumber \\
&&\left.-2e^2|\Phi|^2\mv^m\mv_m-\kappa^2\left(F_{ml}\mv^l\right)^2
\right\}, \label{L03}\\
\mathcal{L}_0(\phi,h)&=&\nabla_lh^{lm}\phi^*\nabla_{m}\Phi+h^{mn}\left(\nabla_m
\phi \right)^*\nabla_n\Phi +c.\,c.\,,\label{L04} \\
\mathcal{L}_0(\phi,\mathcal{V})&=&2ie\mv^m\phi^*\nabla_m\Phi-\kappa^2F^{lm}\mv_m\phi^*\nabla_l\Phi+c.\,c.\,,
\label{L05}\\
\mathcal{L}_0(h,\mathcal{V})&=&\mv^n\left(\nabla_mh_{ln}F^{lm}-h_l^m\nabla_mF^l_{\,\,\,\,n}\right),
\label{L06}\eea
where $\Phi$ and $\phi$ are the background and the fluctuation of
the 6D scalar. In this vanishing-helicity sector, it turns out that
we have not only mixing terms of the order $\eta$ but also mixing
terms of the order $\eta^{1/2}$, coming from $\mathcal{L}_0(\phi,h)$
and $\mathcal{L}_0(\phi,\mathcal{V})$.
 So now we can't neglect
the mixing between the sectors with different values of $l$, as we
did in the helicity $\pm 1$ sector. If we integrate these bilinear
terms over the extra-dimensions we get an infinite dimensional
squared mass matrix. However, we are interested only in the light
masses, therefore we can use the perturbation theory of quantum
mechanics in order to extract the correction of the order $\eta$ to
the masses of the 6 real scalars which are massless for $\eta=0$. We
already used this method for the computation of the mass spectrum in
the scalar theories of Section \ref{general}. We explain now how to
use it in this framework.

Formally we can write the bilinears $\mathcal{L}_0$ of the scalar
fields in this way
\be  \mathcal{L}_0 =\frac{1}{2}S^{\dag}\partial^2
S-\frac{1}{2}S^{\dagger}\mathcal{O} S,\label{formalbilinear}\ee
where $S$ is an array which includes all the scalar fluctuations; we
choose
\be S=\left(\ba {c} \phi \\
  \phi^*
 \\
  h_{++} \\
h_{--}  \\
h_{+-}  \\
  \mathcal{V}_+  \\
\mathcal{V}_- \ea\right). \label{Sdefinition}\ee
 We have just to solve a 2-dimensional eigenvalue
problem for the squared mass operator\footnote{The matrix elements
of  $\mathcal{O}$ can be computed by comparing
(\ref{formalbilinear}) with the explicit expression of
$\mathcal{L}_0$.} $\mathcal{O}$:
\be \mathcal{O}S=M^2S. \ee
In particular we want to find the 6 values of $M^2$ which go to zero
as $\eta $ goes to zero. Since we are working at the order $\eta$ we
decompose $\mathcal{O}$ as follows
\be \mathcal{O}=\mathcal{O}_0+\mathcal{O}_1+\mathcal{O}_2,
\label{Odecomposition}\ee
where $\mathcal{O}_0$ does not depend on $\eta$, $\mathcal{O}_1$ is
proportional to $\eta^{1/2}$ and $\mathcal{O}_2$ is proportional to
$\eta$. From the perturbation theory of quantum mechanics in the
degenerate case we know that the 6 values of $M^2$ we are interested
in are the eigenvalues of the following $6\times 6$
matrix\footnote{Like in Section \ref{general} we use the Dirac
notation; for two states $|S_1>$ and $|S_2>$ and for an operator
$A$, $<S_1|A|S_2>$ represents $\int S_1^{\dagger}AS_2$, where the
integral is performed with the round $S^2$ metric.}:
\be
M^2_{ij}=-\sum_{\tilde{i}}\frac{<i|\mathcal{O}_1|\tilde{i}><\tilde{i}|\mathcal{O}_1|j>}{M^2_{\tilde{i}}}
+<i|\mathcal{O}_2|j>,\label{QMperturbation} \ee
where $|i>$, $i=1,...6$ represent the 6 orthonormal eigenfunctions
of $\mathcal{O}_0$ with vanishing eigenvalue and they have the form
\be  |i>=\left(\ba {c} \phi \\
  \phi^*
 \\
  0 \\
.  \\
.  \\
  .  \\
0 \ea\right). \ee
Moreover $|\tilde{i}>$ are all the remaining orthonormal
eigenfunctions of $\mathcal{O}_0$ and $M^2_{\tilde{i}}$ the
corresponding eigenvalues. We note that the matrix elements
$<i|\mathcal{O}_1|\tilde{i}>$ are non vanishing for
\be |\tilde{i}>=\left(\ba {c} 0 \\
  0
 \\
  h_{++} \\
h_{--}  \\
h_{+-}  \\
  \mathcal{V}_+  \\
\mathcal{V}_- \ea\right). \label{tildei}\ee
Further the operator $\mathcal{O}_1$ modifies the integration
measure just by a factor proportional to the harmonics
$\mathcal{D}^{(1)}$, therefore we need just a finite subset of
${|\tilde{i}>}$ for the evaluation of $M^2_{ij}$, namely those
constructed through the harmonics with $l=0,1,2$, which are given in
Appendix \ref{GR}. An explicit form for $|i>$ and $|\tilde{i}>$, and
the preliminary computations of the 6 eigenvalues we are interested
in, are given in Appendix \ref{spin-0calculation}.

We give here just the final result: we have two unphysical scalar
fields (a real and a complex one) which form the helicity-0
component of the massive vector fields; they have in fact the same
squared masses given in (\ref{M0}) and (\ref{M1}), as it's required
by Lorentz invariance, which is not manifest in the light cone
gauge. Then we have a physical real scalar and a physical complex
scalar, charged under the residual U(1) symmetry, with squared
masses given respectively by (for $\mu^2<0$)

\bea M^2_S&=&-2\mu^2, \nonumber \\
 M^2_{S\pm}&=&-\mu^2\frac{\lambda_H+\lambda_G}{\lambda_H-\lambda_G}.\label{MS} \eea
For $\mu^2>0$, we get a negative value for $M^2_S$, therefore the
corresponding solution is unstable. Note that the squared mass
$M^2_S$ has exactly the same expression as in the 4D effective
theory, for every $c_i$. But for $c_i=0$, which corresponds to
neglecting the heavy mode contribution, the effective theory
prediction for $M_{S\pm}^2$ in (\ref{MS+-}) is not equal to the
correct value (\ref{MS}). We note that this is a physical
inequivalence because the ratio $M^2_S/M^2_{S\pm}$, which is in
principle a measurable quantity, is not correctly predicted by the
4D effective theory without the heavy mode contribution. More
precisely the effective theory prediction for $M^2_S/M^2_{S\pm}$, in
the case $c_i=0$, is always greater than the correct value.

\subsection{Spin-1/2 Spectrum} \label{spin-1/2}

The spin-1/2 spectrum can be calculated by linearizing the EOM
(\ref{fermioneq}): for $n=2$ we get
\bea &&\left(\partial^2 + 2 \nabla_+ \nabla_-
-g_Y^2|\Phi|^2\right)\psi_{+L}
=0, \nonumber \\
&&\left(\partial^2 + 2 \nabla_- \nabla_+
-g_Y^2|\Phi|^2\right)\psi_{-L}
=0, \nonumber \\
&&\left(\partial^2 + 2 \nabla_- \nabla_+
-g_Y^2|\Phi|^2\right)\psi_{+R}
+\sqrt2 g_{Y}\left(\nabla_+\Phi \right)^*\psi_{-R}=0, \nonumber \\
&&\left(\partial^2 + 2 \nabla_+ \nabla_-
-g_Y^2|\Phi|^2\right)\psi_{-R} +\sqrt2 g_{Y}\nabla_+\Phi
\,\psi_{+R}=0, \label{linearizedfermion} \eea
where $\Phi$ represents again the background of the 6D scalar,
namely the third line of (\ref{solution2}), and the covariant
derivatives are evaluated with the background metric and background
gauge field given by the first and the second line of
(\ref{solution2}). These covariant derivatives are in the $\pm$
basis defined by (\ref{epmeps}) and it includes the modified spin
connection when it acts on spinors:
\be \nabla_{\a} \psi_{\pm R} = e_{\a}^m(y,\eta)\left( \partial_m
\pm\omega_m \frac{1}{2} +ie_{\pm} A_m \right)\psi_{\pm R}, \ee
\be \nabla_{\a} \psi_{\pm L} = e_{\a}^m(y,\eta)\left( \partial_m
\mp\omega_m \frac{1}{2}  +ie_{\pm} A_m \right)\psi_{\pm L},
 \ee
where $\omega_\theta = 0$, $\omega_{\varphi}\equiv
\omega_{\varphi\,\,\, +} ^{\,\,+}$ is given in equation
(\ref{spinconn}) and the value of the charges $e_{\pm}$ and the
iso-helicities\footnote{For $\eta\neq0$ we adopt the same harmonic
expansion as in the $\eta=0$ case; this gives the correct result for
the fermionic masses squared at the order $\eta$.} of the fermions
are given at the end of Subsection \ref{su2 x u1}. There we give
also the fermionic massless spectrum for $\eta=0$: an $SU(2)$
singlet from $\psi_{+L}$ and an $SU(2)$ triplet from $\psi_{-R}$.

From (\ref{linearizedfermion}) it's clear that the left handed
sector does not present mixing terms of the order $\eta^{1/2}$ but
only of the order $\eta$. Therefore the calculation of the squared
mass $M_F^2$ of the light fermion coming from $\psi_{+L}$ is quite
easy. The result is
\be M_F^2=\frac{3g_Y^2}{16\pi
a^2}\frac{-\mu^2}{\lambda_H-\lambda_G}.\label{fermionmass}\ee
Instead the evaluation of the right-handed spectrum is complicated
by the presence of mixing terms of the order $\eta^{1/2}$, as in the
scalar sector. Therefore we use the perturbation theory of quantum
mechanics also in the fermion right-handed sector. Formally we can
write the eigenvalue equation for the mass squared operator
$\mathcal{O}$ acting in the right-handed sector as follows
\be \mathcal{O}F_R=M^2 F_R,\label{eigenfermion} \ee
where $F_R$ is an array which includes both the right-handed
fermions; we choose
\be F_R=\left(\ba {c} \psi_{+R} \\
  \psi_{-R}
\ea\right). \label{FRdefinition}\ee
One can easily compute $\mathcal{O}$ acting on $F_R$ by performing
the substitution $\partial^2\rightarrow M^2$ in the last two
equations of (\ref{linearizedfermion}). Then we can proceed as in
the scalar spectrum, performing the decomposition
(\ref{Odecomposition}). However, in this case the matrix $M^2_{ij}$
in (\ref{QMperturbation}) is a 3$\times$3 matrix as the number of
zero modes for $\eta=0$ in the right-handed sector is 3. Like in the
scalar spectrum we need only those $|\tilde{i}>$ vectors made of
harmonics with $l\leq 2$, because the operator $\mathcal{O}_1$
modifies the integration measure just by a factor proportional to
the harmonics $\mathcal{D}^{(1)}$.
 In Appendix \ref{spin-1/2calculation} we give an expression for the $|i>,i=1,-1,0,$
vectors, for the $|\tilde{i}>$ vectors and the $M^2_{\tilde{i}}$
eigenvalues for the relevant values of $l$: $l=1,2$. Here we give
the final result: the right-handed low energy spectrum has a pair of
massless right-handed fermions as in the 4D effective theory, which
have opposite charge under the residual $U(1)$ symmetry, and a
massive right-handed fermion with the same squared mass given in
(\ref{fermionmass}). This right-handed fermion together with the
massive left-handed fermion form a massive Dirac spinor with mass
$M_F$.

Also in the fermionic sector we note that the heavy modes
contribution is needed in order that the effective theory reproduces
the correct 6D result; this sentence is evident if one compares the
effective theory prediction (\ref{efffermion}) with the correct
result (\ref{fermionmass}).

\section{Conclusions and Outlook of Part I}\label{Comments} \setcounter{equation}{0}

The principal result of Chapter \ref{GHBscalar} and \ref{EMS} is
that the contribution of the heavy KK modes to the effective 4D
action is necessary in order to reproduce the correct D-dimensional
predictions concerning the light KK modes. We have calculated such a
contribution for a class of scalar theories in Chapter
\ref{GHBscalar}. However, this result holds in a more general
framework. In order to show this, in this chapter we have studied a
6D gauge and gravitational theory which involves a complex scalar
and, possibly, fermions. In particular we have considered the
compactification over $S^2$, for a particular value of the monopole
number ($n=2$), and the construction of a 4D $SU(2)\times U(1)$
effective theory. The latter contains a scalar triplet of  $SU(2)$
which, through an Higgs mechanism, gives masses to the vector,
scalar and fermion fields. An explicit expressions for these masses
and for the VEV of the scalar triplet was found at the leading order
in the small mass ratio $\mu/M$, where $M$ is the lightest heavy
mass. On the other hand, for $n=2$, we found a simple perturbative
solution of the fundamental 6D EOMs with the same symmetry of the 4D
effective theory in the broken phase. This solution presents a
deformation of the internal space $S^2$ to an ellipsoid, which has
isometry group $U(1)$ instead of $SU(2)$. Moreover we computed the
corresponding vector, scalar and fermion spectrum with quantum
mechanics perturbation theory technique. We have demonstrated by
direct calculation that these quantities, computed in the 6D
approach, are equal to the corresponding predictions of the 4D
effective theory only if the contribution of the heavy KK modes  are
taken into account. In Table \ref{comparison} we give the spectrum
predicted by the 4D effective theory for $c_i=0$, namely, without
heavy KK mode contribution, and the low energy spectrum predicted by
the 6D theory for the stable ($\mu^2<0$) solution, that we gave in
the text.
\begin{table}[top]
\begin{center}
\begin{tabular}{|l|l|l|}
\hline Squared Mass  & 4D Effective Theory &\rule{0.80cm}{0pt} 6D
Theory \quad \quad \quad  \\ \hline
 \raisebox{-0.30cm}{\rule{0pt}{0.80cm}} \quad $M^2_{V}$ & \quad \quad \quad $\frac{3e^2}{8\pi
a^2}\frac{-\mu^2}{\lambda_H}$ & \quad \quad$\frac{3e^2}{8\pi
a^2}\frac{-\mu^2}{\lambda_H-\lambda_G}$ \\ \hline
 \raisebox{-0.30cm}{\rule{0pt}{0.80cm}} \quad $M^2_{V\pm}$ & \quad \quad \quad $\frac{9e^2}{16\pi a^2}\frac{-\mu^2}{\lambda_H}$ &
\quad \quad $\frac{9e^2}{16\pi
a^2}\frac{-\mu^2}{\lambda_H-\lambda_G}$ \\ \hline
\raisebox{-0.30cm}{\rule{0pt}{0.80cm}} \quad $M^2_{S}$ & \quad \quad \quad $-2\mu^2$ & \quad \quad $-2\mu^2$ \\
\hline      \raisebox{-0.30cm}{\rule{0pt}{0.80cm}} \quad
$M^2_{S\pm}$ & \quad \quad \quad $-\mu^2$ & \quad \quad
$-\mu^2\frac{\lambda_H+\lambda_G}{\lambda_H-\lambda_G}$ \\\hline
\raisebox{-0.30cm}{\rule{0pt}{0.80cm}} \quad $M^2_{F}$ & \quad \quad
\quad $\frac{3g_Y^2}{16\pi a^2}\frac{-\mu^2}{\lambda_H}$
& \quad \quad $\frac{3g_Y^2}{16\pi a^2}\frac{-\mu^2}{\lambda_H-\lambda_G}$ \\
\hline  \raisebox{-0.30cm}{\rule{0pt}{0.80cm}} \quad $M^2_{F\pm}$ &
\quad \quad \quad $0$ & \quad \quad $0$ \\ \hline
\end{tabular}
\end{center}\caption{\footnotesize  The spectra predicted by the 4D effective theory without heavy modes
contribution ($c_i=0$) and by the 6D theory.} \label{comparison}
\end{table}
We observe that ratios of masses which involve only vector and
fermion excitations are correctly predicted by the 4D effective
theory even without the heavy KK mode contribution. But the ratios
of masses which involve at least one scalar mode are not correctly
predicted and the error is measured by $\lambda_G/\lambda_H$, where
$\lambda_G$ and $\lambda_H$ are defined in equations (\ref{lGlH}).
We can roughly estimate the magnitude of this disagreement: if we
require $g_1$ and $g_2$ in (\ref{g12}) to be of the order of $1$ and
we consider also the relation  between $\kappa$ and the 4D Planck
length $\kappa_4$
\be \frac{4\pi a^2}{\kappa^2}=\frac{1}{\kappa_4^2}, \ee
we get that $\sqrt {\kappa}$, $e$ and $a$ are all of the order of
$\kappa_4$. So roughly speaking the condition
$\lambda_G/\lambda_H\ll 1$ becomes $\lambda_H\gg 1$, which is a
strong coupling regime. Therefore we can't probably neglect the
heavy KK mode contribution and believe in the perturbation theory of
quantum field theory at the same time.

Finally we note that there is a value of $c_1$ and $c_2$
($c_1=-1/3$, $c_2=1$) such that the effective theory VEV and vector,
scalar and fermion spectrum turn out to be correct, namely, they are
equal to the corresponding quantities given in Section \ref{u13}.
This is a sign of the equivalence between the geometrical approach,
which involves the deformed internal space geometry, to the
spontaneous symmetry breaking and the Higgs mechanism in the 4D
effective theory. In particular the heavy KK mode contribution can
be interpreted in a geometrical way as the internal space
deformation of the 6D solution: in fact if we put $\b=0$ but we keep
$\a \neq 0$ in (\ref{solution2}), which corresponds to neglecting
the $S^2$ deformation, we get exactly the VEV and the spectrum
predicted by the 4D effective theory without heavy KK modes
contribution.

Possible applications can be its extension to the case which
resembles more the standard electro-weak theory. The latter could be
for instance the 6D gauge and gravitational theory presented in this
chapter, compactified over $S^2$ but with monopole number $n=1$; in
this case we have in fact an Higgs doublet in the 4D effective
theory. Other interesting applications could be models without
fundamental scalars, which, in some sense, geometrize the Higgs
mechanism or the context of supersymmetric version of 6D gauge and
gravitational theories. Such supersymmetric theories have been
recently investigated in connection with attempts to find a solution
to the
 cosmological dark energy  problem, a summary of which can be found in
\cite{Burgess:2004ib} and in Section 3 of Chapter \ref{6DSUGRA}.

\addcontentsline{toc}{chapter}{PART II: 6D Supergravity}
\chapter*{ Part II: 6D Supergravity}

In the second part of this thesis we consider supersymmetric and
non-Abelian extensions of the 6D Einstein-Maxwell-Scalar model that
we discussed in Chapter \ref{EMS}. The discussion of chapter
\ref{EMS} was motivated by a theoretical question, concerning the
role of heavy modes, with masses of the order of the Planck scale,
in the low energy dynamics. Here we want to discuss 6D supergravity
from both a theoretical and phenomenological point of view. We know
that models in six dimensions are relevant for several reasons; for
example the attempt to solve the hierarchy problem (between the
electroweak and the Planck scale) by means of the ADD scenario is
phenomenologically viable but also falsifiable in 6D, because it is
subjected to tests of gravity at submillimeter scales, as we
discussed in Section \ref{LED}. On the other hand also supersymmetry
has several motivations, in particular of the theoretical type. One
of them is the fact that superstring theories, the only attempt to
unify fundamental interactions including gravity, are
supersymmetric; another motivation for supersymmetry is the
possibility to solve the hierarchy problem in a supersymmetric
framework. Of course this does not mean that one has to choose among
LED and supersymmetric theories to address the hierarchy problem:
the LED scenario can play a role in addition to supersymmetry,
rather than in competition with it. The aim of Part II is to study
the implications of 6D supersymmetric models including gravitational
interactions and therefore we will deal necessarily with
supergravity.

This part contains two chapters. In Chapter \ref{6DSUGRA} we will
review the general features of 6D supergravities, focusing on the
{\it minimal gauged} supergravity. In particular we shall discuss
the vacua of such models, which {\it spontaneously compactify} from
6D to 4D and share many properties with realistic string
compactifications; moreover we will illustrate the so called {\it
supersymmetric large extra dimensions} scenario in which one {\it
can hope} to solve the cosmological constant problem through a {\it
self tuning} mechanism. However, the embedding of 6D supergravity in
the ADD scenario needs the appearance of 3-branes where the low
energy degrees of freedom physically localize. So we will review
also singular {\it 3-brane solutions} of such models which has to be
interpreted as backgrounds around which physical degrees of freedom
fluctuate. Indeed in Chapter \ref{SUGRAspectrum}, which represents
our original contribution \cite{SUSHA-RD-AS} to this scenario,  we
shall study perturbations around such 3-brane solutions, in
particular focusing on {\it axisymmetric solutions}. These solutions
will turn out to have conical defects, which we have discussed in
Section \ref{C2B}. Our main interest will be the gauge field and
fermion sectors which can contain SM fields and the effect of the
warping and the deficit angles on the KK towers. Moreover, in
Appendix \ref{SUGRAstability} we shall also perform a stability
analysis for the only one known maximally symmetric solution, in the
presently known anomaly-free models. Finally, in the rest of
Appendix \ref{AppSugra}, we will discuss some technical aspects
concerning the gauge field and fermion sector.


\chapter{General Features}\label{6DSUGRA}


In this review chapter we focus on the minimal supersymmetric
version of 6D supergravities, in which we have the minimum number of
supercharges in six dimensions. But we consider the possibility of
gauging a subgroup of the R-symmetry group which rotates the
supercharges; this type of models (\cite{Salam:1984cj},
\cite{Nishino:1984gk}-\cite{Parameswaran:2005mm}) are called gauged
supergravities and they have attracted much interest over the years
for several reasons. A reason motivating such models is that the
flat 6D space-time {\it is not} a solution of the corresponding
equations of motion (EOM) and the most symmetric solution is
$(Minkowski)_4\times S^2$, which has been shown recently to be the
{\it unique} maximally symmetric solution of such models
\cite{Gibbons:2003di}. This phenomenon of {\it spontaneous
compactification} is a good property which is not shared by 10D and
11D supergravities, as the low energy limit of the superstring
theories or the {\it M-theory}; indeed the most symmetric ground
state solutions in all of the higher dimensional supergravities are
the flat 10D manifolds and the pp waves. Moreover 6D gauged
supergravity compactifications share some properties with
superstring realistic compactifications \cite{Aghababaie:2002be}, in
particular they can give rise to chiral fermions in 4D. Futhermore,
like in string theory, the requirement of anomaly freedom is a
strong guiding principle to construct consistent models. Indeed the
minimal version of such gauged supergravity, {\it the Salam-Sezgin
model} \cite{Salam:1984cj}, suffers from the breakdown of local
symmetries due to the presence of gravitational, gauge and mixed
anomalies, which render this model inconsistent at the quantum level
\cite{Alvarez-Gaume:1983ig}; but it can be transformed in an anomaly
free model by choosing the gauge group and the supermultiplet in a
suitable way \cite{Randjbar-Daemi:1985wc, Avramis:2005qt,
Avramis:2005hc, Suzuki:2005vu}. Recently such 6D supergravities have
been proposed as possible frameworks in which one {\it can hope} to
solve the cosmological constant problem. A reason is that, if one
chooses large extra dimensions, in 6D the corresponding KK mass
scale (to the fourth power) is of the order of the observed vacuum
energy density. This numerical coincidence gives hope to get the
correct cosmological constant including both classical and quantum
contributions, through a mechanism of {\it self tuning} of the
cosmological constant. Such scenario is called {\it supersymmetric
large extra dimensions} (SLED) scenario and it has been studied in
recent works\footnote{For a review on this topic see
\cite{Burgess:2004ib}.}
\cite{Aghababaie:2003wz}-\cite{Hoover:2005uf}. However, a complete
proof of this mechanism has not been found and it is not clear what
is the complete effect of the breakdown of supersymmetry in the
bulk, which is needed in order to implement such an idea. Moreover,
in order the extra dimensions to be so large one should find a
mechanism which localizes the low energy degrees of freedom on a
{\it 3-brane}, placed on some singularities of the internal space
\cite{Gibbons:2003di,Aghababaie:2003ar,Burgess:2004dh,Lee:2005az}.

The aim of this chapter is to review such topics in order to prepare
the background for the contents of Chapter \ref{SUGRAspectrum},
which includes the original work on this framework. The composition
of the chapter is as follows. In Section \ref{SUGRAD>4} we discuss
supergravity in diverse dimensions and in Section
\ref{6DgaugedSUGRA} we focus on the minimal 6D {\it gauged}
supergravity by discussing the supermultiplets and the actions for
such models and then by describing the presently known anomaly free
versions. In Appendix \ref{SUGRAstability} we perform the stability
analysis for an $S^2$ compactification of these models. In Section
\ref{SLED} we describe the SLED scenario in more detail, explaining
in particular the self tuning mechanism for the cosmological
constant. Finally in Section \ref{singular} we review the brane
solutions of 6D gauged supergravity models.

\section{Supergravity in Diverse Dimensions}\label{SUGRAD>4}

Here we want to discuss briefly general properties of supergravities
in diverse dimensions, in order to introduce notations and
terminology. For a more complete introduction to such a topic see
for instance Ref. \cite{Salam:1989fm}-\cite{VanProeyen:2003zj}.

The starting point to construct a supersymmetric model, in
particular a supergravity model, is the choice of a {\it
superalgebra} or {\it super-Poincar\'e algebra}. The latter includes
by definition the generators of the Poincar\'e group and a set of
supercharges, which generates supersymmetry. In order to be
consistent such generators have to be spinors, that is objects
transforming under the spinorial representation of the Lorentz group
$SO(1,D-1)$. The generators in such a representation are given by
$\frac{1}{4}[\Gamma^M,\Gamma^N]$, where the matrices $\Gamma^M$obey
the Clifford algebra:
\be
\{\Gamma^M,\Gamma^N\}=2\eta^{MN}.
\ee
 In this way we define a {\it Dirac spinor}, which has real\footnote{In this thesis we refer always to the real spinor dimension.}
dimension $2^{[D/2]+1}$, where $[D/2]$ means the integer part of
$D/2$, and exists in all space-time dimensions. However, when $D$ is
even the Dirac spinor is a reducible representation because one can
define a chirality matrix, which is given by the product of all the
matrices $\Gamma^M$ and commutes with all the generators
$\frac{1}{4}[\Gamma^M,\Gamma^N]$. The eigenvectors of the chirality
matrix are called Weyl spinors. Moreover for some particular value
of $D$ we can impose a {\it reality condition} on spinors and define
{\it Majorana spinors}. In Table \ref{spinors} we summarize the
existence of Weyl, Majorana and Majorana-Weyl spinors in diverse
space-time dimensions.
\begin{table}[top]
\begin{center}
\begin{tabular}{|l|l|l|}
\hline {\bf $D$}  & Spinor & Components  \\ \hline
 2 mod 8 & Maj-Weyl & $2^{D/2-1}$\\ \hline
 3,9 mod 8 & Maj & $2^{(D-1)/2}$ \\ \hline
4,8 mod 8 & Maj or Weyl & $2^{D/2}$
\\
\hline 5,7 mod 8 & Dirac & $2^{(D+1)/2}$
\\ \hline 6 mod 8 & Weyl & $2^{D/2}$
\\ \hline
\end{tabular}
\end{center}\caption{\footnotesize  Existence of Weyl, Majorana and Majorana-Weyl spinors in diverse space-time dimensions.
Here we give also the number of real components.}
 \label{spinors}
\end{table}
Besides generators of the Poincar\'e group and supercharges, the superalgebra can include also a set of gauge bosonic generators.

The requirement that the action functional is invariant under local
supersymmetry implies the presence of gravitational interactions,
due to the generators of translations in the superalgebra. So local
supersymmetry and supergravity are equivalent names for such
theories. Therefore supergravity has to contain a metric tensor
$G_{MN}$ which turns to be necessarily associated to a $\psi_{M}$
called gravitino, which is labeled by both spinor and vector
indices. In Table \ref{supergravity} we give the on shell degrees of
freedom of the metric and the gravitino as function of $D$.
\begin{table}[top]
\begin{center}
\begin{tabular}{|l|l|l|}
\hline Field & Spin & On-shell d.o.f.  \\ \hline
 $G_{MN}$ & 2 & $(D-2)(D-1)/2-1$\\ \hline
 $\psi_{M}$ & 3/2 & $(D-3) \mathcal{I} /2$ \\ \hline
\end{tabular}
\end{center}\caption{\footnotesize  On shell degrees of freedom of the metric $G_{MN}$ and the gravitino $\psi_{M}$, which always
appear in the {\it supergravity multiplet}. The integer $\mathcal{I}$ represents
the number of components of the irreducible spinorial represention.}
 \label{supergravity}
\end{table}

Both for local and global supersymmetry the total number of
supercharges must be a multiple N of the components of the
irreducible spinorial representation of the Lorentz group. Since
there are no consistent quantum field theories including fields
whose spin is greater than 2 and no non-gravitational field theories
including fields with spin greater than one, one can get a
constraint on N. In particular there are no consistent 4D field
theories with N$>$8 and no consistent 4D field theories without
gravity with N$>$4. Since the dimension of the irreducible spinorial
representation depends on $D$ the maximum value of N, which gives
rise to a consistent theory, depends on $D$ as well. For instance
10D supergravity has at most N=2 and 11D supegravity has necessarily
N=1.

\section{The 6D Case}\label{6DgaugedSUGRA}

Now we focus on 6D supergravity which is one of the main topic of
this thesis. Contrary to the 4D case, supercharges with positive
chirality and with negative chirality do not give rise to equivalent
models. Therefore we introduce the notation
\be N=(N_+,N_-), \ee
where $N_{\pm}$ is the number of supercharges with positive (negative) chirality. If $N_+\neq N_-$ the corresponding
supergravity is called chiral. There exist several possibilities: $N=(1,0)$, $N=(1,1)$, $N=(2,0)$, $N=(2,2)$ and $N=(4,0)$.
Although in this thesis we are interested
in the minimal version $N=(1,0)$ of such models, in the following we shall describe briefly all the possibilities for the sake
of completeness.

We start with the {\bf chiral $N=(1,0)$ supergravity}, which has a number of supercharges equal to $N=2$ supersymmetry in 4D.
Indeed the R-symmetry group turns out to be $Sp(1)=SU(2)$ and henceforth it will be denoted by $Sp(1)_R$. The supergravity
multiplet consists of
\be \left(G_{MN}, \psi_M^j, B^+_{MN}\right),\label{gravmult}\ee
where $j$ takes value in the fundamental of $Sp(1)_R$, $\psi_M^j$ has a positive chirality, that is\footnote{In Appendix \ref{GR}
we give our conventions on the gamma matrices in the 6D case.}
\be \Gamma^7 \psi_M^j=\psi_M^j,\ee
  and $B^+_{MN}$ represents a {\it self-dual} field strength.
However, it is well known that field theories with self-dual field
strength do not admit a manifestly Lorentz invariant action
formulation, in dimensions 2 mod 4 \cite{Marcus:1982yu}. This
problem can be avoided if one combines the multiplet
(\ref{gravmult}) with a {\it tensor multiplet}:
\be \left(B^-_{MN}, \chi^j, \sigma \right),\ee
where $B^-_{MN}$ is an {\it antiself-dual} field strength, the spinor $\chi^j$ is called tensorino and it satisfies
\be \Gamma^7 \chi^j=-\chi^j\ee
and $\sigma$ is a real scalar field, which is called dilaton.
Now one can form a generic field strength $B_{MN}=B^+_{MN}+B^-_{MN}$ and give a manifestly Lorentz invariant action formulation.
We shall refer to $B_{MN}$ as a Kalb-Ramond field.

Moreover there exist also Yang-Mills multiplets, corresponding to a gauge group $\mathcal{G}$
\be \left(\mathcal{A}^I_M, \lambda^{Ij} \right),\ee
where $I$ is a Lie algebra index, $\mathcal{A}^I_M$ are the gauge fields and $\lambda^{Ij}$ the gauginos, which
satisfy
\be \Gamma^7 \lambda^{Ij}= \lambda^{Ij}.
\ee
Finally we can introduce also hypermultiplets:
\be \left(\psi^a,\phi^{\a}\right), \ee
where $\psi^a$, $a=1,...,2n_H$, called hyperinos,
satisfy
\be \Gamma^7\psi^a=-\psi^a \label{chiralhyper}\ee
and $\phi^{\alpha}$, $\a=1,...,4n_H$ are called hyperscalars. The
latter parametrize a  manifold which is non-compact and
quaternionic: a quaternionic manifold is a Riemannian manifols with
holonomy group\footnote{On a Riemannian manifold, tangent vectors
can be moved along a path by parallel transport, which preserves
vector addition and scalar multiplication. So a closed loop at a
base point p, gives rise to a invertible linear map of the tangent
space of the manifold in $p$. It is possible to compose closed loops
by following one after the other, and to invert them by going
backwards. Hence, the set of linear transformations arising from
parallel transport along closed loops is a group, called the
holonomy group.} contained in $Sp(n)\times Sp(1)$. In Table
\ref{Quaternionic} we give the quaternionic symmetric spaces
parametrized by hyperscalars for this type of supergravity.

The gauge group $\mathcal{G}$ can be taken to be a direct product of
an arbitrary gauge group, which does not act on hypermultiplets,
times a group $H_s$, defined in Table \ref{Quaternionic}, or
subgroup of $H_s$, which do act on hypermultiplets. We observe that
hypermultiplets are always neutral with respect to $Sp(1)_R$, which
can be either gauged or ungauged. In Section \ref{N=10} we will
discuss more the $N=(1,0)$ model gauging part of the isometry group
of the hyperscalar manifold and giving the explicit expression for
the action functional in this model.
\begin{table}[top]
\begin{center}
\begin{tabular}{|l|l|}
\hline  $G_s/H_s\times Sp(1)_R$  & $H_s$-representation of $\psi^a$  \\ \hline
 $Sp(n,1)/Sp(n)\times Sp(1)_R$ & ${\bf 2n}$ \\
 $SU(n,2)/SU(n)\times U(1) \times Sp(1)_R$ & ${\bf n_q}+{\bf n_{-q}}$  \\
$SO(n,4)/SO(n)\times SO(3)\times Sp(1)_R$ & $({\bf n}, {\bf 2})$ \\
$E_8 / E_7 \times Sp(1)_R$ & ${\bf 56}$\\
$E_7/SO(12)\times Sp(1)_R$ & ${\bf 32}$\\
$E_6/SU(6) \times Sp(1)_R$ & ${\bf 20}$\\
$F_4/Sp(3)\times Sp(1)_R$& ${\bf 14}$  \\
$G_2/Sp(1)\times Sp(1)_R$& ${\bf 4}$  \\
\hline
\end{tabular}
\end{center}\caption{\footnotesize  Quaternionic symmetric spaces parametrized by the hyperscalars. These are coset space
of the form $G_s/H_s\times Sp(1)_R$, where $G_s$ is a group and $H_s$ a subgroup of $G_s$. We give also the corresponding
hyperinos representation.}\label{Quaternionic}
\end{table}

We continue our analysis of 6D supergravities considering the {\bf non chiral $N=(1,1)$ model}. This is an $SU(2)$
gauged supergravity consisting of
\be \left(G_{MN}, \psi_{Mi}, B_{MN}, A_{M}, A_{Mi}^j, \lambda_i, \phi \right) \ee
coupled to a vector multiplet
\be \left(B_M, \chi_i, A_i^j, \xi\right). \ee
Minimal versions of these models has no stable maximally symmetric
ground state, that is Minkowski, de Sitter or anti de Sitter space.
However, it is possible to generalize this framework, by introducing
a mass parameter for the 2-form tensor $B_{MN}$, in a way that there
exist maximally symmetric solutions.

Another possible version is the {\bf chiral $N=(2,0)$ supergravity}. The supergravity multiplet
consist of
\be \left(G_{MN}, \psi^k_{M}, B^{+kl}_{MN}\right), \ee
where $k,l$ label ${\bf 4}$ of $USp(4)$ and the 2-form tensor field is in the ${\bf 5}$ of $USp(4)$. This supergravity
admits couplings to a tensor multiplet which contains
\be \left(B^-_{MN}, \lambda^k, \phi^{kl},\right). \ee
One can introduce an arbitrary number $n$ of tensor multiplets and the scalars parametrize $SO(n,5)/SO(n)\times SO(5)$ and
the $(n+5)$ 2-form fields trasforms as $(n+5)$ of $SO(n,5)$.

Moreover a {\bf non chiral $N=(2,2)$} and {\bf two chiral $N=(4,0)$} supergravities exist.
The chiral versions have exotic field contents which does not include a graviton.

\subsection{The $N=(1,0)$ Gauged Supergravity}\label{N=10}

Now we want to discuss in more detail the action formulation of the
$N=(1,0)$ model, gauging the isometry group of the hyperscalar
manifold. This derivation has been done in \cite{Nishino:1984gk} but
here we give the explicit expression for the action as it is
relevant for the original developments of Chapter
\ref{SUGRAspectrum}.

As a first step we choose one quaternionic manifold targed by the
hyperscalars. The most common choice is $Sp(n,1)/Sp(n)\times
Sp(1)_R$, appearing in the first row of Table \ref{Quaternionic}.
This is also the choice of Ref. \cite{Nishino:1984gk}. We remind
that hypermultiplets transform non-trivially under the isometry
group $Sp(n,1)$ of such manifold. Indeed the hypermultiplets turn
out to be in the representation ${\bf 2n}$ of $Sp(n)$. Here we
consider the gauging of the complete $Sp(n)\times Sp(1)_R$ but,
after giving the action in this case, we will explain what changes
if we gauge only a subgroup of such group. Many geometric properties
of such a manifold can be specified by introducing a representative
$L^{aj}$ of the coset. This is a map from the coset to the group
$Sp(n,1)$ and therefore depend on $\phi^{\a}$. The {\it Maurer
Cartan} 1-form decomposes as \cite{SS}
\be L^{-1}\partial_{\a}L=A^i_{\a}T^i+A^{\hat{I}}_{\a}T^{\hat{I}}+V^{aj}_{\a}T_{a j}, \ee
where $T^i$, $i=1,2,3$, and $T^{\hat{I}}$, $\hat{I}=1,...,n(2n+1)$,
are the anti-hermitian generators of $Sp(1)_R$ and $Sp(n)$
respectively, while $T_{a j}$ are the anti-hermitian coset
generators. The objects $A^i_{\a}$ and $A_{\a}^{\hat{I}}$ transforms
as $Sp(1)_R$ and $Sp(n)$ connection, respectively, while
$V^{aj}_{\a}$ transforms homogeneously under the induced local
tangent space transformations. Thus $A^i_{\a}$ and
$A_{\a}^{\hat{I}}$ can be used in the definition of the $Sp(n)\times
Sp(1)_R$ covariant derivatives and $V^{aj}_{\a}$ can be used as a
frame on the coset space and it is covariantly constant with respect
to the composite $Sp(n)\times Sp(1)_R$ connections and the
Christoffel connection defined on $Sp(n,1)/Sp(n)\times Sp(1)_R$ in
the usual way. As a result, the $Sp(n)\times Sp(1)_R$ connections
can be expressed in terms of the frames $V^{aj}_{\a}$. The gauging
of $Sp(n)\times Sp(1)_R$  can be implemented by using these objects.
For example, the covariant derivative of the hyperinos can be
written as follows
\be \nabla_M \psi=\left(\partial_M+\frac{1}{8}\omega^{[A,B]}_M[\Gamma_A,\Gamma_B]+\partial_{M}\phi^{\a}A_{\a}^{\hat{I}}T^{\hat{I}}
\right)\psi, \ee
where $\omega^{[A,B]}_M$ is the Lorentz connection defined in Appendix \ref{GR}.
Moreover the covariant derivative of hyperscalars is
\be \nabla_M\phi^{\a}=\partial_{M}\phi^{\a}-\mathcal{A}_{M}^{\hat{I}}\xi^{\a \hat{I}}-\mathcal{A}^i_M\xi^{\a i},\ee
where the $\xi^{\a I}$ are defined by
\be \xi^{\a I}=\left(T^I \phi \right)^{\a}. \ee
The elements discussed so far are sufficient to derive an action functional (up to quartic fermionic terms)
for $N=(1,0)$ gauged supergravity:
\bea S &=&\int d^6 X \sqrt{-G}\left[\frac{1}{\kappa^2}R-\frac{1}{4}\partial_M\sigma\partial^M\sigma
-\frac{\kappa^2}{48}e^{\kappa\sigma}G_{MNR}G^{MNR}
\right.\nonumber \\
&&\qquad \qquad \qquad
-\frac{1}{4}e^{\kappa\sigma/2}\left(\frac{1}{\hat{g}^2}\hat{F}^2+\frac{1}{g_1^2}F_1^2\right)
-g_{\a \b}(\phi)\nabla_M\phi^{\a}\nabla^M\phi^{\beta}   \nonumber \\
&&\qquad \qquad \qquad
-\frac{8}{\kappa^4}e^{-\kappa \sigma/2}C^{i I}C^{i I}
+\frac{1}{2}\overline{\psi}_M\Gamma^{MNR}\nabla_N\psi_R\nonumber \\
&&\qquad \qquad \qquad +\frac{1}{2}\overline{\chi}\Gamma^{M}\nabla_M\chi+\frac{1}{2}\overline{\lambda}\Gamma^M\nabla_M\lambda
+ \frac{1}{2}\overline{\psi}_a\Gamma^M\nabla_M\psi^a \nonumber \\
&& \qquad \qquad \qquad +\frac{\kappa}{4}\overline{\chi}\Gamma^N \Gamma^M \psi_N\partial_M\sigma
-\frac{\kappa}{2}\overline{\psi}^j_M\Gamma^N \Gamma^M \psi^a \nabla_N\phi^{\a} V_{\a a j}\nonumber \\
&& \qquad \qquad \qquad
+\frac{\kappa^2}{96}e^{\kappa \sigma/2} G_{MNR}\left(\overline{\psi}^L\Gamma_{[L}\Gamma^{MNR}\Gamma_{T]}\psi^T
+2\overline{\psi}_L \Gamma^{MNR}\Gamma^L \chi \right.\nonumber \\
&& \qquad \qquad \qquad \left.-\overline{\chi}\Gamma^{MNR}\chi+\overline{\lambda}\Gamma^{MNR}\lambda
+\overline{\psi}_a\Gamma^{MNR}\psi^a \right)\nonumber \\
&& \qquad \qquad \qquad -\frac{\kappa}{4\sqrt{2}\hat{g}}e^{\kappa
\sigma/4}\hat{F}_{MN}^{\hat{I}}
\left(\overline{\psi}_L\Gamma^{MN}\Gamma^L\lambda^{\hat{I}}
+\overline{\chi}\Gamma^{MN}\lambda^{\hat{I}}\right)\nonumber \\
&& \qquad \qquad \qquad -\frac{\kappa}{4\sqrt{2}g_1}e^{\kappa
\sigma/4}F_{1MN}^{i}
\left(\overline{\psi}_L\Gamma^{MN}\Gamma^L\lambda^{i}
+\overline{\chi}\Gamma^{MN}\lambda^{i}\right)\nonumber \\
&& \qquad \qquad \left.+\frac{\kappa}{2\sqrt{2}}e^{-\kappa \sigma /4}\left(\overline{\psi}_M\Gamma^M T^i\lambda C^i
- \overline{\chi} T^i \lambda C^i-2 \overline{\psi}^a\lambda^jV_{\a a j}\tilde{\xi}^{\a }\right)\right],\label{completeaction}
\eea
where $(\hat{g},\hat{F})$ and $(g_1,F_1)$ are the gauge constants
and the field strengths of $Sp(n)$ and $Sp(1)_R$ respectively, the
3-form $G_3$ is defined by
\be G_3=dB_2+\frac{1}{g^2}\left(\mathcal{A}\wedge F
-\frac{2}{3}\mathcal{A}\wedge \mathcal{A}\wedge
\mathcal{A}\right),\ee
where $g$ is a generic gauge constant, moreover
\be \tilde{\xi}^{\alpha I}\equiv \left(\hat{g}\xi^{\a \hat{I}},g_1
\xi ^{\a i}\right) \ee
and the $C$-functions are defined by
\be C^{i\hat{I}}=\hat{g} A^i_{\a}\xi^{\a \hat{I}},
C^{ik}=g_1\left(A^i_{\a}\xi^{\a k}-\delta^{ik}\right). \ee
In the second line of (\ref{completeaction}) a trace over the gauge group generators is understood.
An important feature of this action is it has a positive definite potential.

As we have already mentioned, the action in (\ref{completeaction}) corresponds to the gauging of the complete
$Sp(n)\times Sp(1)_R$ group. If we want to gauge just a subgroup, for example $E_7 \times U(1)_R$, only the gauge
field strengths, the gauginos and the $C$-functions of such a subgroup will appear in the action. Moreover one can also introduce
additional vector multiplets containing the gauge fields of an additional {\it group factor} which is not a subgroup
of $Sp(n)\times Sp(1)_R$.
The hypermultilpets are neutral with respect to the additional gauge interactions.
In this case one has to add the corresponding Yang-Mills fields and gauginos in the action as in the explicit
example treated in \cite{Randjbar-Daemi:1985wc}, in particular the additional gauginos will appear
only in line 4 and 8 of (\ref{completeaction}).

The action and the field content that we have discussed so far can be generalized to include all the quartic
fermionic terms \cite{Nishino:1986dc} and a generic number $n_T$ of tensor multiplets
\cite{Nishino:1997ff}, \cite{Ferrara:1997gh}, \cite{Riccioni:2001bg}.

The elements that we discussed in this section are quite general but in the next subsection we shall
discuss some particular examples.

\subsection{Salam-Sezgin Model and Anomaly-Free Versions} \label{SSandAfree}

Now we turn to describe the simplest example of 6D $N=(1,0)$ gauged
supergravity: the so called Salam-Sezgin model \cite{Salam:1984cj}.
In the bosonic sector of this model we have only the metric, the
Kalb-Ramond field, the dilaton and the gauge field associated to a
subgroup $U(1)_R$ of $Sp(1)_R$. This can be considered as a
supersymmetrization of the 6D Einstein-Maxwell model of Chapter
\ref{EMS} as the complete gauge group is $U(1)_R$. Moreover in the
fermionic sector we have the $U(1)_R$ gaugino in the vector
multiplet, the gravitino in the supergravity multiplet and the
tensorino in the tensor multiplet. The complete action for this
system can be obtained by putting equal to zero in
(\ref{completeaction}) all the fields that we have not mentioned. In
particular there are no hyperscalars in the bosonic sector.
Therefore the bosonic part $S_B$ of the action is
\bea S_B &=&\int d^6 X \sqrt{-G}\left[\frac{1}{\kappa^2}R-\frac{1}{4}\partial_M\sigma\partial^M\sigma
-\frac{\kappa^2}{48}e^{\kappa\sigma}G_{MNR}G^{MNR}
\right.\nonumber \\
&&\qquad \qquad \qquad
\left.-\frac{1}{4g_1^2}e^{\kappa\sigma/2}F_1^2
-\frac{8g_1^2}{\kappa^4}e^{-\kappa \sigma/2}
\right],\label{SSaction} \eea
and the corresponding EOM are
\bea &&\frac{1}{\kappa^2}R_{MN}=
\frac{1}{2g_1^2}e^{\kappa\sigma/2}F_{1MP}F^{\,\,P}_{1N}+\frac{1}{4}\partial_{M}\sigma\partial_N\sigma
+\frac{\kappa^2}{16}e^{\kappa \sigma}G_{MPQ}G_N^{\,\,\, PQ}-\frac{1}{4\kappa}G_{MN}\nabla^2\sigma, \nonumber \\
 &&\frac{1}{\kappa}\nabla^2\sigma=\frac{1}{4g_1^2}e^{\kappa\sigma/2}F_1^2+ \frac{\kappa^2}{24} e^{\kappa\sigma} G_{MNP}G^{MNP}
 -\frac{8g_1^2}{\kappa^4}e^{-\kappa\sigma/2},\nonumber \\
&&\nabla_M\left(e^{\kappa\sigma/2}F_1^{MN}\right)=\frac{\kappa^2}{4}
e^{\kappa\sigma} G^{NPQ} F_{1PQ},\nonumber \\
&& \nabla_M\left(e^{\kappa\sigma}G^{MNP}\right)=0.\label{SSEOM0}\eea

We look now for maximally symmetric solutions of this model which
preserve a 4D Poincar\'e invariance. To this end we observe that 4D
Poincar\'e invariance implies $G_{MNR}=0$ at the background level.
By using this constraint, the EOM (\ref{SSEOM0}) reduce to
\bea &&\frac{1}{\kappa^2}R_{MN}=
\frac{1}{2g_1^2}e^{\kappa\sigma/2}F_{1MP}F^{\,\,P}_{1N}+\frac{1}{4}\partial_{M}\sigma\partial_N\sigma
-\frac{1}{4\kappa}G_{MN}\nabla^2\sigma, \nonumber \\
 &&\frac{1}{\kappa}\nabla^2\sigma=\frac{1}{4g_1^2}e^{\kappa\sigma/2}F_1^2
 -\frac{8g_1^2}{\kappa^4}e^{-\kappa\sigma/2},\nonumber \\
&&\nabla_M\left(e^{\kappa\sigma/2}F^{MN}\right)=0. \label{SSEOM}\eea

Since we are looking for a maximally symmetric solution we suppose
$\sigma$ constant and inserting this ansatz in (\ref{SSEOM}) we get,
after some manipulations,
\be R=\frac{16 g_1^2}{\kappa^2}e^{-\kappa \sigma/2},\label{RSS}\ee
that is the Ricci scalar is a non vanishing constant. Therefore the
6D Minkowski space is not a solution of the Einstein equation
appearing in (\ref{SSEOM}). Since $S^2$ is the only orientable 2D
manifold with positive constant curvature an obvious solution of
(\ref{RSS}) is $(Minkowski)_4\times S^2$. Even if a priori we cannot
exclude the presence of a {\it warp factor}, in Ref.
\cite{Gibbons:2003di} it was proved that this is actually the unique
maximally symmetric smooth solution for the metric. In particular
the 6D de Sitter and Anti de Sitter spaces are not solutions of the
EOM. Requiring the other fields to have the same symmetry as the
metric one gets the complete {\it Salam-Sezgin background}:
\bea ds^2 &=&\eta_{\mu \nu}dx^{\mu}dx^{\nu}+
a^2\left(d\theta^2+\sin^2\theta d\varphi^2\right), \nonumber \\
\mathcal{A}&=&\frac{n}{2}(\cos\theta \pm 1)d\varphi, \nonumber \\
\sigma&=&\sigma_0=constant, \quad G_{MNR}=0. \label{SSB}\eea
This is a solution of (\ref{SSEOM}) if
\be \frac{1}{\kappa^2}=\frac{n^2e^{\kappa \sigma_0/2}}{8g_1^2
a^2},\qquad \frac{1}{\kappa^4}=\frac{n^2 e^{\kappa
\sigma_0}}{64g_1^4 a^4}.\label{constraintSS}\ee
We observe that first and second line of (\ref{SSB}) correspond
exactly to equations (\ref{sphere}) and (\ref{monopole}) of Chapter
\ref{EMS}, that is the monopole configuration over $S^2$. Dirac
quantization condition still holds, that is $n$ has to be an
integer. However, supersymmetry of the present model and the
presence of the dilaton convert constraints (\ref{constraint}) into
(\ref{constraintSS}). An interesting consequence is that now the
monopole number $n$ must satisfy
\be n=\pm1, \label{SSn}\ee
which can be proved squaring first constraint in (\ref{constraintSS}) and then using second constraint in (\ref{constraintSS}).
A detailed analysis of the local supersymmetry transformations shows that background (\ref{SSB}) preserves 1/2 of the 6D $N=(1,0)$
supersymmetries of this model \cite{Salam:1984cj}, that is a 4D $N=1$ supersymmetry.

Since the Salam-Sezgin model is the simplest realization of the 6D $N=(1,0)$ gauged supergravity, it was used as a toy
model to study the properties of this class of models and of higher dimensional supergravities. For instance
the authors of \cite{Aghababaie:2002be} constructed the 4D $N=1$ supergravity, which describes
the low energy dynamic of the Salam-Sezgin
model expanded around the background (\ref{SSB}), in order to provide a simple setting sharing
the main properties of realistic string compactifications.
Moreover in \cite{Gibbons:2003di} additional solutions of this model, which
present conical and non conical singularities, were found\footnote{We shall describe these solutions in Section \ref{singular}.}.

On the other hand, like most 6D supergravities, the Salam-Sezgin
model suffers from the breakdown of local symmetries due to the
presence of gravitational, gauged and mixed anomalies. Therefore
such a model must be enlarged to include additional supermultiplets.
It is interesting that the requirement of anomaly freedom is a
strong guiding principle to select consistent models, like in 10D
supergravities. So this is another property shared by the 6D
$N=(1,0)$ gauged models and higher dimensional supergravities.

Until recently the only one known anomaly free model of this type
was the $\mathcal{G}=E_7 \times E_6 \times U(1)_R$ model, where
$E_7$ is a subgroup\footnote{In this case $n=456$.} of the $Sp(n)$
that we discussed in Subsection \ref{N=10}, $E_6$ is an additional
{\it group factor} and $U(1)_R$ is a subgroup of $Sp(1)_R$. As we
discussed in Subsection \ref{N=10} the hypermultiplets are singlets
with respect to $E_6$ and $U(1)_R$ but, in order to cancel the
anomalies by means of the {\it Green-Schwarz mechanism}, they are in
representation {\bf 912} of $E_7$. Recently more example of
anomaly-free models were found. For instance in
\cite{Avramis:2005qt} a $\mathcal{G}=E_7\times G_2 \times U(1)_R$
model was proposed, with $E_7\times G_2\subset Sp(392)$ and the
hypermultiplets in the representation ${\bf (56,14)}$ of $E_7\times
G_2$. Moreover in \cite{Avramis:2005hc} the authors proposed
 $\mathcal{G}=F_4\times Sp(9) \times U(1)_R$ with hypermultiplets in the representation ${\bf(52,18)}$ of $F_4\times Sp(9)$.
  Finally in Ref. \cite{Suzuki:2005vu} a huge number of simple anomaly free models was presented with $\mathcal{G}$ given by
products of $U(1)$ and/or $SU(2)$ and particular hyperinos representations.

Therefore non-Abelian extensions of the Salam-Sezgin model,
including hypermultiplets, are very interesting because they are
needed for the consistency at the quantum level. The gauge group for
these models will be $\mathcal{G}=\tilde{\mathcal{G}}\times
\mathcal{G}_R$, where $\mathcal{G}_R$ is a subgroup of $Sp(1)_R$ and
$\tilde{\mathcal{G}}$ is composed by a subgroup of $Sp(n)$ and in
case by an additional gauge group\footnote{For instance $E_6$ in the
$E_7\times E_6 \times U(1)_R$ model.}. In the following we shall
describe the $S^2$ compactification of such non-Abelian models
\cite{Randjbar-Daemi:1985wc}, \cite{Avramis:2005qt}, with a monopole
embedded in the Cartan subalgebra of $\tilde{\mathcal{G}}\times
\mathcal{G}_R$. This set up is the only compatible with maximal
symmetry.

Like the Salam-Sezgin model the bosonic theory includes the metric, the Kalb-Ramond
field and  the dilaton  but here we have also the gauge fields of the complete gauge group
$\mathcal{G}=\tilde{\mathcal{G}}\times \mathcal{G}_R$ and the hyperscalars $\phi^{\alpha}$, in some representation of
the subgroup of $Sp(n)$. The complete
bosonic action for these fields is
\bea S_B &=&\int d^6 X \sqrt{-G}\left[\frac{1}{\kappa^2}R-\frac{1}{4}\partial_M\sigma\partial^M\sigma
-\frac{\kappa^2}{48}e^{\kappa\sigma}G_{MNR}G^{MNR}
\right.\nonumber \\
&&\qquad \qquad \qquad
-\frac{1}{4}e^{\kappa\sigma/2}\left(\frac{1}{\tilde{g}^2}\tilde{F}^2+\frac{1}{g_1^2}F_1^2\right)
-g_{\a \b}(\phi)\nabla_M\phi^{\a}\nabla^M\phi^{\beta}   \nonumber \\
&&\qquad \qquad \qquad
\left.-\frac{8}{\kappa^4}e^{-\kappa \sigma/2}C^{i I}C^{i I}\right], \label{Baction} \eea
 The symbols have been defined in Subsection \ref{N=10}. In
\cite{Randjbar-Daemi:2004qr} the explicit expression for the
$C$-functions in the last line of (\ref{Baction}) for
$\mathcal{G}=E_7\times E_6 \times U(1)_R$ have been given, but the
result is applicable to any other model of this type. A remarkable
result is that the absolute minimum of the scalar potential is at
$\phi^{\a}=0$ \cite{Nishino:1986dc,Randjbar-Daemi:2004qr}. This
property suggest us to set $\phi^{\a}=0$ at the background level in
order to get a stable solution and henceforth we will assume that.
Moreover in the following we will focus on the case
$\mathcal{G}_R=U(1)_R$. This is true for all the anomaly-free models
with a gauge group that contains the gauge group of the SM. In this
case the bosonic EOM are
\bea
&&\frac{1}{\kappa^2}R_{MN}=\frac{1}{2}e^{\kappa\sigma/2}\left(\frac{1}{\tilde{g}^2}\tilde{F}_{MP}\tilde{F}^{\,\,P}_{N}
+\frac{1}{g_1^2}F_{1MP}F^{\,\,P}_{1N}\right)
+\frac{1}{4}\nabla_{M}\sigma\nabla_N\sigma
\nonumber \\&&\qquad\qquad+\frac{\kappa^2}{16}e^{\kappa \sigma}G_{MPQ}G_N^{\,\,\, PQ}-\frac{1}{4\kappa}G_{MN}\nabla^2\sigma, \nonumber \\
 &&\frac{1}{\kappa}\nabla^2\sigma=\frac{1}{4}e^{\kappa\sigma/2}\left(\frac{1}{\tilde{g}^2}\tilde{F}^2+\frac{1}{g_1^2}F_1^2
+ \frac{\kappa^2}{24} e^{\kappa\sigma} G_{MNP}G^{MNP}\right)-\frac{8g_1^2}{\kappa^4}e^{-\kappa\sigma/2},\nonumber \\
&&\nabla_M\left(e^{\kappa\sigma/2}\tilde{F}^{MN}\right)=\frac{\kappa^2}{4}
e^{\kappa\sigma} G^{NPQ} \tilde{F}_{PQ},\nonumber \\
&&\nabla_M\left(e^{\kappa\sigma/2}F_1^{MN}\right)=\frac{\kappa^2}{4}
e^{\kappa\sigma} G^{NPQ} F_{1PQ},\nonumber \\
&& \nabla_M\left(e^{\kappa\sigma}G^{MNP}\right)=0.\label{EOM1}\eea
We want to discuss the following maximally symmetric solution
\bea && ds^2=\eta_{\mu \nu}dx^{\mu}dx^{\nu} + a^2\left(d\theta^2 +\sin^2\theta d\varphi^2\right),\nonumber \\
&&\mathcal{A}=\frac{n}{2}Q(\cos\theta \pm 1) \quad \sigma=constant,\nonumber \\
&& \sigma=\sigma_0=constant, \quad
G_{MNR}=0\label{spheremonopole}\eea
where $Q$ is a generator of a $U(1)$ subgroup of a simple factor of
$\mathcal{G}$, satisfying $Tr\left(Q^2\right)=1$, and $n$ is a real
number which takes discrete values because of the Dirac quantization
condition. In the following we will denote by $U(1)_M$ the abelian
group generated by $Q$. The configuration (\ref{spheremonopole}) is
the trivial generalization of (\ref{SSB}) to a non-Abelian model.
The background gauge field in (\ref{spheremonopole}) is a monopole
configuration and, from group theory point of view, it breaks
$\mathcal{G}$ to some subgroup $H$, which is generated by the
generators of $\mathcal{G}$ which commute with $Q$. Moreover the
equations (\ref{EOM1}) implies \cite{Randjbar-Daemi:1985wc}
\be \frac{1}{\kappa^2}=\frac{n^2e^{\kappa \sigma_0/2}}{8g^2
a^2},\qquad \frac{g_1^2}{\kappa^4}=\frac{n^2 e^{\kappa \sigma_0}}{64
g^2 a^4},\label{constr1}\ee
where $g$ is the gauge constant corresponding to the background
gauge field. The constraints (\ref{constr1}) represent the
generalization of (\ref{constraintSS}) and they imply the following
equation
\be n^2=\frac{g^2}{g_1^2}. \ee
If $g=g_1$ we have $n=\pm 1$, that is Eq. (\ref{SSn}), but this is
not needed if we embed $U(1)_M$ in $\tilde{\mathcal{G}}$. An
important aspect is that the only monopole embedding which preserves
part of the 6D $N=(1,0)$ supersymmetries is $Q$ along the Lie
algebra of $U(1)_R$ and in this case we have $g=g_1$. No other
embedding ($g\neq g_1$) preserve any residual supersymmetry.

Given the symmetries of the problem, we can expect that
the low energy effective gauge group is ${\cal H}\times \mathcal{G}_{KK}$, where ${\cal H}$ is the subgroup of ${\cal G}$
that commutes with $U(1)_M \subset {\cal G}$ in which the
monopole lies, and is orthogonal to\footnote{Due to the Chern-Simons coupling in supergravity,
the $U(1)$ gauge field in the direction of the monopole eats the axion arising
from the Kalb-Ramond field and acquires a mass \cite{Randjbar-Daemi:1985wc,Aghababaie:2002be}.} $U(1)_M$;
moreover $\mathcal{G}_{KK}$ is the
KK gauge group coming from the isometry of the internal space.

An explicit example is $Q$ along the Lie algebra of $E_6$ in the
$E_7\times E_6 \times U(1)_R$ model \cite{Randjbar-Daemi:1985wc}.
From the purely group theory point of view such an embedding breaks
$E_6$ down to $SO(10)\times U(1)_M$ and leaves $E_7\times U(1)_R$
unbroken. However, the gauge group of the 4D effective theory
contains an additional SU(2) factor, coming from the isometry of
$S^2$. This is a well known mechanism in KK theories and we shall
denote this extra group by $SU(2)_{KK}$. Moreover, due to the {\it
Chern-Simons coupling} in supergravity, the $U(1)_M$ gauge field
eats the axion arising from the Kalb-Ramond field and acquires a
mass \cite{Randjbar-Daemi:1985wc,Aghababaie:2002be}. So the complete
gauge group of the 4D effective theory is $E_7\times SO(10) \times
U(1)_R \times SU(2)_{KK}$. Since the only fermions that interact
with the background gauge field\footnote{There exist no normalizable
fermionic zero modes on $S^2$ satisfying a standard Dirac equation
in the absence of coupling to the background monopole field.} are
the gauginos of $E_6$, the latter sector contains the fermionic zero
modes. The 4D massless chiral fermions comprise $2|n|$ families of
$SO(10)$, in the representation {\bf 16} and no antifamilies. They
belong to the $|n|$-dimensional irreducible representation of
$SU(2)_{KK}$. These fermions are neutral with respect to $E_7$ but
they carry a $U(1)_R$ charge, half the families are positive and the
other half are negative. Such ${\bf 16}$-families can be interpreted
as leptons and quarks of the SM embedded in a grand unification
scenario. Unfortunately the classical stability requires $|n|=1$ and
therefore only 2 families
\cite{Randjbar-Daemi:1985wc,Avramis:2005qt,Avramis:2005hc}. The
bosonic sector of the low energy 4D effective theory contains the
graviton and the gauge fields corresponding to $E_7\times SO(10)
\times U(1)_R \times SU(2)_{KK}$. The gauge fields interacting with
leptons and quarks are only the gauge fields of $SO(10)\times
U(1)_R$ and of $SU(2)_{KK}$ if $|n|>1$. There is not a complete
discussion of the stability issue in the literature. In Appendix
\ref{SUGRAstability} we will give a first step toward this direction
following the lines of \cite{Randjbar-Daemi:1985wc,Avramis:2005qt}.
In addition to \cite{Randjbar-Daemi:1985wc,Avramis:2005qt} we will
include all the present known anomaly free models. The result is
that a stable sphere compactification and the embedding of the SM
gauge group is possible only in the $E_7\times E_6 \times U(1)_R$
case.

\section{Supersymmetric Large Extra Dimensions}\label{SLED}

So far we have analysed 6D supergravities without any assumptions on
the size of the extra dimensions. A physical applications of this
framework can be made in the context of LED. As we have mentioned at
the beginning of the present chapter, SLED give hope to solve the
cosmological constant problem. The aim of this section is to explain
this idea in more detail \cite{Burgess:2004ib} and point out its
shortcomings.

In order to discuss the cosmological constant problem, let us
consider a generic parameter $p$ which describes a physical quantity
and is found to be small when measured in an experiment which is
performed at an energy scale $\mu$. We would like to understand this
in terms of a microscopic theory which is defined at energy
$\Lambda\gg \mu$ and predicts the value of $p(\mu)$ as follows
\be p(\mu)=p(\Lambda)+\d p(\mu,\Lambda), \ee
where $p(\Lambda)$ represents the contribution to $p$ due to the
parameters in the microscopic theory, and $\delta p$ represents the
contributions to $p$ which are obtained as we integrate out all of
the physics in the energy range $\mu<E<\Lambda$. The smallness of
$p(\mu)$ can be understood if both $p(\Lambda)$ and $\d p$ are
small. Although we may not be able to understand why $p(\Lambda)$ is
small until we have a correct microscopic theory (up to some energy
scale), we should be able to understand why ordinary physics at
energies $\mu<E<\Lambda$ do not make $\d p(\mu, \Lambda)$
unacceptably large. If we find $\delta p (\mu,\Lambda)$ to be many
orders of magnitude larger than the measured value $p(\mu)$ then we
suspect that we do not understand the physics at energies
$\mu<E<\Lambda$ as well as we thought. This is actually the case for
the cosmological constant: cosmological observations indicate that
the vacuum's energy density is at present $\rho\sim (10^{-3}eV)^4$
but the theoretical prediction for $\d \rho(\mu,\lambda)$ is many
orders of magnitude greater as a particle of mass $m$ contributes an
amount of $\delta \rho(\mu,\Lambda)\sim m^4$ when it is integrated
out and pratically all of the elementary particles we know have
$m\gg 10^{-3}eV$.

A solution of the cosmological constant problem could involve a
modification of gravity at energy scale\footnote{Other interesting
works \cite{Dvali:2000hr,Dvali:2000xg,Dvali:2002pe} propose a
possible solution of the cosmological constant problem, by
considering modification of gravity at very large distances, as we
have summarized in Section \ref{CCproblem}.} $E>\mu\sim10^{-3}eV$,
but, if this is the case, must not ruin the precise agreement with
all the many non-gravitational experiments which have been performed
so far. A scenario which could have both these properties is the LED
scenario. Indeed LED allows that only gravity propagate in large
extra dimensions, while all the non gravitational interactions
should be confined on a 3-brane within the extra-dimensional space.
Since we have a precise relation between the 4D Planck scale, the
fundamental scale $M$ of the higher dimensional theory and the
volume of the internal space we can have a constraint on the latter
by requiring $M$ to be of the order of $TeV$. This constraint rules
out just one large extra dimension and the simplest choice is
therefore a 6D model. In this case the typical physical size $r$ of
the internal space is of the order of the observed value
$\rho^{-1/4}(\mu)$ of the vacuum energy density. Therefore we expect
that gravity (and also its prediction for $\rho$) is much different
from ordinary 4D Einstein's theory at submillimeter length scale,
which corresponds to energies close to $10^{-3}eV$. The idea of SLED
is to supersymmetrize a LED model in order to have control on the
prediction for $\rho$: indeed we know that a supersymmetric theory,
without explicit and spontaneous symmetry breaking, predicts a
vanishing vacuum's energy. One can hope to get the observed value of
the vacuum's energy performing a small supersymmetry breaking in
such a framework.

In a higher dimensional theory with size $r$ of extra dimensions the
superymmetry breaking scale is expected to be of the order of $1/r$,
and therefore small for large extra dimensions. Of course in order
for this to be phenomenologically viable one has to find a mechanism
which separates the supersymmetry breaking in the bulk and on the
brane. The complete (theoretical) value of $\rho$ in this context
can be computed by performing the following three steps. First one
integrates out at the full quantum level all the brane degrees of
freedom to obtain an effective theory defined on the brane with
tension $T$. After that one can consider such tension as source for
the higher dimensional geometry by inserting in Einstein equation
delta-function sources proportional to $T$. In a 6D supergravity the
sum of the quantum brane contribution and the classical bulk
contribution to $\rho$ cancel exactly and one is left with only the
bulk quantum contribution. In some circumstances this quantum
contribution is of order $m_{sb}^4$, where $m_{sb}\sim M^2/M_{Pl}$,
where $M$ is the 6D Planck mass and $M_{Pl}$ is the 4D Planck mass
\cite{Burgess:2004kd}. The small size of the 4D vacuum energy is in
this way attributed to the very small size with which supersymmetry
breaks in the bulk relative to the scale with which it breaks on the
brane.

Although this is an interesting idea, there are some points that are
not clear and need more investigations, for instance the complete
effect of supersymmetry breaking in the bulk and of the
compactness of the internal space, which is not shared by the models \cite{Dvali:2000hr,Dvali:2000xg,Dvali:2002pe} that we have briefly summarized
at the end of Section \ref{CCproblem}. Moreover, in explicit realizations of the SLED scenario, for example in
the original paper \cite{Aghababaie:2003wz}, a background monopole
is introduced on the bulk to support a compact internal manifold. Consequently a Dirac quantization condition
in general emerges and it is not completely clear if this additional constraint can ruin the SLED argument.
Finally, a detailed study of such a scenario requires the complete analysis of perturabations,
which are actually needed in order to compute sistematically
all the contributions to the vacuum's energy density.

However, in order to implement this idea, one is interested in
explicit brane solutions of the 6D supergravity EOM, which break
supersymmetry. Such solutions are presented in Section
\ref{singular}.

\section{Brane Solutions}\label{singular}

So far we have analysed smooth solutions of minimal gauged
supergravity, which turn out to be also the maximal symmetric one.
In this section we want to study what happens if one relaxes the
maximal symmetry assumption for the complete 6D space-time. As we
will see this necessarily leads to singularities. One interesting
application is interpreting them as 3-branes which support SM
fields. Indeed in order to realize a LED or a SLED scenario brane
solutions are interesting because standard KK compactification are
known to be phenomenological incompatible with LED.

However, we shall assume the following properties of the background
solutions:
\begin{itemize}
\item[(i)] {\it 4D Poincar\'e invariance.  }
\item[(ii)] {\it Axisymmetry of the internal 2D space.}
\item[(iii)] {\it The hyperscalars are not active.}
\end{itemize}

Assumptions (i), (ii) and (iii)  will simplify our calculations, but
(iii) is also motivated by the fact that the potential has a global
minimum in $\phi^{\a}=0$, and therefore such set up will support
stability. The 4D Poincar\'e invariance implies that the Kalb-Ramond
field strength vanishes, and this, together with (iii), leads to the
bosonic EOM (\ref{EOM1}). For later purposes, we remark that those
equations are invariant under the constant classical scaling
symmetry
\be G_{MN}\rightarrow \xi G_{MN},\qquad e^{\kappa \sigma /2}\rightarrow \xi e^{\kappa \sigma /2}.\label{scaling} \ee
Furthermore, the sole effect of the transformation
\be e^{\kappa \sigma /2}\rightarrow \zeta^2  e^{\kappa \sigma /2},
\qquad \left(F_{1MN},\tilde{F}_{MN}\right)\rightarrow
\zeta^{-1}\left(F_{1MN},\tilde{F}_{MN}\right)\label{scaling2}\ee
is the rescaling
\be g_1\rightarrow \zeta g_1.\label{effect}\ee
We have to remember these symmetries when we will count the number of independent parameters for a given solution.
Now we analyse the implications of (i) and (ii) on the background tensors.
The 6D metric has to be of the form
\be ds^2=e^{A(\rho)}\eta_{\mu \nu}dx^{\mu}dx^{\nu}+d\rho^2+e^{B(\rho)}d\varphi^2,\label{metricAB}\ee
where $\rho$ is a {\it radial coordinate}, with range $0\leq
\rho\leq \overline{\rho}$, and $\varphi$ an {\it angular
coordinate}, whose range is assumed to be $0\leq \varphi<2\pi$,
moreover $A$ and $B$ does not depend on $\varphi$ and on the 4D
coordinates $x^{\mu}$. Another equivalent radial coordinate is
\be u(\rho)\equiv \int_0^{\rho}d\rho' e^{-A(\rho')/2}, \label{udef}\ee
whose range is $0\leq u\leq \overline{u}\equiv
\int_0^{\overline{\rho}}d\rho e^{-A(\rho)/2}$, and in this frame the
metric reads
\be ds^2=e^{A(u)}\left(\eta_{\mu \nu}dx^{\mu}dx^{\nu}+du^2\right)
+e^{B(u)}d\varphi^2. \label{metricu}\ee
The coordinate $u$ will be useful when we will discuss fluctuations
in Chapter \ref{SUGRAspectrum}. Indeed $u$ will turn out to be the
independent variable of 1D Schroedinger-like equation governing the
fluctuations. The metric (\ref{metricAB}) represents a warped
geometry because of the warp factor $e^A$. Moreover the dilaton and
the gauge field depend only on $\rho$ and
\be F_{\mu \nu}=0,\quad F_{\mu m}=0, \quad F_{mn}=f(\rho)\epsilon_{mn}, \ee
where $\epsilon_{mn}$ is the 2D anti-symmetric Levi-Civita symbol (with $\epsilon_{\rho \varphi}=1$).

\subsection{General Axisymmetric Solutions}\label{generalGGPsub}

In Ref.  \cite{Gibbons:2003di} Gibbons, Guven and Pope (GGP) found
the most general solution satisfying (i), (ii), and (iii). We do not
prove that the GGP solution is actually a solution because this
would involve a very long and standard calculation but we give now
the explicit expression:
\bea e^{2A}&=& \left(\frac{q \lambda_2}{4g_1
\lambda_1}\right)\frac{\cosh\left[\lambda_1(\eta - \eta_1)\right]}
{\cosh\left[\lambda_2(\eta - \eta_2)\right]},\nonumber \\
e^{-2B}&=&\left(\frac{g_1 q^3}{\lambda_1^3
\lambda_2}\right)e^{-2\lambda_3
\eta}\cosh^3\left[\lambda_1(\eta-\eta_1)\right]
\cosh\left[\lambda_2(\eta - \eta_2)\right]\nonumber \\
F&=&\left(\frac{g q}{\kappa}\right)e^{B-A}e^{-\lambda_3\eta}\,Q d\eta \wedge d\varphi, \nonumber \\
e^{\kappa \sigma}&=&e^{2A}e^{2\lambda_3\eta},\label{generalaxi}\eea
where $\eta\equiv\int^{\rho}d\rho'e^{-2A-B/2}$, $Q$ again denotes a
generator of a $U(1)$ subgroup of a simple factor of $\mathcal{G}$,
satisfying $Tr\left(Q^2\right)=1$, moreover
$q,\eta_{1,2},\lambda_{1,2,3}$ are constants, and the parameters
$\lambda_i$ satisfy
\be \lambda_1^2+\lambda_3^2=\lambda_2^2.\label{constlambda} \ee
Without loss of generality one can take $\lambda_{1,2}\geq 0$ and
Eq. (\ref{constlambda}) implies $\lambda_2\geq \lambda_1$. Moreover
in (\ref{generalaxi}) $F$ represents the gauge field strength which
is not zero in the background and $g$ the corresponding gauge
constant. At face value we have a total of 5 integration constants
$q,\lambda_1,\lambda_2,
 \eta_1,\eta_2$. However, one combination of these five constants correspond to the scaling (\ref{scaling}).
A second combination similarly corresponds to the second rescaling (\ref{scaling2}), whose sole effect is (\ref{effect}),
leaving a total of 3 nontrivial parameters. The metric in (\ref{generalaxi}) has at most 2 singularities because
singularities can occur where the metric components vanish or diverge. Inspection
of (\ref{generalaxi}) shows that this only occurs when $\eta\rightarrow -\infty$ and $\eta\rightarrow +\infty$.
Such singularities are not all of the purely conical type. The metric singularities which are not purely conical
($\lambda_3\neq 0$) come in two
categories \cite{Burgess:2004dh}. Some can be interpreted as describing the fields of two localized 3-brane like objects,
while others are better interpreted as the bulk fields which are sourced by a combination of a 3-brane and 4-brane,
rather than being due to two 3-branes.
An extended discussion on the singularities of the non purely conical type is given in \cite{Burgess:2004dh}. In the
next subsection we shall describe the purely conical case.

The class of solutions that we presented here are not the most
general appearing in the literature. Relaxing the condition of 4D
Poincar\'e symmetry to that of only 4D maximal symmetry should allow
more general solutions to be found \cite{Tolley:2005nu}.  More
general solutions also exist breaking axial symmetry
\cite{Lee:2005az}, or having nontrivial VEVs for the hyperscalars
\cite{Parameswaran:2005mm}. Furthermore, {\it dyonic string}
solutions have been constructed in \cite{Randjbar-Daemi:2004qr}.

\subsection{Conical Singularities}\label{conSing}

Here we focus on the subset of the general solutions (\ref{generalaxi}) having purely
conical singularities ($\lambda_3=0$), which can be interpreted as being sourced by two 3-branes. We shall call
this background conical-GGP solution.

The explicit conical-GGP solution is then\footnote{The coordinate
$u$ is related to the coordinate $r$ in \cite{Gibbons:2003di} by
$r=r_0\cot(u/r_0)$.} \cite{Gibbons:2003di,Burgess:2004dh}:
\bea e^A&=&e^{\kappa \sigma/2}=\sqrt{\frac{f_1}{f_0}}, \quad e^B=\alpha^2 e^A\frac{r_0^2\cot^2(u/r_0)}{f_1^2},\nonumber\\
\mathcal{A}&=&-\frac{4\alpha g}{q\kappa f_1}\, Q\,
d\varphi,\label{GGPsolution}\eea
where $\alpha$ is a real positive number, $q$ is a real number,
 and moreover
\be f_0\equiv 1+\cot^2\left(\frac{u}{r_0}\right), \quad f_1 \equiv
1+\frac{r_0^2}{r_1^2}\cot^2\left(\frac{u}{r_0}\right), \ee
with $r_0^2\equiv \kappa ^2/(2g_1^2)$, $r_1^2\equiv 8/q^2$. The
range of $u$ is from $0$ to $\overline{u}\equiv \pi r_0/2 $ and the
deficit angles are
\bea \delta&=&2\pi \left(1-\alpha \frac{r_1^2}{r_0^2}\right). \\
\overline{\delta}&=&2\pi\left(1-\alpha\right).\eea
Notice then that a non-trivial warping enforces the presence of a 3-brane source.  On the other hand, the parameter $\alpha$ is
not fixed by the EOM and it represents a
modulus.

The expression for the gauge field background in equation
(\ref{GGPsolution}) is well-defined in the limit $u\rightarrow 0$,
but not as $u \rightarrow \overline{u}$.  We should therefore use a
different patch to describe the $u={\overline{u}}$ brane, and this
must be related to the patch including the $u=0$ brane by a
single-valued gauge transformation\footnote{We have already treated
this topological constraint for the sphere compactification of the
Einstein-Maxwell-Scalar system in Subsection \ref{su2 x u1} of
Chapter \ref{EMS}.}.  This leads to a Dirac quantization condition,
which for a field interacting with $\mathcal{A}$ through a charge
$e$ gives
\be \alpha e\frac{4 g}{\kappa q}= \alpha e\frac{r_1}{r_0}
\frac{g}{g_1} = N \label{DiracQ}\ee
where $N$ is an integer. The charge $e$ can be computed once we have
selected the background gauge group, since it is an eigenvalue of
the generator $Q$. Finally, an explicit calculation shows that the
internal manifold corresponding to solution (\ref{GGPsolution}) has
an $S^2$ topology (its Euler number equals 2). Therefore the
internal manifold is compact. Thus we expect a discrete spectrum of
the fluctuations around such a background. This is confirmed by the
calculation of Chapter \ref{SUGRAspectrum}, where we will focus on
bulk sectors that could give rise to SM-like gauge fields and
charged matter.


\chapter{Fluctuations around Brane Solutions}\label{SUGRAspectrum}

In Chapter \ref{6DSUGRA} we have described minimal 6D gauged
supergravity and its solutions which have received much interest
recently for several reasons. From the top down, the theory shares
many features in common with 10D supergravity, whilst remaining
relatively simple, and so it can be used as a toy model for 10D
string theory compactifications.  From the bottom up, it provides a
context in which to extend the well-trodden path of 5D brane world
models to codimension two.  Moreover, as we have discussed in
Section \ref{SLED}, in 6D models, SLED have shown some promise in
addressing the two fine-tuning problems of fundamental physics: the
Gauge Hierarchy and the Cosmological Constant Problems.

In terms of the phenomenological study of brane worlds, one should
ask what are the qualitative differences between 5D and 6D models.
For example, in 5D RS models, that we have summarized in Section
\ref{RS}, the warping of 4D spacetime slices is exponentially
dependent on the proper radius of the extra dimension, whereas in
the six dimensional models of Chapter \ref{6DSUGRA} it is only power
law dependent, at least if we consider the brane solutions given in
Section \ref{singular}. Moreover, the singularities sourced by the
branes are distinct, the codimension one case being a jump and the
codimension two case being conical.

5D RS models with a large (or infinite) extra dimension and the SM
confined to the brane were developed to explain the hierarchy in the
Planck and Electroweak scales.  Although the mass gap in the KK
spectrum goes to zero as usual in the infinite volume limit, 4D
physics is retrieved thanks to the warp factor's localization of the
zero mode graviton - and exponential suppression of higher modes -
close to the brane with positive tension. Subsequently it was found
that the step singularities in the geometry could also localize bulk
fermions \cite{Chang:1999nh}, in much the same way as previously
achieved with scalar fields and kink topological defects, that we
have introduced in Subsection \ref{DWsub}. Models were then
developed in which SM fields all\footnote{Although the Higgs field
should be confined to the brane in order not to lose the gauge
hierarchy.} originate from the bulk as localized degrees of freedom
\cite{5DSM}.

A study of warped brane worlds in 6D supergravity has been given in
Section \ref{singular}, where the focus was on the background
solutions. A general solution with 4D Poincar\'e and 2D axial
symmetry was given, and it was shown in Subsection \ref{conSing}
that warping can lead to conical singularities in the internal
manifold, which can be naturally interpreted as 3-branes sources. It
is certainly interesting to go beyond these background solutions,
and study the dynamics of their fluctuations. The final objective
would be to obtain the effective theory describing 4D physics, and
an understanding of when this effective theory is valid.

Although in these constructions the SM is usually put in by hand,
envisioned on a 3-brane source, the bulk theory is potentially rich
enough to contain the SM gauge and matter fields.  As hinted above,
should the SM arise as KK zero modes of bulk fields, there are two
ways to hide the heavy modes and recover 4D physics.  They may have
a large mass gap, and thus be unattainable at the energy scales
thus-far encountered in our observed universe.  Or they may be light
but very weakly coupled to the massless modes, for example if the
massless modes are peaked near to the brane, and the massive modes
are not.  In any case, whether or not one expects the bulk to give
rise to the SM, one should study its degrees of freedom and
determine under which conditions they are observable or out of
sight.  This could also prove useful for a deeper understanding of
the self-tuning mechanism of SLED, and its quantum corrections.

A complete study of the linear perturbations is a very complicated
problem, involving questions of gauge-fixings and a highly coupled
system of dynamical equations.  Some partial results have been
obtained for the scalar perturbations in \cite{Lee:2006ge}.  In this
chapter we consider sectors within which SM gauge and charged matter
fields might be found.  By some fortune, these also happen to be two
of the least complicated ones.  Much of our discussion is general,
and could easily be applied or extended to other 6D models with
axial symmetry.  We follow the usual KK procedure, and reduce the
equations of motion to an equivalent non-relativistic quantum
mechanics problem, which we are able to solve exactly.  We consider
carefully the boundary conditions that the physical modes must
satisfy, and from these derive the wave function profiles and
complete discrete mass spectra.

Our exact solutions enable us to analyze in detail the effects of
the power-law warping and conical defects that arise in 6D brane
worlds. We find that the warping cannot give rise to zero modes
peaked at the brane, without also leading to peaked profiles for the
entire KK tower.  On the other hand, the conical defects do break
another standard lore of the classical KK theory. Remarkably, even
if the volume of the internal manifold goes to infinity, the mass
gap does not necessarily go to zero.  This decoupling between the
mass gap and volume means that in principle SM fields, in addition
to gravity, could `feel' the extent of large extra dimensions,
whilst still being accurately described by a 4D effective field
theory.

This chapter is organized as follows. We begin in Section
\ref{6dtheory} by giving our set up and additional properties of
axisymmetric solutions, that we have introduced in Section
\ref{singular}. Then, in Section \ref{gaugefields} we analyze the
gauge field fluctuations, deriving the wave functions and masses of
the KK spectrum. A similar analysis is presented in Section
\ref{fermion} for the fermions.  Section \ref{LargeMV} discusses the
physical implications of the results found, and in particular
whether they can be naturally applied to the LED scenario. Finally
we end in Section \ref{conclusions2} with some conclusions and
future directions.

In the appendices we give some results that are useful for the
detailed calculations.  Appendix \ref{deltafn} explains how the
conical defects manifest themselves in the metric ansatz.  In
Appendix \ref{boundcond} we show in detail how the boundary
conditions are applied to obtain a discrete mass spectrum, and we
give the complete fermionic mass spectrum thus derived in Appendix
\ref{fermspectrum}.

\section{The Set Up} \label{6dtheory} \setcounter{equation}{0}

We consider 6D $N=(1,0)$ gauged supergravity, and its warped
braneworld solutions, whose fluctuations we will then study. This
theory has been defined in Section \ref{6DgaugedSUGRA}, in
particular the field content is given in
(\ref{gravmult})-(\ref{chiralhyper}). As we discussed in Subsection
\ref{SSandAfree}, in general the theory has anomalies but for
certain gauge groups and hypermultiplet representations these
anomalies can be cancelled via a Green-Schwarz mechanism.  We will
consider a general matter content, with gauge group of the form
$\mathcal{G}=\tilde{\mathcal{G}}\times U(1)_R$.  For example, we
could take the anomaly free group $\mathcal{G}=E_6 \times E_7 \times
U(1)_R$, under which the fermions are charged as follows: $\psi_M
\sim (1,1)_1$, $\chi \sim (1,1)_1$, $\lambda \sim (78,1)_1 +
(1,133)_1 + (1,1)_1$, $\psi \sim (1, 912)_0$.

We remind that the bosonic action and the EOM take the form
(\ref{Baction}), and (\ref{EOM1}) respectively. In this chapter we
will consider the general class of warped solutions with 4D
Poincar\'e symmetry, and axial symmetry in the transverse dimensions
that we described in Section \ref{singular}. We can summarize those
solutions as follows:
\bea ds^2=G_{MN}dX^M dX^N&=&e^{A(\rho)}\eta_{\mu \nu}dx^{\mu}dx^{\nu}+d\rho^2+e^{B(\rho)}d\varphi^2,\nonumber\\
\mathcal{A}&=&\mathcal{A}_{\varphi}(\rho) Q d\varphi,\nonumber \\
\sigma &=&\sigma(\rho),\nonumber \\
G_{MNP} &=& 0, \label{axisymmetric} \eea
with $0\leq \rho \leq \overline{\rho}$ and $0\leq \varphi<2\pi.$ In
the following we shall also use the radial coordinate $u$ defined in
(\ref{udef}), and in this frame the metric has the form given in
(\ref{metricu}).

Given the above ansatz, the general solution to the equations of
motion (\ref{EOM1}) is given in Subsection \ref{generalGGPsub}.
Although much of our formalism for the perturbation analysis can be
applied to the general ansatz (\ref{axisymmetric}), we will focus on
a subset of this general solution, namely that which contains
singularities no worse than conical. Thus, in addition to the ansatz
(\ref{axisymmetric}), we impose the following asymptotic behaviour
for the metric:
$$e^A\stackrel{\rho\rightarrow0}{\rightarrow}constant \neq 0,\quad e^A\stackrel{\rho\rightarrow \overline{\rho}}{\rightarrow}
constant \neq 0,$$ and
\be \quad
e^B\stackrel{\rho\rightarrow0}{\rightarrow}\left(1-\delta/2\pi\right)^2\rho^2,
\quad e^B\stackrel{\rho\rightarrow \overline{\rho}}{\rightarrow}
\left(1-\overline{\delta}/2\pi\right)^2\left(\overline{\rho}-\rho\right)^2,\label{assumeeB}\ee
that is we assume conical singularities with deficit angle $\delta$
at $\rho=0$ and $\overline{\delta}$ at $\rho=\overline{\rho}$, at
which points the Ricci scalar contains delta-functions (see Appendix
\ref{deltafn}). These singularities can be interpreted as 3-brane
sources with tensions $T=2\delta/\kappa^2$ and
$\overline{T}=2\overline{\delta}/\kappa^2$ as we proved in Section
\ref{C2B}. The explicit expression for these solutions is given in
Subsection \ref{conSing} and we called them conical-GGP solutions.

We end this section by considering the various parameters in the
model, and the phenomenological constraints which can arise when we
give it a brane world interpretation.  There are three free
parameters in the 6D theory, which can be taken to be the gauge
coupling ${\tilde g}$, and two out of the following three
parameters: the 6D Planck scale, $\kappa$, the gauge coupling $g_1$
and the length-scale $r_0 = \kappa/\sqrt{2} g_1$. In the solution
there are two free parameters, $r_1$ (or $q$) and $\alpha$. However,
one combination of all these parameters is constrained by the
quantization condition (\ref{DiracQ}).

The relation between the 6D Planck scale $\kappa$ and our observed
4D Planck scale $\kappa_4$ is
\be \frac{1}{\kappa^2}V_2=\frac{1}{\kappa_4^2}, \label{k4}\ee
where the volume $V_2$ is given by
\be V_2=\int d^2y \sqrt{-G}e^{-A}=2\pi \int du e^{(3A+B)/2}. \ee
This is a particular case of (\ref{kk4}), and (\ref{Vd}). For
solution (\ref{GGPsolution}) we have
\be V_2=4\pi \alpha \left(\frac{r_0}{2}\right)^2. \label{V2}\ee
Notice that this volume does not depend on $r_1$, and so we can keep
it fixed whilst varying the warp factor, namely $e^A$ in
(\ref{GGPsolution}).  Moreover from (\ref{k4}) a phenomenological
constraint follows between the bulk couplings and the brane
tensions, which can be written:
\be \frac{g_1}{\sqrt{\alpha}}=\sqrt{\frac{\pi}{2}}\kappa_4. \ee
This implies that $g_1/\sqrt{\a}$ is very small, of the order of the
Planck length.

Now let us embed the ADD scenario into the present model, in order
to try to explain the large hierarchy between the Electroweak scale
and the Planck scale via the size of the extra dimensions.  Thus
identifying the 6D fundamental scale with the Electroweak scale
$\kappa\sim TeV^{-2},(10TeV)^{-2}$ and constraining the observed 4D
Planck scale $\kappa_4^2\sim 10^{-32}TeV^{-2}$; the above relation
translates to\footnote{We remind that the following conversion
relation holds: $(TeV)^{-1}\sim 10^{-16}mm$.}:
\be \sqrt{\alpha}r_0\sim 0.1 mm.\label{submillim}\ee
Here, the LED corresponds to tuning the bulk gauge coupling and
brane tensions.  However, we can also observe that (\ref{submillim})
fixes just one parameter among $\alpha$, $r_0$ and $r_1$ and we
still have two independent parameters even if we require large extra
dimensions. Later we will see that this novel feature proves to have
interesting consequences for the mass spectrum of fluctuations.

\section{Gauge Fields}\label{gaugefields}
\setcounter{equation}{0}

 Having established the brane world solution
and its properties, we are now ready to examine the fluctuations
about this background, which will represent the physical fields in
our model.  In this section our focus will be on the gauge field
fluctuations.

Normalizable gauge field zero modes in axially symmetric codimension
two branes are known to exist \cite{Giovannini:2001vt,
Giovannini:2002mk, RS}. However, in these known examples there is no
mass gap between the zero and non-zero modes
 which renders an effective
4D description somewhat problematic, especially in non-Abelian case
\cite{Randjbar-Daemi:2003qd}. In contrast to this
 for the axisymmetric solutions studied in this chapter the presence of a mass gap
will be automatic due to the compactness of the transverse space. In
this section we shall give the full spectrum of zero and non-zero
modes.

As we discussed in Subsection \ref{SSandAfree} in a more general
context, given the symmetries of the problem, we can expect that the
gauge fields in the low energy effective theory belong to ${\cal H}
\times U(1)_{KK}$, where ${\cal H}$ is the unbroken subgroup of
${\cal G}$ that commutes with the $U(1)_M  \subset {\cal G}$ in
which the monopole lies\footnote{The gauge group ${\cal H}$ does not
contain $U(1)_M$ because, as we discussed in Subsection
\ref{SSandAfree}, the latter is broken in the 4D effective theory.}.
The $U(1)_{KK}$ arises from the vector fluctuations of the metric,
due to the axial symmetry of the internal manifold, and is promoted
to $SU(2)_{KK}$ in the sphere limit of the background.

The non-Abelian sector of ${\cal H}$ may be rich enough to contain
the SM gauge group.  For example, consider the anomaly free model of
\cite{Randjbar-Daemi:1985wc}, with gauge group ${\cal G} = E_6
\times E_7 \times U(1)_R$, and the monopole background in $E_6$. As
we pointed out in Subsection \ref{SSandAfree}, the surviving gauge
group, $SO(10) \times E_7 \times U(1)_R \times U(1)_{KK}$, then
contains the Grand Unified Group $SO(10)$, and the model also
includes charged matter in the fundamental of $SO(10)$. Therefore
our present interest will be in the fields belonging to various
representations of ${\cal H}$. Specifically, we will consider gauge
field fluctuations orthogonal to the monopole background.  For the
case ${\cal G}=E_6 \times E_7 \times U(1)_R$, the gauge field
sectors that are covered by our analysis are given in Table
\ref{T:summary}.

\subsection{Kaluza-Klein Modes}
Using the background solution in the 6D action (\ref{Baction}), we
can identify the bilinear action for the fluctuations.  This step
requires some care, because to study the physical spectrum we must
first remove the gauge freedoms in the action due to 6D
diffeomorphisms and gauge transformations. The problem has been
studied in a general context in \cite{RS}, where the authors choose
a light-cone gauge fixing.

In the light-cone gauge, the action for the gauge field
fluctuations, orthogonal to the monopole background, at the bilinear
level reads \cite{RS}
\be S_G(V,V)\equiv -\int d^6X
\sqrt{-G}\,\frac{1}{2}e^{\phi}\left(\partial_{\mu}V_j\partial^{\mu}V^j
+e^{-A}\partial_{\rho}V_j\partial_{\rho}V_j +\nabla_{\varphi}V_j
\nabla^{\varphi}V^j\right), \label{SVV}\ee
where $V_j$ is the gauge field fluctuation in the light cone gauge
(j=1,2) and all the indices in (\ref{SVV}) are raised and lowered
with the $\rho$-dependent metric $G_{MN}$ given in
(\ref{axisymmetric}).  Indeed, here and below $G_{MN}$ represents
the background metric.  We have multiplied the formula of \cite{RS}
by an overall $e^\phi$, with $ \phi \equiv \kappa\sigma/2$ and
$\sigma$ in the background, due to the presence of the dilaton in
our theory\footnote{That the dilaton invokes only this simple change
with respect to Ref.~\cite{RS} can be seen as follows. First, notice
that since we are considering fluctuations orthogonal to $U(1)_M$
background, there are no mixings with other sectors, and the
bilinear action is simply $S_G = -1/4 \int d^6X \sqrt{- G }
e^{\phi}G^{MN}G^{PQ}\left(F_{MP}F_{NQ}\right)^{(2)}$. We emphasise
that $G_{MN}$ and $\phi$ now signify the background fields.  Also,
$()^{(2)}$ indicates the bilinear part in the fluctuations.  Next,
make the change of coordinates, $d\rho = e^{-\phi/2} d{\tilde\rho}$,
and rewrite the background metric in (\ref{axisymmetric}) as $ds^2 =
e^{-\phi} \left( e^{{\tilde A}} \eta_{\mu\nu} dx^{\mu} dx^{\nu} +
d{\tilde\rho}^2 + e^{{\tilde B}} d\varphi^2 \right)$, with ${\tilde
A} \equiv A+\phi $ and ${\tilde B} \equiv B+\phi$.  In this way, the
bilinear action, $S_G$, reduces to exactly the same form as that of
Ref.~\cite{RS}, and we can proceed as they do to transform into
light-cone coordinates, fix the light-cone gauge, and eliminate
redundant degrees of freedom using their equations of motion.}.
Notice that since we are looking at the sector orthogonal to the
monopole background, the Chern-Simons term does not contribute, and
the action takes a simple form.

In general, the covariant derivative $\nabla_{\varphi}V_j$ includes
the gauge field background
\be
\nabla_{\varphi}V_j=\partial_{\varphi}V_j+ie_V\mathcal{A}_{\varphi}V_j,\ee
where again the charge $e_V$ can be computed using group theory once
the gauge group $\tilde{\cal G}$ is chosen. The value $e_V=0$
corresponds to the gauge fields in the 4D low energy effective
theory.  However, since we can do so without much expense, we keep a
generic value of $e_V$.  Those fluctuations with $e_V\neq 0$
corresponds to vector fields in a non-trivial representation of the
4D effective theory gauge group. The Dirac quantization condition
(\ref{DiracQ}) then gives $ e_V \, 4\alpha g/(\kappa q)=N_V$, where
$N_V$ is an integer.

Next we perform a KK expansion of the 6D fields.  Since our internal
space is topologically $S^2$, we require gauge fields to be periodic
functions of $\varphi$:
\be V_j(X)=\sum_m V_{jm}(x)f_m(\rho)e^{im\varphi},\label{FourierV}
\ee
where $m$ is an integer.

If we put (\ref{FourierV}) in (\ref{SVV}) we obtain kinetic terms
 for the 4D effective fields proportional to
\be  \int d^4 x \sum_m \eta^{\mu
\nu}\partial_{\mu}V_{jm}^{\dagger}\partial_{\nu}V_{jm}\int d\rho
e^{\phi+B/2}|f_m|^2. \ee
Therefore physical fluctuations, having a finite kinetic energy,
must satisfy the following normalizability condition (NC):
\be \int du |\psi|^2 <\infty, \label{NCV}\ee
where
\be \psi = e^{(2\phi+A+B)/4}f_m. \ee
The quantity $\left|\psi\right|^2$ represents the probability
density of finding a gauge field in $[u,u+du]$.

In fact, this is not the only condition that the physical fields
must satisfy. If we want to derive the EOM from (\ref{SVV}) through
an action principle we have to impose\footnote{Actually we impose
that for every pair of fields $V_j$ and $V'_j$ the condition $\int
d^6X\partial_{M}\left(\sqrt{-G}e^{\phi-A}V_j{\cal
D}^{M}V'_j\right)=0$ is satisfied but in (\ref{HCV}) the prime is
understood.} the following boundary condition
\be \int d^6X\partial_{M}\left(\sqrt{-G}e^{\phi-A}V_j{\cal
D}^{M}V_j\right)=0 \, ,\label{HCV}\ee
where ${\cal D}_M$ is the gauge covariant derivative.  Equation
(\ref{HCV}) represents conservation of current
$J_M=e^{\phi-A}V_j{\cal D}_M V_j$ and it is the generalization of
Eq. (\ref{boundarydomain}) that we studied in the simple domain wall
model. Moreover, since the fields are periodic functions of
$\varphi$, (\ref{HCV}) becomes
\be \left[\sqrt{-G}\,e^{\phi-A}\,V_j \partial_{\rho} V_j
\right]_0^{\overline{\rho}}=0.\label{HC2}\ee

The EOM can then be derived as:
\be \sqrt{-G}e^{\phi-2A}\eta^{\mu
\nu}\partial_{\mu}\partial_{\nu}V_j= -\partial_{\rho}\left(\sqrt{-G}
e^{\phi-A}\partial_{\rho}V_j\right)-\sqrt{-G}e^{\phi-A-B}\nabla_{\varphi}^2V_j.\label
{EOMV}\ee
By inserting (\ref{FourierV}) in (\ref{EOMV}) we obtain
\be
-\frac{e^{-\phi+2A}}{\sqrt{-G}}\partial_{\rho}\left(\sqrt{-G}e^{\phi-A}\partial_
{\rho}f_m\right)
+e^{A-B}\left(m+e_V\mathcal{A}_{\varphi}\right)^2f_m=M_{V,m}^2f_m,\label{EOMm}
\ee
where $M_{V,m}^2$ are the eigenvalues of $\eta^{\mu
\nu}\partial_{\mu}\partial_{\nu}$.

At this stage, we can already identify the massless fluctuation that
is expected from symmetry arguments.  For $e_V=0$, when $m=0$, a
constant $f_0$ is a solution of (\ref{EOMm}) with $M^2_{V,0}=0$.
This solution corresponds to 4D effective theory gauge fields.  It
has a finite kinetic energy, and trivially satisfies (\ref{HC2}).
The fact that such gauge fields have  a constant transverse profile
guarantees charge universality of fermions in the 4D effective
theory (see below and Subsection \ref{Gauge Field Localization}).

To find the massive mode solutions, we can express (\ref{EOMm}) in
terms of $u$ and $\psi$ and obtain a Schroedinger equation:
\be \left(-\partial^2_u+V\right)\psi=M^2_V\psi,
\label{SchroedingerV}\ee
where the ``potential'' is
\be
V(u)=e^{A-B}\left(m+e_V\mathcal{A}_{\varphi}\right)^2+e^{-(2\phi+A+B)/4}
\partial^2_u
e^{(2\phi+A+B)/4}. \ee
We want to find the complete set of solutions to
(\ref{SchroedingerV}) satisfying the NC (\ref{NCV}) and the boundary
conditions (\ref{HC2}), which can be written in terms of $u$ and
$\psi$ as follows
\be \left(\lim_{u\rightarrow \overline{u}}-\lim_{u\rightarrow
0}\right)
\left\{\psi^*\left[-\partial_u+\frac{1}{4}\left(2\partial_u\phi +
\partial_uA+\partial_u
B\right)\right]\psi\right\}=0. \label{HCV1}\ee
In order for (\ref{HCV1}) to be satisfied, both the limits
$u\rightarrow 0$ and $u\rightarrow \overline{u}$ must be finite.
Condition (\ref{HCV1}) ensures that the Hamiltonian in the
Schroedinger equation (\ref{SchroedingerV}) is hermitian, and so has
real eigenvalues and an orthonormal set of eigenfunctions.
Therefore, we shall call it the hermiticity condition (HC).

So far our analysis has been valid for all axially symmetric
solutions of the form (\ref{axisymmetric}).  We will now use these
results to determine the fluctuation spectrum about the conical-GGP
solution (\ref{GGPsolution}). We observe that $V(u)$ then contains a
delta-function contribution, arising from the second-order
derivative of the conical metric function $\partial_u^2B$ (see
Appendix \ref{deltafn} and Eq. (\ref{B2u})). However, we can drop it
because $\partial_u^2B$ also contains stronger singularities at
$u=0$ and $u=\overline{u}$: respectively $1/u^2$ and
$1/(\overline{u}-u)^2$. These singularities are also a consequence
of the behavior of $e^B$ given in (\ref{assumeeB}) and they imply
that the behaviour of the wave functions close to $u=0$ and
$u=\overline{u}$ cannot depend on the mass. In particular, this
immediately implies that if the wave functions of zero modes are
peaked near to one of the branes, then the same will be true also
for the infinite tower of non-zero modes.  In other words, we cannot
hope to dynamically generate a brane world scenario, in which zero
modes are peaked on the brane, and massive modes are not, leading to
weak coupling between the two sectors\footnote{In fact, a similar
singular behaviour for the potential in general arises for the
general axisymmetric solutions given in Subsection
\ref{generalGGPsub} and studied in \cite{Burgess:2004dh}, where the
hypothesis (\ref{assumeeB}) is relaxed.}. If we are to interpret the
zero mode gauge fields as those of the SM, therefore, for the
massive modes to have escaped detection they must have a large mass
gap.

Meanwhile, we note that in contrast to the non-relativistic quantum
mechanics problem, here we cannot deduce qualitative results about
the mass spectrum from the shape of the potential.  This is because
the boundary conditions to be applied in the context of dimensional
reduction are in general different to those in problems of quantum
mechanics.  In particular, the HC (\ref{HCV1}) is a non-linear
condition, contrary to the less general linear boundary conditions
usually encountered in quantum mechanics to ensure hermiticity of
the Hamiltonian.  We will be able to impose the more general case
thanks to the universal asymptotic behaviour of the KK tower.

Returning then to our explicit calculation of the KK spectrum, we
can write $V(u)$ as
\be V(u)=V_0+v\cot^2\left(\frac{u}{r_0}\right)
+\overline{v}\tan^2\left(\frac{u}{r_0}\right),\label{SchroedingerVGGP}\ee
and
\be r_0^2V_0 \equiv 2m\omega (m-N_V)\ob -\frac{3}{2},\quad
r_0^2v\equiv m^2 \omega^2-\frac{1}{4},\quad r_0^2 \overline{v}\equiv
(m-N_V)^2 \ob^2-\frac{1}{4}.\ee
Moreover in this case the expression (\ref{HCV1}) for the HC becomes
\be \lim_{u\rightarrow
\overline{u}}\psi^*\left(-\partial_u+\frac{1}{2}\frac{1}{u-\overline{u}}
\right)\psi -\lim_{u\rightarrow
0}\psi^*\left(-\partial_u+\frac{1}{2u}\right)\psi=0. \label{HCV2}\ee
If we introduce $z$ and $y$ in the following way \cite{manning}
\be z=\cos^2\left(\frac{u}{r_0}\right),\qquad
\psi=z^{\gamma}\left(1-z\right)^{\beta}y(z),\ee
Eq. (\ref{SchroedingerV}) becomes
\be z(1-z)\partial_z^2y+\left[c-(a+b+1)z\right]\partial_zy-ab
y=0\label{hyper},\ee
where
\bea \gamma &\equiv& \frac{1}{4}\left[1+2(m-N_V)\ob\right], \,\,
\beta\equiv \frac{1}{4}\left(1+2m\omega\right),
\,\, c\equiv 1+(m-N_V)\ob, \nonumber \\
a&\equiv&\frac{1}{2}+\frac{m}{2}\omega+\frac{1}{2}(m-N_V)\ob
+\frac{1}{2}\sqrt{r_0^2M^2_{V,m}+1+\left[m\omega-(m-N_V)\ob\right]^2},\nonumber\\
b &\equiv&\frac{1}{2}+\frac{m}{2}\omega+\frac{1}{2}(m-N_V)\ob
-\frac{1}{2}\sqrt{r_0^2M^2_{V,m}+1+\left[m\omega-(m-N_V)\ob\right]^2},
\label{gbetaabcV}\eea
and
\be \omega\equiv(1-\delta/2\pi)^{-1}, \qquad
\ob\equiv(1-\overline{\delta}/2\pi)^{-1}.\ee
Eq. (\ref{hyper}) is the hypergeometric equation and its solutions
are known. For $c\neq 1$ the general solution is a linear
combination of the following functions:
\be y_1(z)\equiv F(a,b,c,z), \quad y_2(z)\equiv z^{1-c}
F(a+1-c,b+1-c, 2-c, z), \label{y12}\ee
where $F$ is Gauss's hypergeometric function.  So for $c\neq 1$ the
general integral of the Schroedinger equation is
\be\psi=K_1\psi_1 + K_2 \psi_2,\label{cnot1}\ee
where
\be \psi_i\equiv z^{\gamma}(1-z)^{\beta}y_i. \label{psii} \ee
and $K_{1,2}$ are integration constants. For $c=1$ we have
$\psi_1=\psi_2$ but we can construct a linearly independent solution
using the Wronskian method and the general solution reads
\be\psi=K_1\psi_1 + K_2
\psi_1\int^u\frac{du'}{\psi_1^2(u')}.\label{wronskian}\ee

Now we must impose the NC (\ref{NCV}) and HC (\ref{HCV2}), to select
the physical modes.  In Appendix \ref{boundcond} we give explicit
calculations; the final result is that the NC and HC give the
following discrete spectrum. The wave functions are
\bea \psi&\propto&  z^{\gamma}(1-z)^{\beta}F(a,b,c,z), \quad
for\quad m\geq
N_V,\label{psi1}\\
\psi&\propto&  z^{\gamma +1-c}(1-z)^{\beta} F(a+1-c,b+1-c, 2-c, z),
\quad for\quad m<N_V.\label{psi2} \eea
and the squared masses are as follows:
\begin{itemize}
\item For  $N_V\leq m < 0$
\be
M_{V\,n,m}^2=\frac{4}{r_0^2}\left\{n(n+1)+\left(\frac{1}{2}+n\right)\left[
-m\omega+(m-N_V)\ob\right]\right\}>0.\label{MV1}\ee
\item For  $m\geq N_V$ and $m\geq 0$
\be
M_{V\,n,m}^2=\frac{4}{r_0^2}\left\{n(n+1)+\left(\frac{1}{2}+n\right)\left[
m\omega+(m-N_V)\ob\right]+m\omega(m-N_V)\ob\right\}\geq
0.\label{MV2}\ee
\item For $m<N_V$ and $m<0$
\be
M_{V\,n,m}^2=\frac{4}{r_0^2}\left\{n(n+1)+\left(\frac{1}{2}+n\right)\left[
-m\omega+(N_V-m)\ob\right]-m\omega(N_V-m)\ob\right\}>
0.\label{MV3}\ee
\item For  $0\leq m <N_V$
\be
M_{V\,n,m}^2=\frac{4}{r_0^2}\left\{n(n+1)+\left(\frac{1}{2}+n\right)\left[
m\omega+(N_V-m)\ob\right]\right\}>0.\label{MV4}\ee
\end{itemize}
The masses given in (\ref{MV1}) and (\ref{MV2}) correspond to the
wave function (\ref{psi1}) whereas the masses given in (\ref{MV3})
and (\ref{MV4}) correspond to the wave function (\ref{psi2}). We
observe that there are no tachyons and that the only zero mode is
for $n=0$, $m=0$ and $N_V=0$ ($e_V=0$), corresponding to gauge
fields in the 4D low energy effective theory.

As a check, we can consider the $S^2$ limit ($\omega,\ob \rightarrow
1$), whose mass spectrum is well-known.  Our spectrum
(\ref{MV1})-(\ref{MV4}) reduces to
\be a^2M^2_V=l(l+1)-\left(\frac{N_V}{2}\right)^2,\quad
multiplicity=2l+1 \, ,\ee
where $a=r_0/2$ is the radius of $S^2$ and\footnote{The number $l$
is defined in different ways in equations (\ref{MV1})-(\ref{MV4}).
For instance we have $l\equiv n+|N_V/2|$ for (\ref{MV1}).}
$l=|\frac{N_V}{2}|+k$, $k=0,1,2,3,...$. This is exactly the result
that one finds by using the spherical harmonic expansion \cite{RSS}
from the beginning.

At this stage we can point towards a novel property of the final
mass spectrum.  Observe that in the large $\a$ (small $\ob$) limit
the volume $V_2$ given in Eq. (\ref{V2}) becomes large but the mass
gap between two consecutive KK states does not reduce to zero as in
standard KK theories\footnote{This is also true for the proper
volume of the 2D internal manifold.}. This a consequence of the
shape of our background manifold and in particular of the conical
defects.  Notice that the large $\a$ limit corresponds to a negative
tension brane at $u=\bar{u}$, but not necessarily at $u=0$.

In Section \ref{fermion} we will show that the same effect appears
also in the fermionic sector, and we will turn to a discussion of
its implications in Section \ref{LargeMV}.

\subsection{4D Effective Gauge Coupling}

Let us end the discussion on gauge fields by briefly presenting the
4D effective gauge coupling.  This can be obtained by dimensionally
reducing the 6D gauge kinetic term.  We consider the zero mode
fluctuations in ${\cal H}$, about the background
(\ref{axisymmetric}), so that the 4D effective gauge kinetic term
is: \bea &&\int d^6X \sqrt{-G}\left\{-\frac{1}{4{g}^2} e^{\kappa
\sigma/2}
TrF_{MN}F^{MN}\right\} \rightarrow \nonumber \\
&& \phantom{0000000000000000} \int d^4x\left\{
-\frac{1}{4{g}^2}\left[\int dud\varphi e^{(3A+B)/2} f^{2}_{0}
\right]TrF_{\mu\nu}F^{\mu\nu} \right\} \,.  \eea
Recalling that $f_0 = const$ and normalizing it to one, we can read:
\be \frac{1}{g_{eff}^2} = \frac{1}{g^2} V_2 \, . \ee

\section{Fermions}\label{fermion}
\setcounter{equation}{0}

We will now consider fermionic perturbations, and in particular our
interest will be in the sector charged under the 4D effective gauge
group, ${\cal H}$, discussed above.  These fields arise from the
hyperinos and the ${\cal H}$ gauginos, for which we also restrict
ourselves to those orthogonal to the $U(1)_R$.  Thus we are
considering matter charged under the non-Abelian gauge symmetries of
the 4D effective theory. For instance, for the anomaly free model
$E_6 \times E_7 \times U(1)_R$, with the monopole embedded in the
$E_6$, the gauginos in the {\bf $78$} of $E_6$ contain a {\bf $16$}
+ {\bf ${\overline{16}}$} fundamental representation of the grand
unified gauge group $SO(10)$, and our analysis will be applicable to
them.  In Tabel \ref{T:summary}, we give the complete list of
fermion fields that are included in our study, for the said example.

We proceed in much the same way as for the gauge field sector of the
previous section, transforming the dynamical equations and necessary
boundary conditions into a Schroedinger-like problem, to obtain the
physical modes and discrete mass spectrum.

The bilinear action for the fluctuations of interest takes a
particularly simple form, comprising as it does of the standard
Dirac action:
\be S_F=\int d^6X \sqrt{-G}\,\,\overline{\lambda}\Gamma^M
\nabla_M\lambda, \label{1/2action}\ee
where\footnote{Our conventions for $\Gamma^A$ and $
\omega^{[A,B]}_M$ are given in Appendix \ref{GR}.}
\be
\nabla_M\lambda=\left(\partial_M+\frac{1}{8}\omega^{[A,B]}_M[\Gamma_A,\Gamma_B]
+ie\mathcal{A}_M\right)\lambda. \label{Dlambda} \ee
Here $e$ is the charge of $\lambda$ under $U(1)_M$, and $G_{MN}$,
$\omega^{[A,B]}_M$ and $\mathcal{A}_M$ are the background metric,
spin connection and gauge field corresponding to an axisymmetric
solution (\ref{axisymmetric}).  Analogously to the gauge field
analysis, in order to derive the Dirac equation
\be \Gamma^M \nabla_M \lambda = 0 \label{DiracEq}\ee
from (\ref{1/2action}) by using an action principle, we require
conservation of fermionic current:
\be \int d^6X
\partial_M\left(\sqrt{-G}\,\,\overline{\lambda}\Gamma^M\lambda\right)=0.
\label{hermiticity}\ee
Eq. (\ref{hermiticity}) implies that the Dirac operator $\Gamma^M
\nabla_M$ is hermitian, and we shall again refer to it as the HC.
This constraint is analogous to Eq. (\ref{HCDMf}), which concerns
fermion fluctuation around the simple kink background. Our aim is to
find the complete fermionic spectrum, that is a complete set of
normalizable solutions of (\ref{DiracEq}) satisfying
(\ref{hermiticity}).

Some care is now needed when discussing the background felt by the
fermionic sector in (\ref{Dlambda}).  As already mentioned, in order
to have a correctly defined gauge connection, it is necessary to use
two patches related by a single-valued gauge transformation.  The
same is true for the spin connection, which must be defined in such
a way as to imply the conical defects in the geometry.
Henceforth we focus on the patch including the $\rho=0$ brane,
chosen to be $0 \leq \rho < {\bar \rho}$.  For this patch a good
choice for the vielbein is
\be e^{a}_{\mu}=e^{A/2}\delta^{a}_{\mu},
\quad \{e^{\a}_m\}=\left(\ba {cc} \cos\varphi & -e^{B/2}\sin\varphi
\\ \sin\varphi & e^{B/2}\cos\varphi \ea \right), \label{vielbein}\ee
where, like in Chapter \ref{EMS}, $a$ is a 4D flat index, $\a=5,6$ a
2D flat index and $m=\rho, \varphi$. The corresponding spin
connection is
\bea
\omega_{\mu}^{[a,5]}&=&\frac{1}{2}A'e^{A/2}\delta^{a}_{\mu}\cos\varphi,
\quad
\omega_{\mu}^{[a,6]}=\frac{1}{2}A'e^{A/2}\delta^{a}_{\mu}\sin\varphi,\nonumber\\
\omega_{\rho}^{[5,6]}&=&0,\qquad
\Omega\equiv\omega_{\varphi}^{[5,6]}=\left(1-\frac{1}{2}B'e^{B/2}\right),
\label{spinconnection} \eea
where $'\equiv \partial_{\rho}$.  It can be checked that this gauge
choice correctly reproduces Stokes' theorem for a small domain
including the conical defect\footnote{See Appendix \ref{deltafn} for
some steps in this calculation.}.

We are now ready to study the Dirac equation (\ref{DiracEq}) for 6D
fluctuations, and write it in terms of 4D effective fields.
 Since $\lambda$ is a 6D Weyl spinor we can represent it by
\be \lambda=\left(\ba {c} \lambda_4 \\ 0 \ea\right), \ee
where $\lambda_4$ is a 4D Dirac spinor:
$\lambda_4=\lambda_R+\lambda_L$, $\gamma^5\lambda_R=\lambda_R,$
$\gamma^5\lambda_L=-\lambda_L$. By using the ansatz
(\ref{axisymmetric}), the vielbein (\ref{vielbein}), the spin
connection (\ref{spinconnection}) and our conventions for $\Gamma^A$
in Appendix \ref{GR}, the Dirac equation (\ref{DiracEq}) becomes
\bea e^{-A/2}\gamma^{\mu}\partial_{\mu}\lambda_L&=&
e^{i\varphi}\left[-\partial_{\rho}-ie^{-B/2}\left(\partial_{\varphi}+ie\mathcal{
A}_{\varphi}\right)
-A'+\frac{1}{2}\Omega e^{-B/2}\right]\lambda_R, \label{EqR}\\
 e^{-A/2}\gamma^{\mu}\partial_{\mu}\lambda_R&=&
e^{-i\varphi}\left[\partial_{\rho}-ie^{-B/2}\left(\partial_{\varphi}+ie\mathcal{
A}_{\varphi}\right) +A'-\frac{1}{2}\Omega
e^{-B/2}\right]\lambda_L.\label{EqL}\eea
Performing the Fourier mode decomposition: \be
\lambda_4(X)=\lambda_R(X)+\lambda_L(X)=\sum_m\left(\lambda_{R,m}(x)f_{R,m}(\rho)
+\lambda_{L,m}(x)f_{L,m}(\rho)\right)e^{im\varphi},\label{Fourier}\ee
where $m$ is an integer, and inserting into (\ref{EqR}) and
(\ref{EqL}) we find:
\bea e^{-A/2}\gamma^{\mu}\partial_{\mu}\lambda_{L,m+1}f_{L,m+1}=
\left[-\partial_{\rho}+e^{-B/2}\left(m+\frac{1}{2}\Omega
+e\mathcal{A}_{\varphi}\right)
-A'\right]\lambda_{R,m}f_{R,m}, \label{EqRm}\\
 e^{-A/2}\gamma^{\mu}\partial_{\mu}\lambda_{R,m-1}f_{R,m-1}=
\left[\partial_{\rho}+e^{-B/2}\left(m-\frac{1}{2}\Omega
+e\mathcal{A}_{\varphi}\right)
+A'\right]\lambda_{L,m}f_{L,m}.\label{EqLm}\eea

For the boundary conditions, analogously to the gauge fields, the NC
can be found to be:
\be \int du \left|\psi\right|^2 <\infty \label{NC}\ee
where
\be \psi\equiv e^{A+B/4}f_{Rm} \label{psi}\ee
and a similar condition for left-handed spinors.
Meanwhile, the HC (\ref{hermiticity}) can be written:
 \be
\left[\sqrt{-G}\,\,f_{L,m+1}f_{R,m}^{*}
\right]_0^{\overline{\rho}}=0.\label{HC3}\ee

Having set up the dynamical equations and the relevant boundary
conditions, we shall now use this information to study the complete
fermionic spectrum in Subsections \ref{zero} and \ref{complete}. In
particular, we will focus on the questions of wave function
localization, and the mass gap problem, crucial to the development
of a phenomenological brane world model.

\subsection{Zero Modes}\label{zero}

We begin by finding the zero mode solutions, for which the problem
simplifies considerably.  Indeed, for the zero modes
$\gamma^{\mu}\partial_{\mu}=0$, and the equations for right- and
left-handed modes (\ref{EqRm}) and (\ref{EqLm}) decouple:
\bea
\left[\partial_{\rho}-e^{-B/2}(m+e\mathcal{A}_{\varphi})+A'-\frac{1}{2}\Omega
e^{-B/2}\right]f_{R,m}&=&0,\label{EqRzero}\\
\left[\partial_{\rho}+e^{-B/2}(m+e\mathcal{A}_{\varphi})+A'-\frac{1}{2}\Omega
e^{-B/2}\right]f_{L,m}&=&0.\label{EqLzero}\eea
By using the expression for $\Omega$ in equation
(\ref{spinconnection}), the solution of (\ref{EqRzero}) is
\be f_{R,m}(\rho)\propto\exp\left[-A-\frac{1}{4}B +\int^{\rho}d\rho'
e^{-B/2}\left(m+\frac{1}{2}+e\mathcal{A}_{\varphi}\right)\right],
\label{zerosolution} \ee
whereas the solution of (\ref{EqLzero}) can be obtained by replacing
$m,e\rightarrow -m,-e$ in (\ref{zerosolution}). The solution
(\ref{zerosolution}) for $e=0$ was found in \cite{Schwindt:2003er}.
Here we give the expression for every $e$ because we want to include
charged fermions. We note that the zero mode solution
(\ref{zerosolution}) automatically satisfies the HC given in
(\ref{HC3}). From (\ref{zerosolution}), (\ref{psi}) and (\ref{udef})
we obtain
\be \psi \propto\exp\left[\int^{u}du'
e^{(A-B)/2}\left(m+\frac{1}{2}+e\mathcal{A}_{\varphi}\right)\right].
\label{psizerosolution} \ee
For the conical-GGP background, (\ref{GGPsolution}), the explicit
expression for $\psi$ is
\be \psi\propto
\sin\left(\frac{u}{r_0}\right)^{\omega(1/2+m)}\cos\left(\frac{u}{r_0}\right)^{
\ob(N-1/2-m)}, \label{0GGP}\ee
where we used (\ref{DiracQ}).
 The NC is satisfied when
\be \frac{\delta}{4\pi}-1<m<N-\frac{\overline{\delta}}{4\pi}.
\label{NCGGP}\ee
From here we retrieve the  result that for the sphere, which has
$\delta=\overline{\delta}=0$, there exist normalizable zero modes
only for $e\neq 0$ ($N\neq0$), that is for a non-vanishing monopole
background \cite{RSS}.  Moreover, as found in
\cite{Schwindt:2003er}, we see that the conical defects also make
massless modes possible, provided that there is at least one
negative deficit angle, even if $N=0$.  However, (\ref{NCGGP})
implies that for positive tension branes, $\delta, {\overline
\delta} >0$, the adjoint of ${\cal H}$, which has $e=0$, is
projected out.  If ${\cal H}$ contains the SM gauge group, this is
appealing since the fermions of the SM are not in adjoint
representations.  In any case, the number of families depends on
$\delta$, ${\overline \delta}$ and $N$.

Let us now consider the wave function profiles, (\ref{0GGP}).
Observe that $\psi$ is peaked on the $u=0$ brane (that is, $\psi
\rightarrow \infty$ as $u \rightarrow 0$, and $\psi \rightarrow 0$
as $u \rightarrow \bar{u}$) when
\be m<-1/2, \qquad and \qquad m<-1/2+N. \label{LCGGP}\ee
By comparing (\ref{NCGGP}) and (\ref{LCGGP}) we understand that we
have normalizable and peaked $\psi$ only for $\delta<0$ (that is for
negative tension brane). If $N-\overline{\delta}/4\pi>0$ we can have
normalizable zero modes for $\delta>0$ (positive tension brane) but
the corresponding $\psi$ are not peaked on the $u=0$ brane (indeed,
$\psi \rightarrow 0$ as $u \rightarrow 0$). On the other hand from
(\ref{0GGP}) we see that $\psi$ is peaked on the $u=\overline{u}$
brane when
\be m>-1/2, \qquad and \qquad m>-1/2+N. \label{LCGGPbar}\ee
By comparing (\ref{NCGGP}) and (\ref{LCGGPbar}) we understand that
we have normalizable and peaked $\psi$ only for
$\overline{\delta}<0$ (that is for negative tension brane).

We can also analyze the chirality structure. Since the left handed
wave functions can be obtained from (\ref{0GGP}) by replacing
$m,N\rightarrow -m,-N$, in order for them to be peaked on the $u=0$
brane we need
\be m>1/2, \qquad and \qquad m>1/2+N, \ee
whereas in order for the left handed wave functions to be peaked on
the $u=\overline{u}$ brane we need
\be m<1/2, \qquad and \qquad m<1/2+N. \ee
So, if we were to require that $\psi$ be peaked on a brane, we
always have a chiral massless spectrum because the chirality index
counts the difference of modes in $f_{R,m}$ and $f_{L,m}$ with given
$m$, $N_R(m)-N_L(m)$ \cite{Schwindt:2003er}. We should point out,
however, that in fact peaked zero modes, $\psi$, may not be
necessary in order to have an acceptable phenomenology. The answer
to this question can be found only after constructing the complete
spectrum, and studying the couplings between different 4D effective
fields.

\subsection{Massive Modes}\label{complete}

We now move on to a study of the complete KK tower for the fermions.
We begin by establishing the corresponding Schroedinger problem. The
two coupled first order ODEs, equations (\ref{EqRm}) and
(\ref{EqLm}), can be equivalently expressed as a single second order
ODE and a constraint equation, as follows\footnote{Eq.(\ref{2nd})
can also be obtained by squaring the 6D Dirac operator, and using
that $[\nabla_M,\nabla_N] = \frac{1}{4}
R_{MN}^{\,\,\,\,\,\,\,\,\,\,AB}\Gamma_{AB} + ie F_{MN}$.}
\bea
&&e^A\left(-\partial_{\rho}^2+h\partial_{\rho}+g_m\right)f_{R,m}=M_{F,m}^2f_{R,m
},\label{2nd}
\\
&&M_{F,m}f_{L,m+1}=e^{A/2}\left[-\partial_{\rho}-A'+\left(m+\frac{1}{2}\Omega
+e\mathcal{A}_{\varphi}\right)e^{-B/2}\right]f_{R,m}, \label{fLR}
\eea
where $M_{F,m}^2$ are the eigenvalues of
$\left(\gamma^{\mu}\partial_{\mu}\right)^2$ and
\bea h &\equiv&-\frac{5}{2}A'+(\Omega-1)e^{-B/2},\\
g_m &\equiv&\left[\frac{1}{2}\Omega'-\frac{1}{4}\Omega
B'-\frac{m}{2}B'
+\frac{5}{4}A'\Omega+\left(\frac{m}{2}-1\right)A'-\frac{e}{2}B'\mathcal{A}_{
\varphi}+ e\mathcal{A}_{\varphi}'+\frac{e}{2} A'
\mathcal{A}_{\varphi}\right]e^{-B/2}\nonumber\\
&&+\left[m(m+1)+\frac{1}{2}\Omega-\frac{\Omega^2}{4}
+(2m+1)e\mathcal{A}_{\varphi}+e^2\mathcal{A}_{\varphi}^2\right]e^{-B}-A''-\frac{
3}{2}(A')^2. \label{g}\eea
Once $f_{R,m}$ is known we can compute $f_{L,m+1}$ by using
(\ref{fLR}), so we can focus on $f_R$ and study the second order ODE
(\ref{2nd}). If we express this equation in terms of $\psi$ and $u$
we obtain the Schroedinger equation
\be
\left(-\partial^2_u+V\right)\psi=M^2_{F,m}\psi,\label{Schroedinger1}\ee
where the ``potential'' $V$ is given by
\bea
V(u)&=&e\partial_u\mathcal{A}_{\varphi}e^{(A-B)/2}+\left(\frac{1}{2}+m
+e\mathcal{A}_{\varphi}\right)\partial_u e^{(A-B)/2}\nonumber \\
&&+\left[\frac{1}{4}+m+e\mathcal{A}_{\varphi}+\left(m+e\mathcal{A}_{\varphi}
\right)^2\right]e^{A-B}.\eea
We observe that transformation (\ref{psi}) exactly removes the
delta-functions which appear in (\ref{g}) through $\Omega'$ (see
Appendix \ref{deltafn}). However, just as for the gauge fields, a
singular behaviour is observed in the potential, so that the
asymptotic behaviour of the wave functions does not depend on their
mass.

Our problem is now reduced to solving equation (\ref{Schroedinger1})
with the conditions NC (\ref{NC}) and HC (\ref{HC3}). By using
(\ref{fLR}) for $M_F\neq 0$
 and
definitions (\ref{udef}) and (\ref{psi}) we can rewrite (\ref{HC3})
as follows
\be \left(\lim_{u\rightarrow \overline{u}}-\lim_{u\rightarrow
0}\right)
\psi^*\left[-\partial_u+\left(m+\frac{1}{2}+e\mathcal{A}_{\varphi}\right)e^{
(A-B)/2}\right]\psi=0.\label{HC5}\ee
 We can now proceed in exactly the same way as for the gauge field
sector.  For the conical-GGP solution (\ref{GGPsolution}) the
explicit expression for $V$ has the form (\ref{SchroedingerVGGP}),
but now with
\bea r_0^2V_0&\equiv& \left(\frac{1}{2}+m\right)\left[\ob
-\omega+2\omega \ob
\left(\frac{1}{2}+m-N\right)\right]-\ob N,\\
r_0^2v&\equiv&
\left(\frac{1}{2}+m\right)\left[-\omega+\omega^2\left(\frac{1}{2}+m\right)\right],\\
r_0^2 \overline{v}&\equiv&\left(\frac{1}{2}+m\right)\left[\ob+\ob^2
\left(\frac{1}{2}+m-2N \right)\right] +\ob N\left(\ob N-1\right).
\eea
Moreover, in this case the explicit expression for the HC is
\be \lim_{u\rightarrow
\overline{u}}\psi^*\left(-\partial_u+\ob\frac{m+1/2-N}{\overline{u}-u}
\right)\psi -\lim_{u\rightarrow
0}\psi^*\left(-\partial_u+\omega\frac{m+1/2}{u}\right)\psi=0.
\label{HC4}\ee
As in the gauge fields sector we introduce $z$ and $y$ in the
following way
\be z=\cos^2\left(\frac{u}{r_0}\right),\qquad
\psi=z^{\gamma}\left(1-z\right)^{\beta}y(z), \label{psiz} \ee
so that equation (\ref{SchroedingerVGGP}) becomes a hypergeometric
equation (\ref{hyper}), with parameters:
\bea \gamma &\equiv&
\frac{1}{2}\left[1+\ob\left(\frac{1}{2}+m-N\right)\right], \,\,
\beta\equiv
\frac{1}{2}\left[1-\omega\left(\frac{1}{2}+m\right)\right],
\,\, c\equiv \frac{3}{2}+\ob \left(\frac{1}{2}+m-N\right), \nonumber \\
a&\equiv&1+\frac{\ob}{2}
\left(\frac{1}{2}+m-N\right)-\frac{\omega}{2}\left(\frac{1}{2}+m\right)+\frac{1}{2}
\sqrt{\Delta},\nonumber\\
b &\equiv&1+\frac{\ob}{2}
\left(\frac{1}{2}+m-N\right)-\frac{\omega}{2}\left(\frac{1}{2}+m\right)-\frac{1}{2}
\sqrt{\Delta},\nonumber\\
\Delta&\equiv& r_0^2 M^2_{F,m}+\left(\ob N\right)^2+
\left(\frac{1}{2}+m\right)
\left[\ob(\ob-2\omega)\left(\frac{1}{2}+m-N\right)-\ob^2N+\omega^2\left(\frac{1}{2}
+m\right)\right].\nonumber\eea
We can construct two independent solutions $\psi_1$ and $\psi_2$ of
the Schroedinger equation (\ref{SchroedingerVGGP}) as in Section
\ref{gaugefields}, and impose the NC (\ref{NC}) and the HC
(\ref{HC4}) to obtain the physical modes.  The resulting wave
functions are:
\be\psi=K_1\psi_1 + K_2 \psi_2,\ee where the integration constants,
$K_{1,2}$, are fixed in Appendix \ref{fermspectrum}.
We plot a few of the wave function profiles in Figure
\ref{fig:wavefns}.

The complete discrete mass spectrum is also given in Appendix
\ref{fermspectrum}.  There it can be seen that the same finiteness
of the mass gap in the large $\alpha$ (hence large volume) limit,
found in the gauge field spectrum, can be observed here.   Moreover,
for $\alpha \sim 1$, the mass gap between the zero modes and the
massive states now goes as:
\be M^2_{GAP}\sim \frac{1}{r_0^2} + \frac{1}{r_1^2}. \ee
Therefore, for the fermions, a finite mass gap in the large volume
limit can also be obtained by taking $r_0 \rightarrow \infty$ and
turning on  $\delta$, thus allowing $r_1$ to remain finite. This
contrasting behaviour to the standard KK picture is a consequence of
the conical defects ($\ob \neq 1$ and $\omega\neq 1$) in our
internal manifold.  Below we shall consider its implications for
phenomenology.

\begin{figure}
\centering
\begin{tabular}{cc}
\epsfig{file=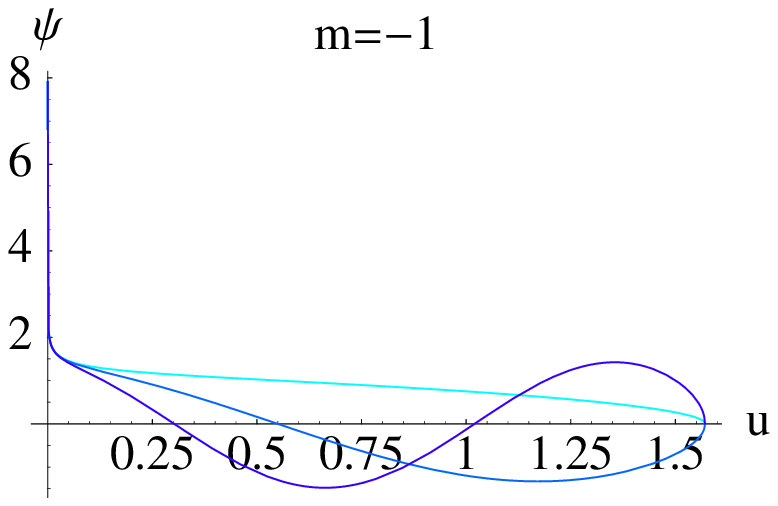,width=0.45\linewidth,clip=} &
\epsfig{file=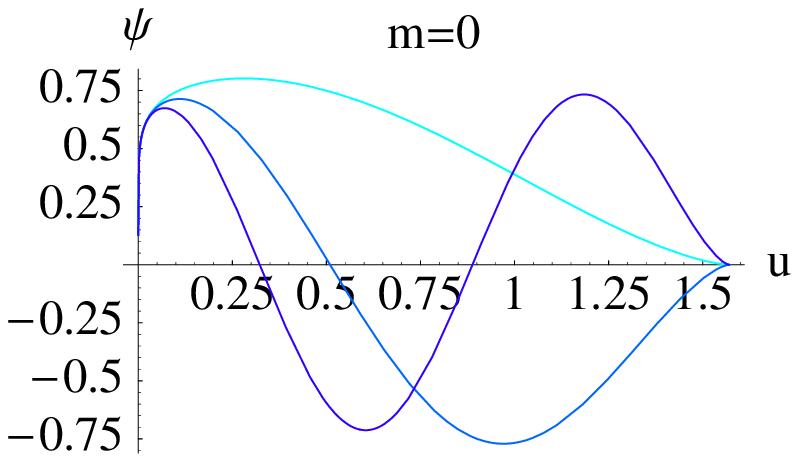,width=0.45\linewidth,clip=}
\end{tabular}
\caption{\footnotesize Fermion Wave Function Profiles: $n=0,1,2$
modes plotted for angular momentum numbers $m=-1,0$ (eqs (\ref{02})
and (\ref{m21}) respectively). The parameters are chosen to be
$(r_0, \omega, \overline{\omega},e)=(1,1/4,1,0)$, corresponding to a
single negative tension brane. Also the normalisation constant is
set to 1. The number of intersections with the $u$-axis equals $n$,
according to quantum mechanics. Notice that the $(m,n)=(-1,0)$ mode
is massless, and that given a localized massless mode, there is also
an infinite KK tower of localized massive modes.}\label{fig:wavefns}
\end{figure}

\subsection{4D Effective Fermion Charges}

Let us first end this section on fermion fluctuations by obtaining
their 4D effective gauge couplings.  This can be calculated by going
beyond their bilinear Lagrangian, and considering the interaction
term:
\be  \int d^6X \sqrt{-G}\,\,\overline{\lambda}\Gamma^M
\nabla_M\lambda = \dots + \int d^6X
\sqrt{-G}\,\,\overline{\lambda}\Gamma^{\mu} (\partial_{\mu} + e
V_{\mu}) \lambda + \dots \, . \ee Using the results for the KK
decomposition found in the preceding sections: \bea \lambda(X) &=&
\sum_{m,n} \lambda_{mn}(x) f^{(\lambda)}_{mn}(\rho) e^{im\varphi} =
\sum_{m,n} \lambda_{mn}(x) \psi^{(\lambda)}_{mn}(u) e^{-A -B/4}
e^{im\varphi}\nonumber \\
V_{\mu}(X) &=& \sum_{m,n} V_{\mu \, mn}(x) f^{(V)}_{mn}(\rho)
e^{im\varphi} = \sum_{m,n} V_{\mu \, mn}(x) \psi^{(V)}_{mn}(u)
e^{-(3A+B)/4} e^{im\varphi} \eea and recalling that the gauge field
zero mode is $f^{(V)}_{00} = 1$, a fermion mode $\lambda_{mn}$ has
the following coupling to the 4D effective gauge group : \bea
e_{eff} &=& \frac{e \, \int d\varphi du \,
\overline{\psi^{(\lambda)}_{mn}} \, f^{(V)}_{00} \,
\psi^{(\lambda)}_{mn}}{\int d\varphi du \,
\overline{\psi^{(\lambda)}_{mn}}\, \psi^{(\lambda)}_{mn}} \nonumber \\
&=& e \eea Since the gauge field zero mode has a constant wave
profile, the effective charges for the fermion modes are universal.
This is a general result, and independent of any possible
localization properties of the fermion modes: massless and massive
fermion modes will always have the same coupling to the massless
gauge fields.  Again, we find that a large mass gap is required in
the fermion spectrum in order to hide the KK tower.  We now consider
this issue in more detail.


\section{Large Volume Compactifications with a Large Mass Gap}\label{LargeMV}
In the previous two sections, we have calculated the complete KK
spectrum for the warped brane world compactification of 6D
supergravity, for two interesting sectors of the gauge and matter
fluctuations. We are now ready to consider the possible implications
of our results.

6D brane world models have long been of interest in the context of
LED, since these may help with the gauge hierarchy problem. In the
conventional ADD picture, Standard Model particles must be confined
to a 4D brane world, in order to explain why the large extra
dimensions have escaped detection.  It would certainly be of
interest to develop a dynamical description of this localization,
within the context of low energy effective field theory.

This could be achieved, for instance, if the zero mode wave profiles
were peaked near to a brane, and the heavy modes suppressed there
\cite{Randjbar-Daemi:2003qd}. However, we have found that zero mode
fermions can be peaked near to negative tension branes, only at the
price of localizing the whole KK tower (see Figure
\ref{fig:wavefns}). Therefore, strong couplings are expected between
light and heavy modes.  If the zero mode bulk fermions are to be
interpreted as matter in the SM, then apparently the only way to
explain why we do not observe all the KK modes is by insisting that
their mass gap is larger than the $100GeV$ scale probed to date.

Usually, this would bring us back to the classical KK scenario, with
the extra dimensions required to be very small  (at least
$(100GeV)^{-1}$ scale), and the generation of the gauge hierarchy
lost. However, in our framework we have seen that a large mass gap
can occur even if the volume $V_2$, defined as the ratio
$\kappa^2/\kappa_4^2$, becomes large. In the fermionic sector this
set up can be achieved with the conical defect associated with the
warping ($\omega\neq 1$ and $\overline{\omega}=1$), by taking the
parameter $r_1$ to be small.
 Another way is turning on the other defect ($\ob\neq
1$) and then taking the large $\alpha$ limit, that is small $\ob$
limit, corresponding to a negative tension brane.  The volume $V_2$
in (\ref{V2}) becomes large but, for both the fermions and gauge
fields, the mass gap does not reduce to zero. We observe that the
latter mechanism works also for $r_0=r_1$, that is
$\omega=\overline{\omega}$, which corresponds to the unwarped
``rugby ball'' solution with branes of equal tension at each of the
two poles \cite{Carroll:2003db,Aghababaie:2003wz}.

The general idea of relaxing the phenomenological constraints on the
size of the extra dimensions by deforming the shape of the internal
space was proposed in \cite{Dienes:2001wu}.  There, it was shown
that the presence of shape moduli can imply that there is no
experimental limit on the size of the largest extra dimension.
However, requiring a large KK mass gap still constrained the overall
volume of the extra dimensions. Here we give an explicit model which
allows arbitrary large values for both $V_2$ and $M_{GAP}^2$, at
least for the fermions and vectors. This could have an interesting
application in the ADD scenario because we can have both $\kappa\sim
TeV^{-2}$ and small effects from the massive modes by setting a
large enough value of $M_{GAP}$.

In terms of hiding Large Extra Dimensions from our four dimensional
universe, another possible approach is to interpret all the bulk
fields that we have found (massless modes and massive ones) as a
hidden sector, only gravitationally coupled to the SM. At this
level, the SM must then be introduced by hand, confined on the
delta-function brane.  It seems that this is the approach to take if
embedding the SLED scenario in our calculations, proposed in
\cite{Burgess:2004ib}, to resolve the Cosmological Constant Problem.
This proposal relates the hierarchy in the Electroweak scale with
that of the Cosmological Constant. The Electroweak scale is set by
the size of the extra dimensions, $r$, and the Cosmological Constant
is given by the KK mass gap, here fixed by the same
scale\footnote{However, the breakdown of SUSY in the bulk, as in the
solutions studied here, may lead to a larger prediction for the
Cosmological Constant, see \cite{Burgess:2004ib}.} $1/r$. Both may
have their observed values when the 6D fundamental scale is $TeV$,
and $r \sim 0.1mm$. For the mass gap to be this small, SM particles
must be localized to the brane.

Let us end by considering the tunings involved, when constructing a
model with large volume (say, $\sqrt{V_2} \sim 0.1mm$) and large
mass gap (say, $M_{GAP} \sim TeV^{-1}$).  Consider first a large
mass gap for the fermions.  If we set $\alpha \sim 1$ and $r_1 \ll
r_0$,
then the Dirac quantization (\ref{DiracQ}) implies: \be e\,
\frac{r_1}{r_0} \frac{g}{g_1} \sim N \ee
If we then assume $e \sim 1$ (which is natural from group theory)
and $N \sim 1$ (which is required for a small number of families),
the large volume - large mass gap condition requires a large
hierarchy in the bulk gauge couplings:
\be \frac{g}{g_1} \sim 10^{15} \, .\ee

Alternatively, we could set $r_1 \sim r_0$ and $\alpha \gg 1$,
allowing a large mass gap for both fermions and gauge fields. In
this case, requiring a large mass gap, $M_{GAP} \sim 1/r_0$, as well
as large volume, $\sqrt{V_2} \sim \sqrt{\alpha} r_0$, requires
$\alpha \sim 10^{30}$.  Then, again, the Dirac quantization
condition (\ref{DiracQ}) reveals a large hierarchy in the bulk gauge
couplings:
\be \frac{g}{g_1} \sim 10^{-30} \ee
In both scenarios we cannot embed the background monopole in
$U(1)_R$.

There are of course other combinations, for example with both
$\alpha \gg 1$ and $r_1 \ll r_0$, in which these hierarchies may be
relaxed.  However, we should say that these tunings do not appear to
be very natural or promising.  For example, choosing the large
dimensionless number $\alpha \gg 1$ corresponds to heavy negative
tension branes, and deficit angles orders of magnitude less than
zero. On the other hand, independently of trying to embed the Large
Extra Dimension scenario into the present model, we have found an
explicit example in which the KK mass gap does not go to zero as the
volume goes to infinity, contrary to standard lore.

\section{Conclusions and Outlook of Part II} \label{conclusions2}

\begin{table}[top]
\begin{center}
\begin{tabular}{|l|l|}
\hline  Gauge Fields  & $\qquad$ Fermions  \\ \hline
 $V\sim \,\,\,\,({\bf 45},{\bf 1})_0$ & $\lambda \sim \,\,\,\,\,({\bf 45},{\bf
1})_1$ \\
 $\qquad +({\bf 16},{\bf 1})_0$ & $\qquad +({\bf 16}+ \overline{{\bf 16}},{\bf
1})_1$  \\
$\qquad +({\bf 1},{\bf 133})_0$ & $\qquad +({\bf 1},{\bf 133})_1$
\\
 $\qquad +({\bf 1},{\bf 1})_0$ & $\psi \sim ({\bf 1},{\bf 912})_0$
 \\ \hline
\end{tabular}
\end{center}\caption{\footnotesize The gauge and fermion fields whose
KK spectrum is given by our work, for the illustrative example of
the anomaly free model $E_6 \times E_7 \times U(1)_R$, when the
monopole is embedded in $E_6$.  We give the quantum numbers under
${\cal H} = SO(10) \times E_7 \times U(1)_R$, which is the unbroken
subgroup of the 6D gauge group.\label{T:summary}}
\end{table}

In this part of the thesis we have analyzed an interesting subsector
of gauge field and fermion fluctuations, in the warped brane world
solutions of 6D minimal gauged supergravities.  In particular, we
have focused on bulk components which could give rise to SM or Grand
Unified gauge and charged matter fields.

We performed a Fourier decomposition of 6D fields, and transformed
the resulting field equations into a Schroedinger-like problem.  We
were then able to find the exact solutions for the KK modes, in
terms of hypergeometric functions. We considered in detail the
boundary conditions that the physical modes must satisfy.  In
addition to the normalizability constraint, consistency also
required a hermiticity condition, which can be interpreted as
demanding current conservation.  We were able to implement this in
its general, quadratic form.  Together, these conditions selected
the physical modes, and gave rise to a discrete mass spectrum, which
we presented in full.  The discreteness of the spectrum is of course
to be expected given the compact topology of the internal manifold,
whose Euler number is two.

 Our study can be applied to several sectors of the 6D supergravities.  In Table
\ref{T:summary} we summarise the 6D fields that are covered by our
analysis, for the illustrative example of the anomaly free model
$E_6 \times E_7 \times U(1)$.
 Moreover, the corresponding spectra for the non-supersymmetric model of
\cite{Wetterich:1984rv} (at least for the unwarped 4D Poincar\'e
invariant case) and \cite{Carroll:2003db}, generalized to
Einstein-Yang-Mills with fermions, can be straightforwardly
extracted from those given above simply by setting the warp factor
to one, that is $r_1=r_0$.

The exact results presented in this chapter enabled us to study the
effects of the conical defects, sourced by codimension two branes,
on the KK wave profiles and mass gaps.  As usual, the gauge fields
have a zero mode with constant wave profile.  For the fermions, we
found that some zero modes can be peaked on a negative tension
brane, but in this case the whole KK tower is peaked there too.
Therefore, in order to interpret the bulk zero modes as 4D effective
fields of the SM, the mass gap must be large.

Intriguingly, this does not necessarily drive us to the conventional
KK picture, with small compact dimensions.  It does not, because the
conical defects allow a novel behaviour in the mass gap, which can
be decoupled from the volume of the compactification, defined by
$\kappa^2/\kappa_4^{\,2}$.  This continues to be observed in the
unwarped limit, where the rugby ball model of
\cite{Carroll:2003db,Aghababaie:2003wz} is retrieved.  Contrary to
standard lore, a finite mass gap can be obtained, even if the volume
goes to infinity.

For example, a large volume could be arranged in order to generate
the Electroweak hierarchy, whilst maintaining a large mass gap
between the zero modes and the KK tower. This picture does not seem
to provide a realisation of the SLED scenario, where the volume and
mass gap should be related. However, in this way, SM fields could
arise from bulk fields, and along with gravity propagate through the
large extra dimensions, perfectly consistently with observation.
Moreover, for better or worse, this picture seems to render the LED
scenario less falsifiable than previously thought, since we do not
have to expect that the bulk KK modes are accessible at $TeV$
scales. However, arranging for both a large volume and large mass
gap seems to require a large degree of fine-tuning in bulk
couplings. Furthermore, for a more complete idea, we would have to
consider the KK spectrum for the remaining bulk sectors, and in
particular the gravitational fluctuations to know the effect of LED
on post-Newtonian tests\footnote{Massless gravitational fluctuations
in the non-supersymmetric rugby ball model were considered in
\cite{Graesser:2004xv} and, for thick branes, in
\cite{Peloso:2006cq}.  Moreover, some results for the Kaluza-Klein
spectra in non-supersymmetric warped brane models (with rugby ball
limit) have been found in \cite{Yoshiguchi:2005nn} and, for thick
branes, in \cite{deRham:2005ci}.}.

Indeed, our analysis of 4D effective gauge fields and charged
fermions is only a first step towards a complete analysis of
fluctuations about the warped brane world background in 6D
supergravity.  A final objective would be to derive the full 4D
effective field theory describing light fluctuations, and an
understanding of when 6D physics comes into play.  The sectors that
we have studied here are the simplest ones, in terms of the mixings,
but it should be possible to continue the project to other fields by
extending our work.

Of the remaining sectors to be analyzed, the scalar perturbations
have special importance, since they can contain Higgs fields and can
have implications for the stability of the background solution.
Indeed, in the round sphere limit of the model that we have studied,
tachyons in general emerge from the internal components of the 6D
gauge field orthogonal to the gauge field background
\cite{Randjbar-Daemi:1983bw}, as discussed in Appendix
\ref{SUGRAstability}. A first step in the study of the scalar
fluctuations has been presented in \cite{Lee:2006ge}, and it would
certainly be interesting if general results can be
found\footnote{The analysis of \cite{Lee:2006ge} is restricted to
the case in which the gauge group is simply $U(1)_R$.}. The
difficulty may be in the mixings of different scalar fluctuations,
which lead to a complicated system of coupled ODEs.

In another direction, much of our analysis was general, and could
also be used to study other theories and other backgrounds with 4D
Poincar\'e-2D axial symmetry.

Finally, it would be interesting to investigate whether there exist
other mechanisms, which lead to the same decoupling between the mass
gap and internal volume that we have found here.  Indeed, in our set
up we have been able to show that the decoupling arises due to the
conical defects, but it may be possible to find other sources in
different frameworks.  In this way, our explicit example may be a
realisation of a more general mechanism.

\addcontentsline{toc}{chapter}{Concluding Remarks}
\chapter*{Concluding Remarks}

The 4D effective theory associated to a higher dimensional model
must be derived carefully in order to obtain the correct physical
predictions. In this thesis we have studied two important issues
concerning the 4D interpretation of models with extra dimensions.
The first one pertains to the role of heavy KK modes in the low
energy dynamics. The second one is related to the dependence of the
KK towers on possible conical defects of the internal manifold. In
both cases we have analyzed examples which are interesting from the
physical point of view. Indeed, in Chapter \ref{EMS} we have studied
a 6D non supersymmetric gauge and gravitational theory which leads
to a 4D chiral effective theory similar to the electroweak part of
the SM. Moreover, in Chapters \ref{6DSUGRA} and \ref{SUGRAspectrum}
we have treated a 6D supergravity expanded around a non
supersymmetric and singular solution, which could give rise to the
SM or to a grand unified theory in the low energy limit. The latter
calculation is also a first step towards the analysis of SLED as a
scenario in which one can hope to solve the cosmological constant
problem. Indeed, one can compute carefully all the contributions to
the vacuum energy density only after constructing the complete 4D
spectrum.

The main original results and outlook of Parts I and II have been
extensively discussed in Sections \ref{Comments} and
\ref{conclusions2}, and so here we consider further possible
outlooks which are shared by Parts I and II.

We observe that an interesting outlook could be the study of the
scalar sector of 6D gauged supergravities expanded around an
axisymmetric solution, for instance the conical-GGP solutions that
we have presented in Subsection \ref{conSing} and analyzed in
Chapter \ref{SUGRAspectrum}. On the one hand this will give us
information on the stability of these solutions and on sectors that
contain candidates for the Higgs field. On the other hand the
contribution of the heavy KK modes to the scalar couplings may be
non vanishing, as in the 6D Einstein-Maxwell-Scalar model,
manifesting the underlying 6D physics. Moreover it would be
interesting to know the form of the heavy mode contribution in the
presence of warping and singularities in the internal manifold,
which are both properties of the conical-GGP solutions.

Another possible outlook is the analysis of the spin-2 fluctuations
of the 6D gauged supergravities of Part II. Indeed this sector is
responsible for the mediation of the gravitational interaction and
therefore it is relevant for cosmological applications. A first
interesting study can be the calculation of the heavy mode
contribution to this sector, which can give rise to modification of
gravity at small length scales. It would also be interesting to know
if the warping and the singularities, which allows the decoupling
between the volume of the internal space and the mass gap as
discussed in Section \ref{LargeMV}, can have physically relevant
effects in the spin-2 sector as well.

\newpage
\phantom{}\vskip 1cm \centerline{ {\LARGE \bf
Acknowledgements}}\vskip 1cm

I would like to thank my supervisor Prof. Seif Randjbar-Daemi for
countless and illuminating discussions and for his helpful advice
throughout this work.

Moreover I am very grateful to Prof. Mikhail Shaposhnikov for very
stimulating discussions and valuable suggestions.

I would also like to thank my collaborator Susha Parameswaran for
her friendly help and availability.

Finally I am grateful to Katarzyna Zuleta for valuable
correspondence and to Martin O'Loughlin, Giulio Bonelli, Tony
Gherghetta, Antonios Papazoglou, Emanuele Macr\`{i}, Carlo
Maccaferri, Alessandro Michelangeli, Elisabetta Majerotto, Luca
Ferretti, Federico Minneci, Francesco Benini, Stefano Cremonesi,
Roberto Valandro, Giuseppe Milanesi, Alessio Provenza and  Giuliano
Panico for several discussions about physics and mathematics.

\newpage

\appendix

\chapter{Conventions and Notations}\label{GR}\setcounter{equation}{0}

We choose the signature $-,+,+,+,...$ for the metric $G_{MN}$. The
Riemann tensor is defined as follows
\be R_{MNS}^{R}=\partial_M \Gamma_{NS}^R -\partial_N \Gamma_{MS}^R +
\Gamma_{MP}^R \Gamma_{NS}^P -\Gamma_{NP}^R \Gamma_{MS}^P,  \ee
where the $\Gamma 's$ are the Levi-Civita connection. Whereas the
Ricci tensor and the Ricci scalar
\be R_{MN}=R_{PMN}^{P}, \ \ \ \ R=G^{MN}R_{MN}.  \ee
Here $M,N,...$ run over all space-time dimensions.

Our choice for the 6D constant gamma matrices $\Gamma^A$,
$A=0,1,2,3,5,6,$ is
\be \Gamma^{\mu}=\left(\bacc 0 & \gamma^{\mu} \\
\gamma^{\mu} & 0 \ea \right), \quad  \Gamma^5=\left(\bacc 0 & \gamma^5 \\
\gamma^5 & 0 \ea \right),\quad  \Gamma^6=\left(\bacc 0 & -i \\
i & 0 \ea \right), \label{Gamma}\ee
where the $\gamma^{\mu}$ are the 4D constant gamma matrices and
$\gamma^5$ the 4D chirality matrix:
\be \gamma^5 = -i\gamma^0 \gamma^1 \gamma^2 \gamma^3. \ee
We define also the 6D chirality matrix $\Gamma^7$ by
\be \Gamma^7 = -\Gamma^0 \Gamma^1 \Gamma^2 \Gamma^3 \Gamma^5 \Gamma^6 = \left(\bacc 1 & 0 \\
0 & -1 \ea \right).\ee

Moreover the spin connection is
\be
\omega^{[A,B]}_M=\eta^{BC}\omega_{M\,\,\,C}^{\,\,A}=\eta^{BC}\left(e^A_N\Gamma_{MR}^Ne^R_B
+e^A_N\partial_Me^N_B\right),\ee
where $e^A_M$ is the vielbein. In the text $e^A_M$ denotes the
background spin connection and $E^A_M$ the complete dynamical spin
connection.

To study compactifications we split the D-dimensional space-time
coordinates $X^M$, $M=0,1,...,D-1$ in two sets: 4D coordinates
$x^{\mu}$, $\mu=0,1,2,3$, and internal coordinates $y^m$,
$m=4,...,D-1$. A background metric with 4D Poincar\'e invariance
reads
\be ds^2=e^{A(y)}\eta_{\mu \nu}dx^{\mu}dx^{\nu}+g_{mn}(y)dy^{m}dy^n, \ee
where $e^A$ is called warp factor and $g_{mn}$ is the metric of the internal space.

\vskip1.0cm
We use the following further symbols.

\vskip0.5cm
{\small $(Minkowski)_D$: D-dimensional Minkowski space-time.

$d$: number of extra dimensions ($d\equiv D-4$).

$K_d$: internal d-dimensional space.

$\kappa$: $D$-dimensional Planck scale, the Einstein-Hilbert term in the lagrangian being $R/\kappa$.

$\kappa_4$: 4D Planck scale.

$\tilde{V}_d$: proper volume of $K_d$.

$V_d$: the ratio $\kappa^2/\kappa_4^2$; for unwarped geometries it
equals $\tilde{V}_d$.

$r\equiv (V_d)^{1/d}$.

$\mathcal{M}^*$: complex conjugate of a matrix $\mathcal{M}$.

 $\mathcal{O}^{\dagger}$: hermitian conjugate of an
operator $\mathcal{O}$.

$<\mathcal{O}>$: vacuum expectation value of $\mathcal{O}$.

$\nabla_M$: covariant derivative, including gravitational and gauge
connections.

 $a$: radius of $S^2$. }

 $\delta^{(n)}(x)$: n-dimensional Dirac $\d$-function ($\int d^n x
 \delta^{(n)}(x)=1$).

\vskip1.0cm

We also use the following abbreviations.
 \vskip0.5cm
\small{ SM: Standard Model of Particle Physics.

GR: General Relativity.

KK: Kaluza-Klein.

VEV: Vacuum Expectation Value.

SSB: Spontaneous Symmetry Breaking.

EOM: Equation of Motion.

ADD: Arkani-Hamed, Dimopoulos and Dvali.

LED: Large Extra Dimensions.

RS: Randall-Sundrum

SLED: Supersymmetric Large Extra Dimensions.

}

\chapter{Spectrum from 6D Einstein-Maxwell-Scalar
Model}\label{AppGHB}

\section{Spin-1 Mass Terms from $S_F$ and $S_R$}\label{SFSR}\setcounter{equation}{0}

\subsection{ $S_F$ Contribution}\label{S_F}

In this subsection we write the contribution of
\be S_F\equiv -\frac{1}{4}\int d^6X \sqrt{-G}F^2 \ee
to the bilinear terms of $V$, $U$ and $W$. By direct computation we
get kinetic terms for $V$ and $U$ and some mass terms for $U$ and
$W$:
\bea &&-\frac{1}{4}\int d^2y \e F^2=
-\frac{1}{4}V_{\mu\nu}V^{\mu\nu}K -\frac{1}{6}U_{\mu\nu}^{\hat{\alpha}}U^{\mu\nu \hb}K_{\ha \hb} \nonumber\\
     && -\frac{2}{3}U_{\mu}^{\ha}U^{\mu \hb}M^{(1)}_{\ha \hb}+
\frac{4}{3}U_{\mu}^{\ha}W^{\mu \hb}M^{(2)}_{\ha \hb}
-\frac{2}{3}W_{\mu}^{\ha}W^{\mu \hb}M^{(3)}_{\ha \hb}+..., \eea
where the 4D curved indices $\mu$ and $\nu$ are contracted with the
4D metric $g_{\mu \nu}$, the dots are constant terms and interaction
terms, moreover
\be V_{\mu\nu}=\partial_{\mu}V_{\nu}-\partial_{\nu}V_{\mu},\,\,\,
U_{\mu\nu}^{\ha}=\partial_{\mu}U_{\nu}^{\ha}-\partial_{\nu}U_{\mu}^{\ha}
\ee
and

\bea K&=&\frac{1}{4\pi a^2}\int d^2y \,\e, \,\,\,\, K_{\ha
\hb}=\frac{3}{4\pi}\left(\frac{\kappa}{\sqrt2 ea^2}\right)^2
\int d^2 y \,\e\,\D^3_{\ha}\D^3_{\hb}, \nonumber \\
M^{(1)}_{\ha \hb}&=&\frac{3}{8\pi}\left(\frac{\kappa}{\sqrt2
ea^2}\right)^2
\int d^2 y \,\e\,g^{mn}\partial_m \D_{\ha}^3\partial_n\D_{\hb}^3 \nonumber \\
M^{(2)}_{\ha \hb}&=&-\frac{3 \kappa^2}{16 \pi ea^3}\int d^2 y \,\e\,
\partial_m\D_{\ha}^3\D_{\hb}^{\alpha}e_{\alpha}^ng^{mq}F_{nq} \nonumber \\
M^{(3)}_{\ha \hb}&=&\frac{3\kappa^2}{16\pi a^2} \int d^2 y
\,\e\,\D_{\ha}^{\alpha}e_{\alpha}^m\D_{\hb}^{\beta}e_{\beta}^pF_p^{\,\,n}F_{mn}.\label{F2bil}
\eea
The results (\ref{F2bil}) are valid for all background $e^{\alpha}$
and $e^3$. We use the $SU(2)\times U(1)$ background in the
Subsection \ref{S^2simm}, the $U(1)_3$ background in the subsection
\ref{S^2}.

\subsection{$S_R$ Contribution} \label{S_R}

In this subsection we write the contribution of
\be S_R\equiv \int d^6X \sqrt{-G}\frac{1}{\kappa^2}R \ee
to the bilinear terms of $W$. The complete contribution of $S_R$ to
the 4D action is given in \cite{SS} in the case of non deformed
background solutions. Here we need explicit expressions, at least
for the bilinears, which are also valid for deformed solutions. We
get a kinetic term and a mass term of $W$: up to a total derivative
we have

\bea &&\int d^2y \frac{1}{\kappa^2}\,\e\,R= -\frac{1}{6}W_{\mu\nu}^{\hat{\alpha}}W^{\mu\nu \hb}K'_{\ha \hb} \nonumber\\
     && +W_{\mu}^{\ha}W^{\mu \hb}M^{(4)}_{\ha \hb}+...,
\eea
where the dots include constant and interaction terms; moreover
\be
W_{\mu\nu}^{\ha}=\partial_{\mu}W_{\nu}^{\ha}-\partial_{\nu}W_{\mu}^{\ha},
\ee
and
\bea K'_{\ha \hb}&=&\frac{3}{8\pi a^2}\int d^2 y \,\e\,\D^{\alpha}_{\ha}\D^{\beta}_{\hb}g_{\alpha \beta}, \nonumber \\
M^{(4)}_{\ha \hb}&=&\frac{1}{4\pi a^2}\int d^2 y
\,\e\left[\partial_n\D_{\ha}^{\a}\D_{\hb}^{\b}\left(-e^m_{\a}\omega_{m\,\,\,
\beta} ^{\,\,\,\c}e^n_{\c}-g_{\a\d}g^{nm}\omega_{m\,\,\, \beta}
^{\,\,\,\d}+2e^n_{\a}e^m_{\c}\omega_{m\,\,\,
\beta} ^{\,\,\,\c} \right)\right.+ \nonumber \\
&&+D_{\ha}^{\a}\D_{\hb}^{\b}\left(-\frac{1}{2}\omega_{n\,\,\, \a}
^{\,\,\,\c}e^m_{\c}\omega_{m\,\,\, \beta}
^{\,\,\,\d}e^n_{\d}-\frac{1}{2}\omega_{n\,\,\, \a}
^{\,\,\,\d}g^{nm}\omega_{m\d \beta}+\omega_{n\,\,\, \a}
^{\,\,\,\d}e^n_{\d}\omega_{m\,\,\,
\beta} ^{\,\,\,\c}e^m_{\c}\right)\nonumber \\
&&\left.+
\partial_n\D_{\ha}^{\a}\partial_m\D_{\hb}^{\b}\left(-\frac{1}{2}e^m_{\a}e^n_{\beta}+e^n_{\a}e^m_{\beta}
-\frac{1}{2}g_{\a \b}g^{nm}\right)\right], \label{SRbil}\eea
where $\omega_{n\,\,\,\beta} ^{\,\,\,\a}$ is the 2-dimensional spin
connection for $e_n^{\a}$. The results (\ref{SRbil}) are also valid
for every background $e^{\alpha}$ and $e^3$. We use the $SU(2)\times
U(1)$ background in the Subsection \ref{S^2simm}, the $U(1)_3$
background in the Subsection \ref{S^2}.

\subsection{The Case of $SU(2)\times U(1)$ background} \label{S^2simm}

We use now the $SU(2)\times U(1)$ background, that is $\eta =0$.
This computation is performed in \cite{RSS}. We have the following
bilinear terms for $V$, $U$ and $W$:
\bea && -\frac{1}{4}V_{\mu\nu}V^{\mu\nu}
-\frac{1}{6}U_{\mu\nu}^{\hat{\alpha}}U^{\mu\nu}_{ \ha}-\frac{1}{6}W_{\mu\nu}^{\hat{\alpha}}W^{\mu\nu}_{ \ha} \nonumber\\
&&-\frac{2}{3a^2}\left(U_{\mu \hat{\alpha}}-W_{\mu
\hat{\alpha}}\right)\left(U^{\mu \hat{\alpha}}-W^{\mu
\hat{\alpha}}\right). \label{bilinearV}\eea
 If we define
\bea \mathcal{A}&=&\sqrt{\frac{1}{3}}(W+U), \nonumber \\
X&=&\sqrt{\frac{1}{3}}(W-U), \label{def2}\eea
we can write (\ref{bilinearV}) as follows
\bea
&&-\frac{1}{4}V_{\mu\nu}V^{\mu\nu}-\frac{1}{4}\mathcal{A}_{\mu\nu}^{\hat{\alpha}}\mathcal{A}^{\mu\nu}_{ \ha} \nonumber \\
&&-\frac{1}{4}X_{\mu\nu}^{\hat{\alpha}}X^{\mu\nu }_{\ha}
-\frac{2}{a^2}X_{\mu\hat{\alpha}}X^{\mu\hat{\alpha}},\eea
So $\mathcal{A}$ is a massless field, in fact it's the $SU(2)$
Yang-Mills field \cite{RSS}, while $X$ is a massive field which can
be neglected in the low energy limit.

\subsection{The Case of $U(1)_3$ Background} \label{S^2}

Let us consider now the solution (\ref{solution2}).
 First we note that $S_R$ and $S_F$ do not give mass terms
for $V$; so the only source for the mass of $V$ is $S_{\phi}$.

We want to prove now that also the $SU(2)$ Yang-Mills fields masses
do not receive contributions from $S_R$ and $S_F$. First we give the
bilinears for $U$ and $W$, which come from $S_R$ and $S_F$:
\bea &&-\frac{1}{6}U_{\mu\nu}^{\hat{\alpha}}U^{\mu\nu \hb}g_{\ha
\hb}\left(1+|\eta| \beta k_{\ha}\right)
-\frac{1}{6}W_{\mu\nu}^{\hat{\alpha}}W^{\mu\nu \hb}g_{\ha \hb}\left(1+|\eta| \beta k'_{\ha}\right) \nonumber\\
 &&-\frac{2}{3}U_{\mu}^{\ha}U^{\mu \hb}g_{\ha \hb}\left(1+|\eta| \beta m^{(1)}_{\ha}\right)
+\frac{4}{3}U_{\mu}^{\ha}W^{\mu \hb}g_{\ha \hb}\left(1+|\eta| \beta m^{(2)}_{\ha}\right) \nonumber\\
&&-\frac{2}{3}W_{\mu}^{\ha}W^{\mu \hb}g_{\ha \hb}\left(1+|\eta| \beta m^{(3)}_{\ha}\right),\label{bilinearsu13} \nonumber\\
\eea
where
\begin{displaymath} k_+=k_-=\frac{2}{5}, \,\,\, k_3=\frac{1}{5}, \,\,\,
k'_+=k'_-=\frac{3}{10}, \,\,\, k'_3=\frac{2}{5}, \nonumber\ed
\bd m^{(1)}_+=m^{(1)}_-=\frac{1}{5}, \,\,\, m^{(1)}_3=-\frac{2}{5},
\nonumber\ed
\bd m^{(2)}_+=m^{(2)}_-=-\frac{1}{20},\,\,\,
m^{(2)}_3=-\frac{2}{5},\,\,\, m^{(3)}_+=m^{(3)}_-=-\frac{3}{10},
\,\,\, m^{(3)}_3=-\frac{2}{5}. \nonumber\ed

In order to prove (\ref{bilinearsu13}) it's useful to use the
following formula for the background spin connection:
\be \omega_{\varphi\,\,\, +} ^{\,\,+}=-\omega_{\varphi\,\,\, -}
^{\,\,-}=\frac{i}{a}(\cos\theta-1 -\frac{1}{2}|\eta|\b\cos\theta
\sin^{2}\theta) \label{spinconn}\ee
and $\omega_{\varphi\,\,\, +} ^{\,\,-}=\omega_{\varphi\,\,\, -}
^{\,\,+}=0$.

Now we define $X$ and $A$ as follows
\bea \left(1+\frac{|\eta| \beta}{2}k'_{\ha}\right)W^{\ha}=
\sqrt{\frac{3}{2}}\left(\cos\theta_{\eta}^{\ha}X^{\ha}+\sin\theta_{\eta}^{\ha}\mathcal{A}^{\ha}\right), \nonumber\\
\left(1+\frac{|\eta| \beta}{2}k_{\ha}\right)U^{\ha}=
\sqrt{\frac{3}{2}}\left(-\sin\theta_{\eta}^{\ha}X^{\ha}+\cos\theta_{\eta}^{\ha}\mathcal{A}^{\ha}\right),
\label{defeps} \eea
where the angle $\theta_{\eta}^{\ha}$ is defined by
\be \cos\theta_{\eta}^{\ha}=\frac{1+|\eta| \beta
\delta^{\ha}}{\sqrt2},\,\,\, \sin\theta_{\eta}^{\ha}=\frac{1-|\eta|
\beta \delta^{\ha}}{\sqrt2},\ee
and the quantities $\d^{\ha}$ are not still fixed. It's simple to
check that the kinetic terms for $X$ and $\mathcal{A}$ are in the
standard form for every $\d^{\ha}$ up to $O(\eta^{3/2})$. The
definition (\ref{defeps}) reduce to (\ref{def2}) for $\eta=0$.

If we choose
\be
\d^{\ha}=\frac{1}{8}\left(m^{(3)}_{\ha}-k'_{\ha}-m^{(1)}_{\ha}+k_{\ha}\right)
\ee
we have no mass terms for $\mathcal{A}$ coming from $S_R+S_F$.

So the only source for the spin-1 low energy spectrum is $S_{\phi}$
and the result is given in equations (\ref{M0}) and (\ref{M1}).

\section{Explicit Calculation of Spin-0 Spectrum}\label{spin-0calculation}\setcounter{equation}{0}

As we pointed out in the text, in order to find the spin-0 spectrum
the expression of the $|i>$, $i=1,...,6$, vectors is needed; these
are defined by $\mathcal{O}_0|i>=0$, which is equivalent to
$\nabla^2\phi+\phi/a^2=0$, where $\nabla^2\phi$ is the Laplacian
over the charged scalar $\phi$, calculated with the round $S^2$
metric. Our choice for the orthonormal vectors\footnote{We express a
generic vector as in (\ref{Sdefinition}).}  $|i>$ is

 $$|1>=\frac{1}{\sqrt{2}}\left(\ba {c} \sqrt{\frac{3}{4\pi}}\mathcal{D}^{(1)}_{-1,1}\\
  \sqrt{\frac{3}{4\pi}}\left(\mathcal{D}^{(1)}_{-1,1}\right)^*
 \\
  0  \\
.\\
. \\
 .  \\

 0
\ea\right),\,|2>=\frac{1}{\sqrt{2}}\left(\ba {c} i\sqrt{\frac{3}{4\pi}}\mathcal{D}^{(1)}_{-1,1}\\
  -i\sqrt{\frac{3}{4\pi}}\left(\mathcal{D}^{(1)}_{-1,1}\right)^*
 \\
  0  \\
.\\
. \\
 .  \\
 0
\ea\right), $$
 $$|3>=\frac{1}{\sqrt{2}}\left(\ba {c} \sqrt{\frac{3}{4\pi}}\mathcal{D}^{(1)}_{-1,0}\\
  \sqrt{\frac{3}{4\pi}}\left(\mathcal{D}^{(1)}_{-1,0}\right)^*
 \\
  0  \\
.\\
. \\
 .  \\

 0
\ea\right), \,|4>=\frac{1}{\sqrt{2}}\left(\ba {c} i\sqrt{\frac{3}{4\pi}}\mathcal{D}^{(1)}_{-1,0}\\
  -i\sqrt{\frac{3}{4\pi}}\left(\mathcal{D}^{(1)}_{-1,0}\right)^*
 \\
  0  \\
.\\
. \\
 .  \\
 0
\ea\right), $$
$$|5>=\frac{1}{\sqrt{2}}\left(\ba {c} \sqrt{\frac{3}{4\pi}}\mathcal{D}^{(1)}_{-1,-1}\\
  \sqrt{\frac{3}{4\pi}}\left(\mathcal{D}^{(1)}_{-1,-1}\right)^*
 \\
  0  \\
.\\
. \\
 .  \\

 0
\ea\right), \,|6>=\frac{1}{\sqrt{2}}\left(\ba {c} i\sqrt{\frac{3}{4\pi}}\mathcal{D}^{(1)}_{-1,-1}\\
  -i\sqrt{\frac{3}{4\pi}}\left(\mathcal{D}^{(1)}_{-1,-1}\right)^*
 \\
  0  \\
.\\
. \\
 .  \\
 0
\ea\right). $$

Another ingredient for the calculation of the spin-0 spectrum is an
explicit expression of the vectors $|\tilde{i}>$ and of the
eigenvalues $M^2_{\tilde{i}}$; the latter are given in
\cite{Schellekens1,Schellekens2}, however, we need here also a
correspondence between eigenvalues and eigenvectors. As we explained
in the text, only the $|\tilde{i}>$ like (\ref{tildei}) and made of
$l=0,1,2$ harmonics are needed. The $|\tilde{i}> $ vectors must
satisfy the following eigenvalue equations\footnote{We derive
(\ref{eigen}) evaluating (\ref{L02}), (\ref{L03}) and (\ref{L06}) in
the basis (\ref{epm}) and performing the redefinition $h_{\pm
\pm}\rightarrow \sqrt{2}\kappa h_{\pm \pm}$ and $h_{+-}\rightarrow
h_{+-}\kappa/\sqrt{2}$, which normalizes the kinetic terms in the
standard way.}:
\bea &&-\nabla^2h_{++}+2R_{+-+-}h_{++}-2\kappa^2F_{+-}^2h_{++}-\sqrt{2}\kappa\nabla_{+}\mv_{+}F_{-+}=M^2h_{++},\nonumber \\
&&-\nabla^2h_{--}+2R_{+-+-}h_{--}-2\kappa^2F_{+-}^2h_{--}+\sqrt{2}\kappa\nabla_{-}\mv_{-}F_{-+}=M^2h_{--},\nonumber \\
&&-\nabla^2h_{+-}-R_{+-+-}h_{+-}-\frac{\kappa}{\sqrt2}\nabla_+\mv_-F_{-+}+
\frac{\kappa}{\sqrt2}\nabla_-\mv_+F_{-+}=M^2h_{+-},\nonumber \\
&&-\nabla^2\mv_{+}+R_{+-}\mv_{+}-\kappa^2\mv_+ F^2_{+-}+
\frac{\kappa}{\sqrt2}\nabla_+h_{+-}F_{-+}-\sqrt2 \kappa\nabla_-h_{++}F_{-+}=M^2\mv_{+},\nonumber \\
&&-\nabla^2\mv_{-}+R_{+-}\mv_{-}-\kappa^2\mv_- F^2_{+-}-
\frac{\kappa}{\sqrt2}\nabla_-h_{+-}F_{-+}+\sqrt2 \kappa\nabla_+h_{--}F_{-+}=M^2\mv_{-},\nonumber \\
\label{eigen}\eea
where the background objects ($\nabla^2$, $R_{+-+-}$,...) correspond
to the background (\ref{sphere}), (\ref{monopole}) and (\ref{phi0}).
We can transform the differential problem (\ref{eigen}) into an
algebraic one by using the expansion (\ref{lexpansion}). We get an
eigenvalue problem for every value of $l$ and we give now an
explicit expression for the $|\tilde{i}>$ vectors for the relevant
value of $l$, namely $l=0,1,2$. For $l=0$ we get just one
eigenvector $|\tilde{1}>$ with $M^2=1/a^2$:
\be |\tilde{1}>=\left(\ba {c} 0 \\
  0
 \\
  0 \\
0  \\
1/\sqrt{4\pi}  \\
  0  \\
0 \ea\right). \ee
For $l=1$ we get three different eigenvalues: $M^2=2/a^2, 4/a^2,
5/a^2$ . The eigenvectors which correspond to $M^2=2/a^2$ are
\be |\tilde{2}_0>, \quad
\frac{1}{\sqrt2}\left(|\tilde{2}_{1}>+|\tilde{2}_{-1}>\right),\quad
\frac{1}{\sqrt2 i}\left(|\tilde{2}_{1}>-|\tilde{2}_{-1}>\right), \ee
where
\be |\tilde{2}_{m}>\equiv\frac{1}{\sqrt6}\left(\ba {c} 0 \\
  0
 \\
  0 \\
0  \\
2\sqrt{\frac{3}{4\pi}}\mathcal{D}^{(1)}_{0,m}  \\
  -\sqrt{\frac{3}{4\pi}}\mathcal{D}^{(1)}_{1,m}  \\
-\sqrt{\frac{3}{4\pi}}\mathcal{D}^{(1)}_{-1,m} \ea\right). \ee
Instead the eigenvectors which correspond to $M^2=4/a^2$ are
\be i|\tilde{3}_0>, \quad \frac{1}{\sqrt2
i}\left(|\tilde{3}_{1}>+|\tilde{3}_{-1}>\right),\quad
\frac{1}{\sqrt2 }\left(|\tilde{3}_{1}>-|\tilde{3}_{-1}>\right), \ee
where
\be |\tilde{3}_{m}>\equiv\frac{1}{\sqrt2}\left(\ba {c} 0 \\
  0
 \\
  0 \\
0  \\
0  \\
  -\sqrt{\frac{3}{4\pi}}\mathcal{D}^{(1)}_{1,m}  \\
\sqrt{\frac{3}{4\pi}}\mathcal{D}^{(1)}_{-1,m} \ea\right). \ee
Moreover the eigenvectors which correspond to $M^2=5/a^2$ are
\be |\tilde{4}_0>, \quad \frac{1}{\sqrt2
}\left(|\tilde{4}_{1}>+|\tilde{4}_{-1}>\right),\quad \frac{1}{\sqrt2
i}\left(|\tilde{4}_{1}>-|\tilde{4}_{-1}>\right), \ee
where
\be |\tilde{4}_{m}>\equiv\frac{1}{\sqrt3}\left(\ba {c} 0 \\
  0
 \\
  0 \\
0  \\
 \sqrt{\frac{3}{4\pi}}\mathcal{D}^{(1)}_{0,m} \\
  \sqrt{\frac{3}{4\pi}}\mathcal{D}^{(1)}_{1,m}  \\
\sqrt{\frac{3}{4\pi}}\mathcal{D}^{(1)}_{-1,m} \ea\right). \ee
Finally, for $l=2$ the values of $M^2$ are given by
\be a^2M^2=6,\,\,2(3-\sqrt3),\,\,
2(3+\sqrt3),\,\,\frac{1}{2}(13-\sqrt{73})
,\,\,\frac{1}{2}(13+\sqrt{73}).\ee
The eigenvectors with $a^2M^2=6$ are
\bea &&|\tilde{5}_0>,\quad
\frac{1}{\sqrt2}\left(|\tilde{5}_{1}>-|\tilde{5}_{-1}>\right),\quad
\frac{1}{\sqrt2i}\left(|\tilde{5}_{1}>+|\tilde{5}_{-1}>\right),\nonumber \\
&&\frac{1}{\sqrt2}\left(|\tilde{5}_{2}>+|\tilde{5}_{-2}>\right),\quad
\frac{1}{\sqrt2i}\left(|\tilde{5}_{2}>-|\tilde{5}_{-2}>\right), \eea
where
\be |\tilde{5}_{m}>\equiv\frac{1}{3\sqrt2}\left(\ba {c} 0 \\
0 \\
-\sqrt2 \sqrt{\frac{5}{4\pi}}\mathcal{D}^{(2)}_{2,m} \\
  -\sqrt2 \sqrt{\frac{5}{4\pi}}\mathcal{D}^{(2)}_{-2,m}
 \\
  -2\sqrt3 \sqrt{\frac{5}{4\pi}}\mathcal{D}^{(2)}_{0,m} \\
  -\sqrt{\frac{5}{4\pi}}\mathcal{D}^{(2)}_{1,m}  \\
\sqrt{\frac{5}{4\pi}}\mathcal{D}^{(2)}_{-1,m} \ea\right). \ee
For $a^2M^2=2(3-\sqrt3)$ we have the eigenvectors
\bea &&i|\tilde{6}_0>,\quad
\frac{1}{\sqrt2}\left(|\tilde{6}_{1}>+|\tilde{6}_{-1}>\right),\quad
\frac{1}{\sqrt2i}\left(|\tilde{6}_{1}>-|\tilde{6}_{-1}>\right),\nonumber \\
&&\frac{1}{\sqrt2}\left(|\tilde{6}_{2}>-|\tilde{6}_{-2}>\right),\quad
\frac{1}{\sqrt2i}\left(|\tilde{6}_{2}>+|\tilde{6}_{-2}>\right), \eea
where
\be |\tilde{6}_{m}>\equiv\frac{1}{\sqrt{2(3+\sqrt3)}}\left(\ba {c} 0 \\
0 \\
-\frac{1+\sqrt3}{\sqrt2} \sqrt{\frac{5}{4\pi}}\mathcal{D}^{(2)}_{2,m} \\
 \frac{1+\sqrt3}{\sqrt2} \sqrt{\frac{5}{4\pi}}\mathcal{D}^{(2)}_{-2,m}
 \\
 0 \\
  \sqrt{\frac{5}{4\pi}}\mathcal{D}^{(2)}_{1,m}  \\
\sqrt{\frac{5}{4\pi}}\mathcal{D}^{(2)}_{-1,m} \ea\right). \ee
For $a^2M^2=2(3+\sqrt3)$ we have the eigenvectors
\bea &&i|\tilde{7}_0>,\quad
\frac{1}{\sqrt2}\left(|\tilde{7}_{1}>+|\tilde{7}_{-1}>\right),\quad
\frac{1}{\sqrt2i}\left(|\tilde{7}_{1}>-|\tilde{7}_{-1}>\right),\nonumber \\
&&\frac{1}{\sqrt2}\left(|\tilde{7}_{2}>-|\tilde{7}_{-2}>\right),\quad
\frac{1}{\sqrt2i}\left(|\tilde{7}_{2}>+|\tilde{7}_{-2}>\right), \eea
where
\be |\tilde{7}_{m}>\equiv\frac{1}{\sqrt{2(3-\sqrt3)}}\left(\ba {c} 0 \\
0 \\
-\frac{1-\sqrt3}{\sqrt2} \sqrt{\frac{5}{4\pi}}\mathcal{D}^{(2)}_{2,m} \\
 \frac{1-\sqrt3}{\sqrt2} \sqrt{\frac{5}{4\pi}}\mathcal{D}^{(2)}_{-2,m}
 \\
 0 \\
  \sqrt{\frac{5}{4\pi}}\mathcal{D}^{(2)}_{1,m}  \\
\sqrt{\frac{5}{4\pi}}\mathcal{D}^{(2)}_{-1,m} \ea\right). \ee
Then for $a^2M^2=(13-\sqrt{73})/2$:
\bea &&|\tilde{8}_0>,\quad
\frac{1}{\sqrt2}\left(|\tilde{8}_{1}>-|\tilde{8}_{-1}>\right),\quad
\frac{1}{\sqrt2i}\left(|\tilde{8}_{1}>+|\tilde{8}_{-1}>\right),\nonumber \\
&&\frac{1}{\sqrt2}\left(|\tilde{8}_{2}>+|\tilde{8}_{-2}>\right),\quad
\frac{1}{\sqrt2i}\left(|\tilde{8}_{2}>-|\tilde{8}_{-2}>\right), \eea
where
\be |\tilde{8}_{m}>\equiv\frac{1+\sqrt{73}}{\sqrt{438+30\sqrt{73}}}\left(\ba {c} 0 \\
0 \\
\frac{13\sqrt2+\sqrt{146}}{2(1+\sqrt{73})} \sqrt{\frac{5}{4\pi}}\mathcal{D}^{(2)}_{2,m} \\
  \frac{13\sqrt2+\sqrt{146}}{2(1+\sqrt{73})} \sqrt{\frac{5}{4\pi}}\mathcal{D}^{(2)}_{-2,m}
 \\
  -\frac{4\sqrt3}{1+\sqrt{73}} \sqrt{\frac{5}{4\pi}}\mathcal{D}^{(2)}_{0,m} \\
  -\sqrt{\frac{5}{4\pi}}\mathcal{D}^{(2)}_{1,m}  \\
\sqrt{\frac{5}{4\pi}}\mathcal{D}^{(2)}_{-1,m} \ea\right). \ee
Finally for $a^2M^2=(13+\sqrt{73})/2$:
\bea &&|\tilde{9}_0>,\quad
\frac{1}{\sqrt2}\left(|\tilde{9}_{1}>-|\tilde{9}_{-1}>\right),\quad
\frac{1}{\sqrt2i}\left(|\tilde{9}_{1}>+|\tilde{9}_{-1}>\right),\nonumber \\
&&\frac{1}{\sqrt2}\left(|\tilde{9}_{2}>+|\tilde{9}_{-2}>\right),\quad
\frac{1}{\sqrt2i}\left(|\tilde{9}_{2}>-|\tilde{9}_{-2}>\right), \eea
where
\be |\tilde{9}_{m}>\equiv\frac{1-\sqrt{73}}{\sqrt{438-30\sqrt{73}}}\left(\ba {c} 0 \\
0 \\
\frac{13\sqrt2-\sqrt{146}}{2(1-\sqrt{73})} \sqrt{\frac{5}{4\pi}}\mathcal{D}^{(2)}_{2,m} \\
  \frac{13\sqrt2-\sqrt{146}}{2(1-\sqrt{73})} \sqrt{\frac{5}{4\pi}}\mathcal{D}^{(2)}_{-2,m}
 \\
  -\frac{4\sqrt3}{1-\sqrt{73}} \sqrt{\frac{5}{4\pi}}\mathcal{D}^{(2)}_{0,m} \\
  -\sqrt{\frac{5}{4\pi}}\mathcal{D}^{(2)}_{1,m}  \\
\sqrt{\frac{5}{4\pi}}\mathcal{D}^{(2)}_{-1,m} \ea\right). \ee

We can now calculate the $6\times6$ matrix $M^2_{ij}$ given in
(\ref{QMperturbation}). In order to do that we need just the matrix
elements $<i|\mathcal{O}_1|\tilde{i}>$ and $<i|\mathcal{O}_2|j>$,
which can be computed by evaluating\footnote{For the background
solution (\ref{sphere}), (\ref{monopole}) and (\ref{phi0}) we have
$\mathcal{L}_0(\phi,\mv)=0$.} $\mathcal{L}_0(\phi,h)$ and
$\mathcal{L}_0(\phi,\phi)$, which appears in (\ref{L01}) and
(\ref{L04}), in the $\pm$ basis given in (\ref{epm}). After the
redefinitions $h_{\pm \pm}\rightarrow \sqrt{2}k h_{\pm \pm}$ and
$h_{+-}\rightarrow h_{+-}k/\sqrt{2}$, which normalize the kinetic
terms in the standard way, we get (for $n=2$)
\bea \mathcal{L}_0(\phi,h)&=&\sqrt2
\kappa\nabla_{+}\Phi\nabla_+h_{--}\phi^{*}
+\frac{\kappa}{\sqrt2}\nabla_{+}\Phi\nabla_-h_{+-}\phi^{*} \nonumber \\
&&+\sqrt2 \kappa\nabla_{+}\Phi h_{--}\left(\nabla_-\phi\right)^{*}
+\frac{\kappa}{\sqrt2}\nabla_{+}\Phi
h_{+-}\left(\nabla_+\phi\right)^{*}+c.c. \, ,\nonumber \\
\mathcal{L}_0(\phi,\phi)&=&\phi^*\partial^2\phi-\phi^*\left[-\nabla^2+m^2+(e^2+4\xi)\Phi^*
\Phi
+\kappa^2\nabla_+\Phi \left(\nabla_+\Phi\right)^*\right]\phi \nonumber \\
&&-\frac{1}{2}\left[\phi(2\xi-e^2)\left(\Phi^*\right)^2\phi+c.c.
\right].\eea
By using these expressions and the values of $|i>$ and $|\tilde{i}>$
given before, we find the following expression for $M^2_{ij}$: \be
\{M^2_{ij}\}= \left(\ba {cccccc} a_1 & 0 &
0 & 0 & a_4 & 0 \\
 0
 & a_1
 & 0 & 0 & 0 & -a_4 \\
0 & 0 & a_2 & 0 & 0 & 0 \\
0 & 0 & 0 & a_3 & 0 & 0 \\
a_4 & 0 &
0 & 0 & a_1 & 0 \\
0
 & -a_4
 & 0 & 0 & 0 & a_1 \ea\right),\label{M2ij} \ee
where
\bea a_1&=&\frac{|\eta|}{a^2}\left(-sign(\eta) +\frac{3}{10}\b + \frac{12}{5}\frac{\b \xi a^2}{\kappa^2}\right), \nonumber \\
 a_2&=&\frac{|\eta|}{a^2}\left(-sign(\eta) -\frac{6}{5}\b + \frac{24}{5}\frac{\b \xi a^2}{\kappa^2}\right), \nonumber \\
a_3&=&\frac{|\eta|}{a^2}\left(-sign(\eta) +\frac{4}{15}\b + \frac{8}{5}\frac{\b \xi a^2}{\kappa^2}\right), \nonumber \\
a_4&=&\frac{|\eta|}{a^2}\b\left(\frac{3}{10}- \frac{4}{5}\frac{ \xi
a^2}{\kappa^2}\right).\eea
By diagonalizing $M^2_{ij}$, we found exactly the spectrum that we
discussed in the Subsection \ref{spin-0}: the squared masses of the
vector particles are reproduced\footnote{In order to see that we use
the background constraints (\ref{constraint}).}, as required by the
light cone gauge; moreover we get the two masses squared given in
(\ref{MS}).

\section{Explicit Calculation of Spin-1/2 Spectrum}\label{spin-1/2calculation}\setcounter{equation}{0}

Here we concentrate on the right-handed sector, which is the non
trivial one because it presents $\eta^{1/2}$ mixing terms.

The eigenvalue equations for the unperturbed ($\eta=0$) mass squared
operator $\mathcal{O}_0$, acting on the right-handed sector, are
\bea - 2 \nabla_- \nabla_+\psi_{+R}
=M^2\psi_{+R}, \nonumber \\
- 2\nabla_+ \nabla_- \psi_{-R} =M^2\psi_{-R},
\label{unperturbedfermion} \eea
The differential equation (\ref{unperturbedfermion}) can be
transformed in an algebraic one through the harmonic expansion,
remembering the iso-helicities of $\psi_{+R}$ and $\psi_{-R}$:
$\lambda_{+R}=\lambda_{-R}=1$.
 Therefore an explicit expression for the vectors $|i>$, which satisfies by definition $\mathcal{O}_0|i>=0$, is given by
\be |i>=\left(\ba {c} 0 \\
  \sqrt{\frac{3}{4\pi}}\mathcal{D}^{(1)}_{-1,i}
\ea\right),\quad \quad i=1,-1,0. \ee
We give also an expression for the vectors $|\tilde{i}>$ and the
corresponding non vanishing eigenvalues $M^2_{\tilde{i}}$. For
$l=1$, we have just one eigenvalue $M^2=2/a^2$ and the corresponding
eigenvectors are
\be |\tilde{1}_m>=\left(\ba {c}
  \sqrt{\frac{3}{4\pi}}\mathcal{D}^{(1)}_{-1,m} \\
0 \ea\right). \ee
For $l=2$ we have an eigenvalue $M^2=6/a^2$, which corresponds to
the eigenvectors
\be |\tilde{2}_m>=\left(\ba {c}
  \sqrt{\frac{5}{4\pi}}\mathcal{D}^{(2)}_{-1,m} \\
0 \ea\right), \ee
and an eigenvalue $M^2=4/a^2$, which corresponds to the eigenvectors
\be |\tilde{3}_m>=\left(\ba {c} 0 \\
  \sqrt{\frac{5}{4\pi}}\mathcal{D}^{(1)}_{-1,m}
\ea\right). \ee 
By inserting these eigenvectors and eigenvalues in the expression
(\ref{QMperturbation}) we get
\be
M^2_{ij}=diag\left(\,\,0,\,\,0,\,\,\frac{2}{3}|\eta|g^2_Y\frac{\b}{\kappa^2}\right),
\ee
which corresponds to the spectrum we discussed at the end of section
\ref{spin-1/2}.

\chapter{Spectrum from 6D Gauged Minimal Supergravity}\label{AppSugra}

\section{Stability Analysis for Sphere Compactification}\label{SUGRAstability}\setcounter{equation}{0}

Here we consider the present known anomaly free 6D gauged minimal supergravities that we discussed in
Subsection \ref{SSandAfree}.  The bosonic action for this class of supergravities is given in (\ref{Baction}) and we
restrict to the case in which $\phi^{\alpha}=0$. We expect that this set up supports stability as we discussed
in Subsection \ref{SSandAfree}. In this section we study
when the monopole $(Minkowski)_4\times S^2$ compactification, given in (\ref{spheremonopole}), is stable,
in the sense that there are no tachyons.
The only cases in which we can have
stability are the old $E_7\times E_6 \times U(1)_R$ model and a new $SU(2)\times U(1)_R$ model with
a particular hyperinos representation (up to the trivial case in which the monopole is embedded in a $U(1)$ factor of the gauge
group). So a stable compactification of this type and the embedding of the SM gauge group is possible
only in the $E_7\times E_6 \times U(1)_R$ case.

\subsection{The Light Cone Gauge}

A first step toward the stability analysis is deriving the
lagrangian for the small fluctuations around the background. Here we
give explicitly such lagrangian for the smooth sphere
compactification and its form in the light cone gauge\footnote{For
an introduction to the light cone gauge in higher dimensional field
theories see \cite{RS,RSS2,RS2}.}. We do not include hyperscalars in
our analysis as they do not mix with the rest and they cannot
contain tachyons: this a consequence of the fact that the 6D
potential has a global minimum in $\phi^{\a}=0$
\cite{Nishino:1986dc,Randjbar-Daemi:2004qr} and the Laplacian on the
internal space gives a positive contribution to the squared
hyperscalar masses.

We denote by $h_{MN}$, $V_M$ and $\sigma'$ the metric, gauge field and dilaton
fluctuations\footnote{The fluctuations are properly normalized in a way that their kinetic terms are canonical.}
 around the background. The expression of $G_3$ at the linear level in the fluctuations is
\be G_3=dV_2+2\bar{F}\wedge V,\ee
where the 2-form fluctuation $V_2$, whose components are $V_{MN}$, is defined by
\be V_2\equiv \kappa \left(B_2-\bar{\mathcal{A}}\wedge V\right)\ee
and $\bar{F}$ is the field strength of $\bar{\mathcal{A}}$ defined in (\ref{spheremonopole}).
The bilinear bosonic action for the fluctuations
$h_{MN}$, $V_M$, $V_{MN}$ and $\sigma'$ around the background (\ref{spheremonopole}) reads
\bea S_{B2}&=&\int d^6X \sqrt{-G}\left\{\frac{1}{4}h_{MN}\nabla^2h^{MN}-\frac{1}{8}h\nabla^2 h \right.
\nonumber \\
&&+\frac{1}{2}\nabla^{N}\left(h_{MN}-\frac{1}{2}G_{MN}h\right)\nabla_R\left(h^{MR}-\frac{1}{2}G^{MR}h\right)
+\frac{1}{2}R_{MN}h^{MR}h^N_R\nonumber \\
&&+\kappa \bar{F}_{MN}\left[-\frac{1}{2}\nabla^M V^N (h+\sigma') + \left( \nabla_RV^N-\nabla^NV_R\right)h^{RM}\right]\nonumber\\
&&+\frac{1}{2}V_M\nabla^2 V^M+\frac{1}{2}R_{MN}V^{M}V^N+\frac{1}{2}\left(\nabla_M V^M\right)^2
-\bar{g}\bar{F}_{MN}V^M\times V^N\nonumber\\
&&-\frac{1}{48}\left(\nabla_{[M}V_{NR]}\right)^2-\frac{\kappa}{12}\nabla_{[M}V_{NR]}V^{[M}\bar{F}^{NR]}
-\frac{\kappa^2}{12}\left(V_{[M}\bar{F}_{NR]}\right)^2\nonumber\\
&&\left.+\frac{1}{4}\sigma'\left(\nabla^2-\frac{1}{a^2}\right)\sigma'-\frac{1}{2}\sigma' R_{MN}h^{MN}\right\}, \label{SB2}\eea
where the indices are raised and lowered by the background metric
$G_{MN}$, moreover $h\equiv h^M_M$ and we have introduced the
following notations
\be \nabla_{[M}V_{NR]}\equiv \nabla_M V_{NR}+ \nabla_N V_{RM} +\nabla_R V_{MN},\ee
\be (V_M\times V_N)^I=-f^{JKI}V^J_MV_N^K,\ee
where $f^{JKI}$ are the structure constants\footnote{We take the generators $T^I$ of $\mathcal{G}$ satisfying
$[T^I,T^J]=if^{IJK}T^K$ and $Tr(T^I T^J)=\delta^{IJ}$.} of $\mathcal{G}$.

To study the physical spectrum in a proper way we have to remove the
gauge freedom of $S_{B2}$ under 6D diffeomorphisms and gauge
transformations. This can be achieved by fixing the light cone
gauge,
which is defined by $h_{-M}=V_{-}=V_{-M}=0$, where the $\pm$
components of a vector $A^M$ are $A^{\pm}\equiv
\frac{1}{\sqrt{2}}(A^0\pm A^3)$. The light cone gauge advantage is
the sectors with different helicities decouple and one can easily
find the physical degrees of freedom. In the light cone gauge the
action (\ref{SB2}) becomes
\bea S_{B2}&=&\int d^6X \sqrt{-G}\left\{\frac{1}{4}h_{ij}^t\nabla^2 h_{ij}^t
+\frac{1}{2}h_{i\alpha}\left(\nabla^2-\frac{1}{a^2}\right)h_{i\alpha}+\frac{1}{2}V_i\nabla^2 V_i\right.\nonumber \\
&&+\frac{1}{8}V_{i\a}\left(\nabla^2-\frac{1}{a^2}\right)V_{i\a}-\kappa \bar{F}_{\a\b}\nabla_{\b}V_i h_{i\a}
-\frac{\kappa}{2}\bar{F}_{\a\b}V_i\nabla_{\b}V_{i\a}\nonumber \\
&&-\frac{\kappa^2}{4}\left(V_i\bar{F}_{\a\b}\right)^2+\frac{1}{4}h_{\a\b}\left(\nabla^2-\frac{2}{a^2}\right)h_{\a\b}
+\frac{1}{8}h_{\a\a}\nabla^2 h_{\b\b}\nonumber \\
&&-\frac{\kappa^2}{2}\left(\bar{F}_{\a\b}V_{\beta}\right)^2-\kappa \bar{F}_{\a\b}\nabla_{\beta}V_{\gamma}h_{\a\gamma}
-\frac{\kappa}{2}\bar{F}_{\a\b}\nabla_{\a}V_{\b} \sigma' \nonumber \\
&&+\frac{1}{2}V_{\a}\left(\nabla^2-\frac{1}{a^2}\right)V_{\a}-\bar{g}\bar{F}_{\a\b}\left(V_{\a}\times V_{\b}\right)
-\frac{\kappa^2}{32}\left(V_{\a\b}\bar{F}_{\a\b}\right)^2\nonumber \\
&&+\frac{1}{16}V_{ij}\nabla^2V_{ij}+\frac{1}{16}V_{\a\b}\nabla^2V_{\a\b}
-\frac{\kappa}{2}\bar{F}_{\beta \gamma}V_{\a}\nabla_{\gamma}V_{\a\b}\nonumber \\
&&\left.+\frac{1}{4}\sigma'\left(\nabla^2-\frac{1}{a^2}\right)\sigma'+\frac{1}{2a^2}\sigma' h_{\a\a}\right\},\label{Slc}\eea
where $i,j,k,...$ label the transverse 4D coordinates ($i,j,k=1,2$), $\alpha, \b, \gamma,...$ label
a local orthonormal basis in the internal space and $h^t_{ij}\equiv h_{ij}-\frac{1}{2}\delta_{ij}h_{kk}$.
Apart from the hyperscalars, the complete bosonic 4D spectrum coming from the compactification (\ref{spheremonopole})
can be computed through (\ref{Slc}) by using
the harmonic expansion over $S^2$ \cite{RSS}, that we discussed in Subsection \ref{su2 x
u1}. The main result is the presence of a massless graviton,
 4D gauge fields of the group ${\cal H}$, which is defined in Subsection \ref{SSandAfree}, 4D gauge fields of the internal
manifold isometries and some scalar fields. Our aim is to
study the latter sector which is of course the possible source of tachyons.

\subsection{Stability Analysis}

To study the stability of background (\ref{spheremonopole}) we focus
on the helicity-0 terms of the action $S_{B2}$ in the light cone
gauge:
\bea S_0&=&\int d^6X\sqrt{-G}\left\{\frac{1}{4}h_{\a\b}\left(\nabla^2-\frac{2}{a^2}\right)h_{\a\b}
+\frac{1}{8}h_{\a\a}\nabla^2 h_{\b\b}\right.\nonumber \\
&&-\frac{\kappa^2}{2}\left(\bar{F}_{\a\b}V_{\beta}\right)^2-\kappa \bar{F}_{\a\b}\nabla_{\beta}V_{\gamma}h_{\a\gamma}
-\frac{\kappa}{2}\bar{F}_{\a\b}\nabla_{\a}V_{\b} \sigma' \nonumber \\
&&+\frac{1}{2}V_{\a}\left(\nabla^2-\frac{1}{a^2}\right)V_{\a}-\bar{g}\bar{F}_{\a\b}\left(V_{\a}\times V_{\b}\right)
-\frac{\kappa^2}{32}\left(V_{\a\b}\bar{F}_{\a\b}\right)^2\nonumber \\
&&+\frac{1}{16}V_{\a\b}\nabla^2V_{\a\b}
-\frac{\kappa}{2}\bar{F}_{\beta \gamma}V_{\a}\nabla_{\gamma}V_{\a\b}+\frac{1}{4}\sigma'\left(\nabla^2-\frac{1}{a^2}\right)\sigma'
\nonumber \\
&&\left. +\frac{1}{16}V_{ij}\nabla^2V_{ij}+\frac{1}{2a^2}\sigma' h_{\a\a}\right\}. \label{S0}\eea
The action (\ref{S0}) contains all the helicity-0 fields, including the helicity-0 components of spin-1 and spin-2 objects.
In particular (\ref{S0}) includes all the physical scalar fields which a priori could be tachyonic.

We observe that the field $V_{ij}$ do not contain tachyons because
its spectrum is simply $a^2M^2=l(l+1)$, $l=0,1,2,3$. The scalars
$V_{\a}$, coming from the 6D gauge field fluctuations, can be
decomposed in the following three pieces
\be V_{\a}=\left(V'_{\alpha},\,V^{0}_{\a},\,v_{\a}\right)\ee
where $V'_{\a}$ is along the generators of ${\cal H}$, $V^0_{\alpha}$ is along the monopole and $v_{\a}$ is along the
generators which do not commute with $Q$.
The simplest sector is $V'_{\a}$ since it does not mix with other
fields as it is clear from (\ref{S0}). The bilinear action for these
fields is simply
$\frac{1}{2}V'_{\a}\left(\nabla^2-1/a^2\right)V'_{\a}$
and the squared masses are $a^2M^2=l(l+1)$, with $l=1,2,3,...$; therefore here we do not have
tachyons. The sector including $V^0_{\a}$ is much more complicated as this field mixes with other degrees of freedom: the complete
sector is given by
\be (V^0_{\a},\,h_{\a\b},\, V_{\a\b},\, \sigma'), \label{V0}\ee
in the sense that the fields given in (\ref{V0}) mix each other but
do not mix with additional helicity-0 fields, as it can be easily
deduced from (\ref{S0}). So in general we have a mixing between the
dilaton, the scalars coming from the 6D gauge field along the
monopole and from the 2-form $B_2$ and the {\it graviscalars}. The
complete spectrum of this sector is given in Table \ref{V0table}. We
observe that there are no tachyons but we have a massless scalar
field which corresponds to the first row of Table \ref{V0table}.
\begin{table}[top]
\begin{center}
\begin{tabular}{|l|l|l|}
\hline {\bf $a^2M^2$}  & Multiplicity & range of l  \\ \hline
 $0$ & $1$ & $l=0$\\ \hline
 $2$ & $2$ & $l=0$ \\ \hline
$2$ & $3$ & $l=1$
\\
\hline $6$ & $2$ & $l=1$
\\ \hline $l(l+1)$ & $3$ & $l \geq 2$
\\ \hline
$l(l-1)$ & $2$ & $l\geq 2$ \\ \hline
$l(l+1)+2(l+1)$ & $2$ & $l\geq 2$ \\\hline
\end{tabular}
\end{center}\caption{\footnotesize  The spectrum of the $(V_{\a}^0,\,h_{\a\b},\, V_{\a\b},\, \sigma')$ sector, including
all the helicity-0 fields. The multiplicity in the second column is given in unit of $2l+1$ where $l$ is a non negative integer.
 \label{V0table}}
\end{table}

To study the stability properties of solution (\ref{spheremonopole})
the most interesting sector is $v_{\alpha}$ because in general it
can contain tachyons \cite{Avramis:2005qt}. Let us summarize the
result of \cite{Avramis:2005qt} as it will be useful for our
analysis. From formula (\ref{S0}) one can easily understand that
$v_{\a}$ does not mix with the rest and its bilinear action is
\be \int d^6 X
\sqrt{-G}\left\{\frac{1}{2}v_{\a}\left(\nabla^2-\frac{1}{a^2}\right)v_{\a}
-\bar{g}\bar{F}_{\a\b}\left(v_{\a}\times v_{\b}\right)\right\}. \ee
We denote by $T^i$ the generators of $\mathcal{G}$ which do not commute with $Q$ and we choose $T^i$ to be a basis of eigenvectors
of the adjoint representation of $Q$:
\be [Q,T^i]=q_iT^i.\ee
The $q$s represent the charges of $v_{\a}$ under\footnote{We remind
that $U(1)_M$ is the Abelian group in the direction of the
monopole.} $U(1)_M$. By using again the $S^2$ harmonic expansion one
finds the following KK tower
\be a^2M^2=l(l+1)-\left(\frac{nq}{2}\right)^2, \ee
where $l=|1\pm|nq/2||+k$, with $k=0,1,2,3,...$. Therefore we have
tachyons whenever $|nq|>1$. We observe that, for a given model, the
product $nq$ has to be an integer for each representation because of
the Dirac quantization condition. This implies that we have tachyons
whenever $|q|$ assumes more than one value.

Now we want to apply these results to all the present anomaly free models of this type
\cite{Randjbar-Daemi:1985wc,Avramis:2005qt,Avramis:2005hc,Suzuki:2005vu}. We summarize their structure as follows.

\begin{description}

\item[I] The first anomaly free model was given in \cite{Randjbar-Daemi:1985wc} and it has
$\mathcal{G}=E_7\times E_6 \times U(1)_R$ and $R_H={\bf(912,1)}$, where $R_H$ is the hyperinos representation.
\item[II] Another example is given in \cite{Avramis:2005qt} where $\mathcal{G}=E_7\times G_2 \times U(1)_R$
and $R_H={\bf(56,14)}$.
\item[III] In Ref. \cite{Avramis:2005hc} we have $\mathcal{G}=F_4\times Sp(9) \times U(1)_R$ and $R_H={\bf(52,18)}$.
\item[IV] Finally in Ref. \cite{Suzuki:2005vu} a huge number of anomaly free models was found with $\mathcal{G}$ given by
products of $U(1)$ and/or $SU(2)$ and particular hyperinos representations.

\end{description}
We observe that models I,II and III have an enough large
$\mathcal{G}$ to include the standard model gauge group, whereas the
models IV have not. However, the results of \cite{Suzuki:2005vu}
prove that $\mathcal{G}$ do not have to include exceptional groups
from the pure mathematical point of view.

The stability analisis for models I and II was already done in
\cite{Randjbar-Daemi:1985wc,Avramis:2005qt,Avramis:2005hc} so we
briefly summarize the result. If we consider model I, the only
stable embedding of $Q$ in a non-Abelian algebra\footnote{The
embedding of $Q$ in an abelian algebra, that is $Q=Lie(U(1))$, is
obviously stable because all the $q_i$ vanish.} is $Q\subset
Lie(E_6)$. Whereas, in model II, we have no non-Abelian stable
embeddings.

The stability analysis for the remaining models III and IV was never performed. So here we study these cases in detail.
Let us consider first model III. In this case the non-abelian embeddings can be $Q\subset Lie(F_4)$ or $Q\subset Lie(Sp(9))$.
In the case $Q\subset Lie(F_4)$ we have instability because the system of roots of $F_4$, which is nothing but all the possible
value of q by definition of roots, is given by
\bea &&(\pm1,\pm1,0,0)\nonumber \\
     &&(\pm1,0,\pm1,0)\nonumber \\
     &&(\pm 1,0,0,\pm1)\nonumber \\
     &&(0,\pm 1,\pm1,0)\nonumber \\
&&(0,\pm 1,0,\pm1)\nonumber \\
&&(0,0,\pm 1,\pm1)\nonumber \\&&(\pm 1,0,0,0)\nonumber \\&&(0,\pm 1,0,0)\nonumber \\&&(0,0,\pm 1,0)\nonumber \\
&&(0,0,0,\pm1)\nonumber \\&&(\pm \frac{1}{2},\pm \frac{1}{2},\pm \frac{1}{2},\pm \frac{1}{2}).\label{rootsF4}
\eea
Each column of system (\ref{rootsF4}) corresponds to a choice of $Q$ in the Cartan subalgebra of $F_4$, which is
indeed 4-dimensional. Each row of (\ref{rootsF4}) correspond to a value of $q$. We note that for every embedding of
$Q$ in the Cartan subalgebra of $F_4$ we have at least the values $q=1$ and $q=1/2$,
which is enough to conclude that the embedding $Q\subset Lie(F_4)$
is unstable.

We consider now  $Q\subset Lie(Sp(9))$ and we prove that this embedding is unstable as well.
The Lie algebra\footnote{For a more complete discussion
on the Lie algebra of $Sp(n)$ see for example \cite{Georgi}.} of $Sp(n)$ is generated by
\be I_2\times A, \quad \sigma_1 \times S_1, \quad \sigma_2\times S_2, \quad \sigma_3\times S_3, \ee
where $I_2$ is the $2\times 2$ identity matrix, $\sigma_i$ the Pauli matrices, $A$ a generic
antisymmetric $n\times n$ matrix, and $S_i$ are generic symmetric $n\times n$ matrices. We observe
that $I_2\times A$ and $\sigma_3\times S_3$ generate an $SU(n)\times U(1)$ subalgebra and we can take, without loss
of generality, the Cartan subalgebra of $Sp(n)$ equal to the Cartan subalgebra of $SU(n)\times U(1)$.
Of course we have several possibilities to embed $Q$ in such Cartan subalgebra. We consider first $Q=Lie(U(1))$.
In this case we observe that the representation ${\bf 18}$, which appears in the hyperinos representations,
 is the fundamental of $Sp(9)$ and we have the following tensor product
\be {\bf 2n}\times {\bf 2n}= {\bf Adj}+{\bf D^2}+{\bf  1 }, \label{tensorP}\ee
where ${\bf Adj}$ is the adjoin representation of $Sp(n)$ and ${\bf D^2}$ and ${\bf 1}$ are irriducible representations
coming from the antisymmetric part of ${\bf 2n}\times {\bf 2n}$. On the other hand we have the following branching
rules with respect to $Sp(n)\rightarrow SU(n)\times U(1)$
 \be {\bf 2n}\rightarrow {\bf n_1}+{\bf \bar{n}_{-1}}\label{branch1}\ee
and
\be {\bf Adj}\rightarrow {\bf Adj}_{SU(n)}+ {\bf R},\label{branch2}\ee
where ${\bf 1}$ and ${\bf -1}$ in (\ref{branch1}) represent the values of $q$ for the representation ${\bf 2n}$ in a particular
normalization\footnote{Of course the normalization of the generators is conventional and it cannot change
the final result.} and ${\bf R}$ in (\ref{branch2}) is some representation of $SU(n)$ which can be reducible or irreducible.
By putting (\ref{branch1}) and
(\ref{branch2}) in (\ref{tensorP}) we get
\be {\bf R}+ {\bf D^2}={\bf n_1}\times {\bf n_1}+{\bf \bar{n}_{-1}}\times {\bf \bar{n}_{-1}}+{\bf \bar{n}_{-1}}\times {\bf n_1}.\ee
Since
\be dim({\bf D^2})<dim({\bf n_1}\times {\bf n_1}+{\bf \bar{n}_{-1}}\times {\bf \bar{n}_{-1}}),\ee
necessarily ${\bf R}$ contains some representation with charge $|q|=2$ and therefore this embedding is unstable. The case
in which $Q\subset Lie(SU(n))$ can be studied in a similar way and this embedding turns out to be unstable as well.

Finally we consider the models IV and we focus on the case in which
$\mathcal{G}$ contains a non-Abelian $SU(2)$ subgroup. All the
hyperinos representations of models IV belong to a
$(2l+1)$-dimensional representations of $SU(2)$. By using a similar
argument we find a stable $Q\subset SU(2)$ embedding only if we have
no even $(2l+1)$-dimensional representation; this is a consequence
of the fact that even representations correspond to half-integer
spin, whereas the adjoint representation of $SU(2)$ has spin 1. In
Ref. \cite{Suzuki:2005vu} one model which satisfies this property is
given and it has $\mathcal{G}=SU(2)\times U(1)_R$ and the following
hyperinos representations: 7 representations ${\bf 3}$, 2
representations ${\bf 5}$ and 31 representations ${\bf 7}$.

\section{Delta-Function Singularities} \label{deltafn}\setcounter{equation}{0}
In this appendix we briefly review how the Ricci scalar acquires a
delta-function contribution in the presence of a deficit angle, and
examine what this implies for our choice of metric function $e^B$ in
(\ref{axisymmetric}). We are going to use some results presented in
Section \ref{C2B}.

 Let us consider the ansatz (\ref{axisymmetric},\ref{assumeeB}), and
illustrate the case for the deficit angle $\delta$ at $\rho=0$. Near
 $\rho=0$ the metric $ds_2^2$ of the 2D internal space can be
written as follows
\be
ds^2_2=d\rho^2+\left(1-\frac{\delta}{2\pi}\right)^2\rho^2d\varphi^2
\, .\ee
By using the change of coordinate
$r^{1-\delta/2\pi}/(1-\delta/2\pi)=\rho$, this metric becomes
\be
ds^2=r^{-\delta/\pi}\left(dr^2+r^2d\varphi^2\right).\label{s2}\ee
From (\ref{s2}) and (\ref{hE}) one can show
\be R=2\, \delta \, r^{\delta/\pi}
\delta^{(2)}\left(y\right)+...,\label{Rdelta}\ee
where the 2D vector $ y$ is defined by $y = (r \cos \varphi, r \sin
\varphi)$, $\delta^{(2)}$ is the 2D Dirac delta-function and the
dots are the smooth contributions\footnote{Eq. (\ref{Rdelta})
describes the asymptotic behaviour in the vicinity of the brane
\cite{Randjbar-Daemi:2004ni}. It seems that for positive $\delta$
the Ricci scalar vanishes at the origin.
 However, the first term on the
right hand side of (B.3) should be interpreted in a distributional
sense. The effect of the Ricci scalar as a distribution on a scalar
test function $f(y)$ is then $\int d^2y \sqrt{g} R f(y) = 2 \,
\delta f({\bf 0}) + \dots$.}. On the other hand, near to $\rho=0$
the Ricci scalar, $R$, can be expressed in terms of derivatives of
$B$:
\be R=-B''-\frac{1}{2}(B')^2, \label{RB}\ee
where $' \equiv \partial_{\rho}$, and from (\ref{Rdelta}) and
(\ref{RB}) it follows that
\be B''=-2\, \delta \, r^{\delta/\pi}
\delta^{(2)}\left(y\right)+...,\label{Bdelta}\ee That is, the metric
function $e^B$ contains a delta-function contribution in its second
order derivative with respect to $\rho$. In coordinate system
(\ref{udef}), (\ref{Bdelta}) becomes
\be \partial^2_u B=-2\, \delta \, r^{\delta/\pi}
\delta^{(2)}\left(y\right)+...,\label{B2u}\ee 

The delta-function in the curvature also gives rise to a
delta-function in the derivative of the spin connection
(\ref{spinconnection}).  The Riemann tensor is defined in terms of
the spin connection as: \be
 R_{\,\,\,\,B}^{\,\,A}=d\omega_{\,\,\,\,B}^{\,\,A}+\omega_{\,\,\,\,C}^{\,\,A}
\wedge\omega_{\,\,\,\,B}^{\,\,C} \,. \label{Riemann-conn}\ee Near to
the brane $\rho=0$, relation (\ref{Riemann-conn}) gives $R_{\rho
\varphi\,\,\,6}^{\,\,\,5}=\partial_{\rho}\omega_{\varphi\,\,\,6}^{\,\,5}$,
and from the expression for the spin-connection
(\ref{spinconnection}):
\be \Omega'=-\left(\frac{1}{2}B''+\frac{1}{4}(B')^2\right)e^{B/2},
\label{Oprime}\ee
thus leading to the delta-function behaviour from (\ref{Bdelta}).

These results must be recalled when obtaining the Schroedinger-like
equations that govern the fluctuations.

\section{Imposing Boundary Conditions } \label{boundcond}\setcounter{equation}{0}

Here we study the implications of the NC and the HC for gauge field
fluctuations.  To impose the boundary conditions the following
properties will be useful:
\bea &&\quad F(a,b,c,z)\stackrel{z\rightarrow0}{\rightarrow}1,\label{z1} \\
&&F(a,b,c,z)=\Gamma_1 F(a,b,a+b-c+1,1-z)\nonumber \\
&&+\Gamma_2 (1-z)^{c-a-b}F(c-a,c-b,c-a-b+1,1-z),\label{1-z} \eea
where
\be \Gamma_1\equiv
\frac{\Gamma(c)\Gamma(c-a-b)}{\Gamma(c-a)\Gamma(c-b)}, \quad
\Gamma_2\equiv \frac{\Gamma(c)\Gamma(-c+a+b)}{\Gamma(a)\Gamma(b)},
\label{G11}\ee
and $\Gamma$ is the Euler gamma function. The relation (\ref{1-z})
is valid if $c-a-b$ is not an integer \cite{whittaker} and $y_1$ and
$y_2$ in (\ref{y12}) are both well defined when $c$ is not an
integer. In general $c-a-b$ and $c$ are not integers for generic
$\omega$ and $\ob$; so we can consider $\omega$ and $\ob$ as
regulators to use (\ref{y12}) and (\ref{1-z}) and at the end we can
take the limits in which $c-a-b$ and $c$ go to an integer, which
will turn out to be well defined.

We first consider the behaviour of $\psi$ for $u\rightarrow
\overline{u}$, that is $z\rightarrow 0$ because of the definition
$z=\cos^2\left(\frac{u}{r_0}\right)$. For $c\neq 1$ we use the
expression for $\psi$ given in (\ref{cnot1}) and property (\ref{z1})
gives us
\be \psi \stackrel{u\rightarrow
\overline{u}}{\rightarrow}K_1(\overline{u}-u)^{2\gamma}+K_2(\overline{u}-u)^{
1-2\gamma},\label{uub}\ee
where we used $c=1/2+2\gamma$. So the NC (\ref{NCV}) implies $K_1=0$
when $\gamma\leq -1/4$ and $K_2=0$ when $\gamma\geq 3/4$. On the
other hand the HC (\ref{HCV2}) implies\footnote{In fact, for the
gauge field sector, each term in the HC (\ref{HCV2}) is separately
zero if one requires their finiteness.  Therefore (\ref{HCa}) and
its counter-part for $u=0$ are sufficient to ensure (\ref{HCV2}).
On the other hand, for the fermions, there are some cases in which
they are each finite and non-zero, and so (\ref{HCV2}) requires that
they cancel.}
\be \lim_{u\rightarrow
\overline{u}}\psi^*\left(-\partial_u+\frac{1}{2}\frac{1}{u-\overline{u}}
\right)\psi<\infty\label{HCa}\ee
and by using the behaviour (\ref{uub}) this limit becomes
\be \left(2\gamma -\frac{1}{2}\right)\lim_{u\rightarrow
\overline{u}}\left[|K_1|^2(\overline{u}-u)^{4\gamma-1}
-K_1^*K_2+K_1K_2^*-|K_2|^2(\overline{u}-u)^{-4\gamma+1}\right],\ee
so the HC implies $K_1=0$ when $\gamma <1/4$ and $K_2=0$ when
$\gamma >1/4$. The case $\gamma=1/4$ corresponds to $c=1$ and so we
have to use the expression of $\psi$ given in (\ref{wronskian}). We
have then
\be \psi \stackrel{u\rightarrow
\overline{u}}{\rightarrow}K_1(\overline{u}-u)^{1/2}-K_2(\overline{u}-u)^{1/2}
\ln(\overline{u}-u)\ee
for which (\ref{HCa}) implies $K_2=0$. Therefore we obtain
(\ref{psi1}) and (\ref{psi2}).

The discreteness of the spectrum emerges when we impose the NC and
HC for $u\rightarrow 0$. For instance for $m\geq N_V$, up to an
overall constant, the behaviour of $\psi$ is given by properties
(\ref{z1}) and (\ref{1-z}):
\be  \psi \stackrel{u\rightarrow
0}{\rightarrow}\Gamma_1u^{2\b}+\Gamma_2u^{1-2\b},\label{psib}\ee
where $\Gamma_{1,2}$ are defined in (\ref{G11}) and we used
$c-a-b=1/2-2\b$. Behaviour (\ref{psib}) is similar to (\ref{uub})
but $\gamma$ is replaced by $\beta$. So, following the same steps as
above, the NC and the HC imply that $\Gamma_1=0$ for $\beta<1/4$ and
$\Gamma_2=0$ for $\beta>1/4$. Let us study the case $m\geq N_V$ and
$\b<1/4$, that is
\be N_V\leq m < 0.\label{assumem}\ee
 We then have
\be 0=\Gamma_1\equiv
\frac{\Gamma(c)\Gamma(c-a-b)}{\Gamma(c-a)\Gamma(c-b)}.\ee
Since the Euler gamma function never vanishes we require that
$\Gamma(c-a)=\infty$ or $\Gamma(c-b)=\infty$ and this is possible
only when $c-a=-n$ or $c-b=-n$, where $n=0,1,2,3,...$. By using the
definitions (\ref{gbetaabcV}) both conditions lead to the following
squared masses
\be
M_{V\,n,m}^2=\frac{4}{r_0^2}\left\{n(n+1)+\left(\frac{1}{2}+n\right)\left[
-m\omega+(m-N_V)\ob\right]\right\}\ee
which are positive because of (\ref{assumem}) and $n\geq 0$. When
$m\geq N_V$ and\footnote{The $\b =1/4$ case is recovered by taking
the limit $\omega \rightarrow 0$.} $\beta\geq 1/4$, that is
\be m\geq N_V \quad and \quad  m \geq 0,\label{assumem2}\ee
we have
\be 0=\Gamma_2\equiv
\frac{\Gamma(c)\Gamma(-c+a+b)}{\Gamma(a)\Gamma(b)}\ee
and this implies $a=-n$ or $b=-n$, where $n=0,1,2,3,...$. The
corresponding squared masses are
\be
M_{V\,n,m}^2=\frac{4}{r_0^2}\left\{n(n+1)+\left(\frac{1}{2}+n\right)\left[
m\omega+(m-N_V)\ob\right]+m\omega(m-N_V)\ob\right\}\ee
which are positive or vanishing. We can study the case $m<N_V$ in a
similar way. The complete result for the gauge fields sector is
given in equations (\ref {MV1})-(\ref{MV4}). We observe that
(\ref{HCV2}) is now automatically satisfied by every pair of wave
functions $\psi$ and $\psi'$, for a given quantum number $m$, since
the asymptotic behaviour of the wave function cannot depend on the
quantum number $n$: this is a consequence of the $1/u^2$ and
$1/(\overline{u}-u)^2$ singularities of the potential $V$ in
(\ref{SchroedingerVGGP}).

\section{Complete Fermionic Mass Spectrum} \label{fermspectrum}\setcounter{equation}{0}

In this appendix we give the complete fermionic spectrum which is
also labeled by an integer quantum number $n=0,1,2,...$. Although
much longer, the calculation proceeds in exactly the same way as for
the gauge field sector, outlined in the previous
appendix\footnote{There is one additional subtlety.  Here, for the
values of $m$ which allow a zero mode, we must impose a mixed HC
between the massless mode and massive modes, in addition to the
diagonal HC. In general the HC involving distinct wave functions
$\psi_{m n}$ and $\psi_{m n'}$ does not lead to additional
constraints, because the asymptotic behaviour of the modes is
independent of $n$. However, the massless modes are more strongly
constrained than the massive ones, obeying as they do a decoupled
Dirac equation in addition to the Schroedinger equation.}.

\begin{description}

\item[{\Large $m \geq -\frac{1}{2}+N+\frac{1}{2\ob}$}]: in this
case $K_2=0$ and we get the following squared masses.

\begin{itemize}
\item For $m>-\frac{1}{2}+\frac{1}{2\omega}$
\be
M^2_{F\,n,m}=\frac{4}{r_0^2}\left[\frac{1}{2}+n+\omega\left(\frac{1}{2}+m\right)\right]
\left[\frac{1}{2}+n +\ob\left(\frac{1}{2}+m-N\right)\right]>0.
\label{m1} \ee
\item For
$-\frac{1}{2}-\frac{1}{2\omega}<m<-\frac{1}{2}+\frac{1}{2\omega}$
\be
M^2_{F\,n,m}=\frac{4}{r_0^2}\left[\frac{1}{2}+n+\omega\left(\frac{1}{2}+m\right)\right]
\left[\frac{1}{2}+n +\ob\left(\frac{1}{2}+m-N\right)\right]>0.
\label{m21} \ee
or
\be
M^2_{F\,n,m}=\frac{4}{r_0^2}(1+n)\left[1+n+\ob\left(\frac{1}{2}+m-N\right)-\omega\left(\frac{1}{2}+m\right)\right]>0.\ee
\item For $m\leq -\frac{1}{2}-\frac{1}{2\omega}$
\be
M^2_{F\,n,m}=\frac{4}{r_0^2}(1+n)\left[1+n+\ob\left(\frac{1}{2}+m-N\right)-\omega\left(\frac{1}{2}+m\right)\right]>0.
\label{m4}\ee
\end{itemize}
\item[{\Large $m \leq -\frac{1}{2}+N-\frac{1}{2\ob}$}]: in this
case $K_1=0$ and we get the following squared masses.

\begin{itemize}
\item For $m>-\frac{1}{2}+\frac{1}{2\omega}$
\be
M^2_{F\,n,m}=\frac{4}{r_0^2}n\left[n-\ob\left(\frac{1}{2}+m-N\right)
+\omega\left(\frac{1}{2}+m\right)\right]\geq 0.  \ee
\item For
$-\frac{1}{2}-\frac{1}{2\omega}<m<-\frac{1}{2}+\frac{1}{2\omega}$
\be
M^2_{F\,n,m}=\frac{4}{r_0^2}n\left[n-\ob\left(\frac{1}{2}+m-N\right)
+\omega\left(\frac{1}{2}+m\right)\right]\geq 0. \label{02} \ee
%
%
%
\item For $m\leq -\frac{1}{2}-\frac{1}{2\omega}$
\be
M^2_{F\,n,m}=\frac{4}{r_0^2}\left[\frac{1}{2}+n-\omega\left(\frac{1}{2}+m\right)\right]\left[
 \frac{1}{2}+n-\ob\left(\frac{1}{2}+m-N\right)\right]>0. \ee
\end{itemize}

\item[{\Large $-\frac{1}{2}+N-\frac{1}{2\ob}<m <
-\frac{1}{2}+N+\frac{1}{2\ob}$}]: this case is possible only when
$\overline{\delta}<0$.

\begin{itemize}
\item For $m>-\frac{1}{2}+\frac{1}{2\omega}$ we have $K_1=0$ and
%
%
%
\be
M^2_{F\,n,m}=\frac{4}{r_0^2}n\left[n-\ob\left(\frac{1}{2}+m-N\right)
+\omega\left(\frac{1}{2}+m\right)\right]\geq 0.\ee
\item For $m\leq -\frac{1}{2}-\frac{1}{2\omega}$ we have two
possibilities. We have $K_2=0$ and
\be
M^2_{F\,n,m}=\frac{4}{r_0^2}(1+n)\left[1+n+\ob\left(\frac{1}{2}+m-N\right)-\omega\left(\frac{1}{2}+m\right)\right]>0\ee
or $K_1=0$ and
\be
M^2_{F\,n,m}=\frac{4}{r_0^2}\left[\frac{1}{2}+n-\omega\left(\frac{1}{2}+m\right)\right]\left[
 \frac{1}{2}+n-\ob\left(\frac{1}{2}+m-N\right)\right]>0\ee
\item $-\frac{1}{2}-\frac{1}{2\omega}<m<-\frac{1}{2}+\frac{1}{2\omega}$:
this case is possible only when $\delta<0$ and we get $K_1=0$ and
\be
M^2_{F\,n,m}=\frac{4}{r_0^2}n\left[n-\ob\left(\frac{1}{2}+m-N\right)
+\omega\left(\frac{1}{2}+m\right)\right]\geq 0. \label{mf}\ee
\end{itemize}
\end{description}

Again, we can perform a check of our results by considering the
$S^2$ limit ($\omega\rightarrow 1$, $\ob\rightarrow 1$).  In this
case, the mass spectrum (\ref{m1})-(\ref{mf}) reduces correctly to
\be
a^2M^2_F=\left(l+\frac{1+N}{2}\right)\left(l+\frac{1-N}{2}\right),\quad
multiplicity=2l+1\ee
where $a=r_0/2$ is the radius of $S^2$ and\footnote{The number $l$
is defined in different ways in equations (\ref{m1})-(\ref{mf}). For
instance we have $l\equiv n+m+(1-N)/2$ for (\ref{m1}) and $l\equiv
1/2+n-N/2$ for (\ref{m4}).} $l=\frac{|N|-1}{2}+k$ and
$k=0,1,2,3,...$.

\end{document}